\documentclass[twocolumn]{aastex62}

\usepackage{float}
\usepackage{amsmath}
\usepackage{afterpage}

\newcommand{\fdeg}{.\!\!^\circ}
\def\etal{{et~al.\null}}

\graphicspath{{./}{figures/}}

\newcommand{\GALEX}{{\it GALEX}}
\newcommand{\Gaia}{{\it Gaia}}

\newcommand{\HST}{{\it HST}}

\newcommand{\Teff}{T_{\rm eff}}

\newcommand{\uBVI}{{\it uBVI}}

\def\etal{{et~al.\null}}

\newcommand{\oCen}{$\omega$~Cen}

\def\tauhb{\tau_{\rm HB}}

\submitjournal{ApJS}

\shorttitle{Above-Horizontal-Branch Stars in Galactic Globular Clusters}

\shortauthors{Davis \etal}

\begin{document}

\title{A Census of Above-Horizontal-Branch Stars in Galactic Globular Clusters\footnote{H.E.B. dedicates this paper to the memory of George Wallerstein (1930 January 13--2021 May 13), pioneer in the study of Population II Cepheids and stars above the horizontal branch in globular clusters, and a friend and mentor for a half century.}}



\correspondingauthor{Brian D. Davis}
\email{bdavis@psu.edu}

\author[0000-0002-8994-6489]{Brian D. Davis}
\affil{Department of Astronomy \& Astrophysics, The Pennsylvania
State University, University Park, PA 16802, USA}

\author[0000-0003-1377-7145]{Howard E. Bond}
\affil{Department of Astronomy \& Astrophysics, The Pennsylvania
State University, University Park, PA 16802, USA}
\affil{Space Telescope Science Institute, 3700 San Martin Dr., Baltimore, MD 21218, USA}
\affil{Visiting astronomer, Cerro Tololo Inter-American Observatory and Kitt Peak National Observatory, National Optical Astronomy Observatory, which are operated by the Association of Universities for Research in Astronomy under a cooperative agreement with the National Science Foundation.}

\author[0000-0003-1817-3009]{Michael H. Siegel}
\affil{Department of Astronomy \& Astrophysics,
The Pennsylvania
State University, University Park, PA 16802, USA}

\author[0000-0002-1328-0211]{Robin Ciardullo}
\affil{Department of Astronomy \& Astrophysics,
The Pennsylvania
State University, University Park, PA 16802, USA}
\affil{Institute for Gravitation and the Cosmos, The Pennsylvania
State University, University Park, PA 16802, USA}

\begin{abstract}

We have carried out a search for above-horizontal-branch (AHB) stars---objects lying above the horizontal branch (HB) and blueward of the asymptotic giant branch (AGB) in the color-magnitude diagram---in 97 Galactic and seven Magellanic Cloud globular clusters (GCs). We selected AHB candidates based on photometry in the \uBVI\/ system, which is optimized for detection of low-gravity stars with large Balmer jumps, in the color range $-0.05\le(B-V)_0\le1.0$. We then used \Gaia\/ astrometry and Gaussian-mixture modeling to confirm cluster membership and remove field interlopers. Our final catalog contains 438 AHB stars, classified and interpreted in the context of post-HB evolution as follows: (1)~AHB1: 280 stars fainter than $M_V=-0.8$, evolving redward from the blue HB (BHB) toward the base of the AGB\null. (2)~Post-AGB (PAGB): 13 stars brighter than $M_V\simeq-2.75$, departing from the top of the AGB and evolving rapidly blueward. (3)~AHB2: 145 stars, with absolute magnitudes between those of the AHB1 and PAGB groups. This last category includes a mixture of objects leaving the extreme BHB and evolving toward the AGB, and brighter ones moving back from the AGB toward higher temperatures. Among the AHB1 stars are 59 RR~Lyrae interlopers, observed by chance in our survey near maximum light. PAGB and AHB2 stars (including W~Virginis Cepheids) overwhelmingly belong to GCs containing BHB stars, in accordance with predictions of post-HB evolutionary tracks. We suggest that most W~Vir variables are evolving toward lower temperatures and are in their first crossings of the instability strip. Non-variable yellow PAGB stars show promise as a Population~II standard candle for distance measurement.

\end{abstract}


\section{Introduction}
\label{sec:intro}

\subsection{Stars Above the Horizontal Branch in Globular Clusters} 
\label{subsec:ahb_intro}

\begin{figure*}[ht]
\centering
\includegraphics[width=4.5in]{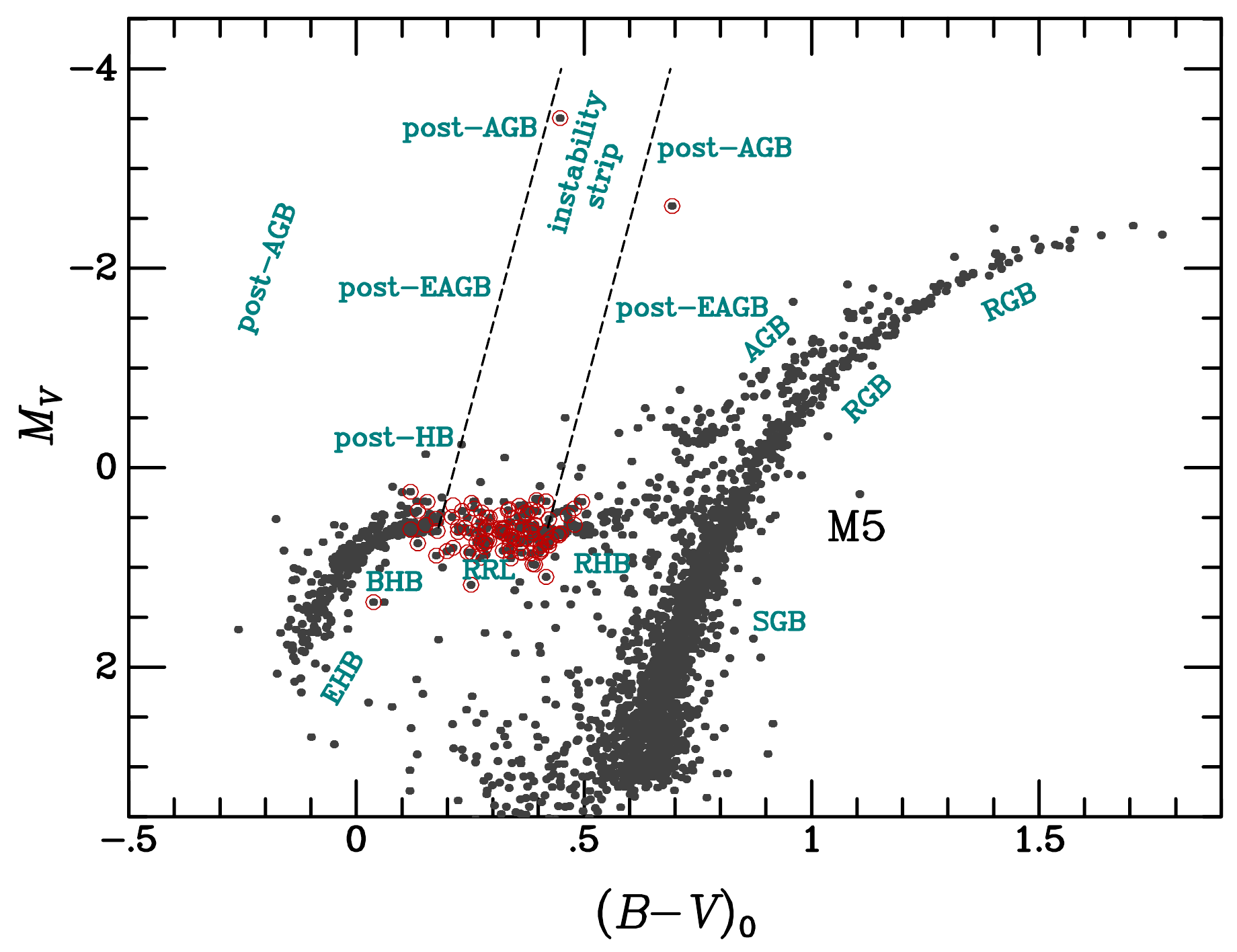}
\caption{
The $M_V$ versus $B-V$ color-magnitude diagram for the globular cluster M5, showing only probable cluster members, based on data from this paper.  Our photometry has been corrected for a foreground reddening of $E(B-V)=0.03$, and assumes a distance modulus of $(m-M)_V=14.46$ (H10). The nominal location of the pulsational instability strip is indicated. Known variable stars \citep{Clement2001} are circled in red. Evolutionary stages are labelled as follows: subgiant branch (SGB), red-giant branch (RGB), horizontal branch [subdivided into the extreme (EHB), blue (BHB), RR Lyr (RRL), and red (RHB) regions], and asymptotic giant branch (AGB\null).  The nominal CMD locations of rare stars that are in three different stages of post-HB evolution are labeled: the post-HB region, which is inhabited by stars that have just left the HB and are evolving redward towards the AGB; the post-early AGB (post-EAGB) region, which can contain stars in a variety of evolutionary states and includes objects that have left the AGB before undergoing helium shell pulses; and the post-AGB stage, where stars that have evolved off the AGB tip are located.  M5 contains two post-AGB stars (an RV~Tauri variable and a W~Vir Cepheid), numerous RRL variables, and several post-HB stars, but no post-EAGB objects. Since our data were taken at a small number of epochs, rather than averaged over time, some variable stars appear to lie outside the instability strip.  Non-variable stars within the instability strip are most likely produced by the blended light of two objects with different effective temperatures.
\label{fig:schematic}
}
\end{figure*}

Color-magnitude diagrams (CMDs) of the bright members of Galactic globular clusters (GCs) are dominated by stars lying on the subgiant branch (SGB), red-giant branch (RGB), horizontal branch (HB), and asymptotic giant branch (AGB)\null. Figure~\ref{fig:schematic} illustrates an observational CMD (visual absolute magnitude, $M_V$, versus $B-V$ color) for a typical GC, M5 (NGC\,5904), based on data from our work; non-member field stars have been excluded using the techniques described below. As explained in the figure caption, the data points have been corrected for reddening and distance using values taken from \citet[][hereafter H10]{Harris2010}.\footnote{The Harris compilation of GC properties, 2010 December version, is available online at \url{http://physwww.mcmaster.ca/~harris/mwgc.dat}} Labels in the figure indicate the positions in the CMD of the stars in the various evolutionary stages just mentioned. The nominal location of the pulsational instability strip is shown; it is based on Figure~3 of \citet{Harris1983}, and is only approximate. We chose M5 for this illustration because of the wide range of colors seen on its HB\null. Stars in this phase can be subdivided on the basis of color and temperature into the extreme HB (EHB), blue HB (BHB), RR~Lyrae variables (RRL), and red HB (RHB).

In this paper we present an observational search of the Galactic GC system for rare luminous stars that {\it do not\/} lie on these principal sequences. It has been known for many decades that the CMDs of GCs occasionally exhibit stars lying above the HB and blueward of the AGB\null. Among early and well-known examples of these objects are K\"ustner~648, the central star of the planetary nebula Ps~1 in M15 \citep{Pease1928}; the bright blue stars von~Zeipel 1128 in M3 and Barnard 29 in M13\footnote{To our knowledge, \citet{Popper1947} was the first to discuss the early-type spectrum of Barnard~29.  However, almost five decades earlier, \citet{Barnard1900}  himself had pointed out the extremely blue color of the star, based on his comparison of blue-sensitive photographs with direct visual examination.}; and the luminous, low-gravity F-type star HD~116745 (``Fehrenbach's Star,'' ROA~24) in $\omega$~Centauri \citep{Harding1965,Sargent1965}, which is the only GC object bright enough to be listed in the \textit{Henry Draper Catalogue}. The Type~II Cepheids (BL~Herculis, W~Virginis, and RV~Tauri stars) in GCs also belong to a category of intermediate-temperature GC stars that are brighter than the HB.   

Following \citet{Strom1970}, \citet{Sandage2006}, and other authors, we call these luminous objects ``above-horizontal-branch'' (AHB) stars.\footnote{Here we are using the designation ``AHB'' for {\it all\/} GC stars lying more than $\sim$0.5~mag above the HB and blueward of the AGB\null. \citet{Strom1970} actually distinguished AHB objects from the hotter and brighter ``von~Zeipel 1128-like'' stars. AHB stars have also been called ``supra-horizontal-branch'' or ``UV-bright'' stars in the literature. Later in this paper (\S\ref{subsec:classification}) we will subdivide the AHB stars into several classification boxes.} In Figure~\ref{fig:schematic} we assign AHB stars to three subgroups, as indicated by labels in the figure: (1)~post-HB stars, which lie within $\sim$1 mag of the HB; (2)~post-AGB (PAGB) stars, which are located at least 3~mag above the HB; and (3)~post-early-asymptotic-giant-branch (PEAGB) stars, which lie in between the post-HB and PAGB, and blueward of the AGB\null. (However, later in this paper, we will argue that most objects in the ``PEAGB'' region are actually stars evolving off the blue end of the HB and {\it toward\/} the AGB.) Figure~\ref{fig:schematic} shows that there are two luminous PAGB stars in M5, both of them variables, as described in the figure caption.\footnote{M5 also contains a hot (44,300~K) PAGB star, ZNG~1 \citep{Dixon2004}; this object is not plotted in our figure, as it is optically faint and blended with a bright AGB star lying only $0\farcs52$ away. Based on its luminosity and effective temperature, ZNG~1 would lie at about $B-V=-0.31$ and $M_V=0.0$.} M5 also hosts several post-HB stars, but does not contain any objects in the PEAGB region of the CMD.

The stellar sequences in GCs are interpreted in terms of the evolution of stars with initial masses of about $0.8\,M_\odot$. After fusing hydrogen in their cores for long intervals, the stars leave the main sequence and ascend the SGB and RGB, as hydrogen is exhausted in their centers. At the tip of the RGB, the core helium ignites, lifting its degeneracy, and the star moves onto the zero-age horizontal branch (ZAHB)\null. Since this ignition occurs at virtually the same core mass in all low-mass stars \citep[$\sim\!\!0.47\,M_{\odot}$;][]{Sweigart1978}, the star's location on the ZAHB depends almost exclusively on its envelope mass at the RGB tip. ZAHB stars with the highest envelope masses fall onto the RHB, while stars with slightly lower envelope masses land within the pulsational instability strip, becoming RRL variables. At lower envelope masses, the stars' effective temperatures are higher, and the ZAHB position falls onto the BHB ($\Teff\lesssim20,000$~K) or the EHB\null. The distribution of stars along the ZAHB varies widely from cluster to cluster, depending primarily on metallicity: relatively metal-rich clusters generally (but not always) have red HBs, while metal-poor GCs often have HBs dominated by hot BHB stars. The existence of GCs that do not follow this paradigm, as first pointed out by \citet{Sandage1960}, indicates that a ``second parameter'' other than metallicity can influence the morphology of cluster HBs. For reviews of HB stars and post-HB evolution, see, for example, \citet{Dorman1995}, \citet{Greggio1999}, \citet{Moehler2001}, \citet{Catelan2009}, \citet{Lagioia2015}, \citet{Heber2016}, \citet[][hereafter M+19]{Moehler2019}, and \citet{Bono2020}. General reviews of AGB and PAGB stars are given by \citet{vanWinckel2003}, \citet{Herwig2005}, and \citet{vanWinckel2011}. 

\subsection{Post--Horizontal-Branch Stellar Evolution\label{subsec:post-hb_evolution}}

The 1960s brought the realization that the ZAHB is the locus of low-mass, post-RGB stars burning helium in their cores and hydrogen in a surrounding shell (e.g., \citealt{Faulkner1966}; \citealt{IbenRood1970}; \citealt{Strom1970} and references therein). The basic features of the subsequent evolution of these objects are as follows. When the core helium of ZAHB stars is exhausted, the stars' luminosities begin to increase, and they enter the AHB region in the CMD\null. The hottest EHB objects burn through their envelope during this phase and evolve directly to the white-dwarf (WD) cooling sequence, becoming so-called ``AGB-manqu\'e'' stars. Cooler BHB objects increase their luminosity, cross the AHB region in the CMD, and begin to ascend the giant branch a second time, becoming AGB stars. If the envelope mass is sufficiently low, shell burning reaches the surface before the onset of thermal pulsing, and the star evolves back to higher temperatures as a PEAGB object. At still higher envelope mass, stars begin to undergo thermal pulses (TPs), and increase their mass-loss rate to $\sim$$10^{-5} M_{\odot}$~yr$^{-1}$.  Finally, when the envelope mass falls below $\sim$1\% of the total mass, these stars leave the AGB at a high luminosity and rapidly cross the CMD to higher temperatures as PAGB stars.  Eventually, the envelope hydrogen is exhausted, and the stars, which are now at the top of the WD sequence, begin to cool. They join the other post-HB stars in spending the rest of eternity descending WD cooling tracks.

Extensive grids of post-ZAHB evolutionary tracks have been computed by several authors over the past many decades, including, among others, \citet{Paczynski1971}, \citet{IbenRood1970}, \citet{Sweigart1976}, \cite{Sweigart1987}, \citet{Castellani1989}, \citet{Lee1990}, \citet{Dorman1993}, \citet{Brown2008}, and M+19. The evolution of PAGB stars that have already ascended to the tip of the AGB and are evolving blueward at high luminosity has been modeled in the classical papers of \citet{Schoenberner83}, \citet{Bloecker1995}, \citet{Vassiliadis1994}, and more recently by \citet{MillerBertolami2016}.

\begin{figure*}[htb]
\centering
\includegraphics[width=4.5in]{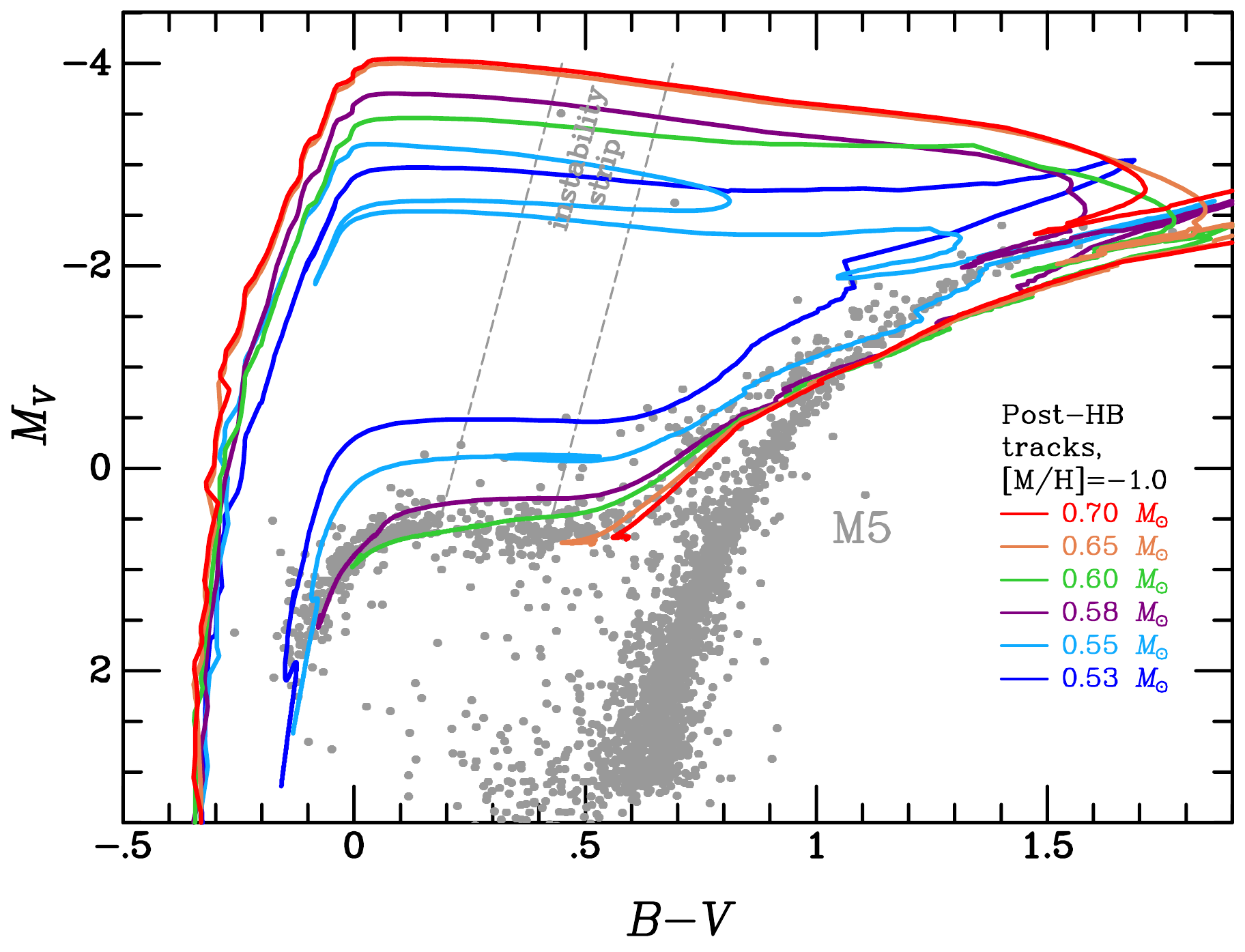}
\caption{
The M5 CMD of Figure~\ref{fig:schematic} superposed with theoretical post-HB evolutionary tracks from \citet{Moehler2019}. These tracks have a metallicity of $\rm[M/H]=-1.0$, and are for the six ZAHB masses indicated in the legend.  The AHB stars within $\sim$1~mag of the ZAHB are seen to be post-BHB and post-EHB objects evolving toward the AGB, and the two luminous variable stars at the top of the CMD are on post-AGB tracks. The AHB region lying between the post-HB and post-AGB stars could theoretically be populated by stars evolving from the hot end of the EHB, but M5 contains few such stars. Note that the tracks near the top of the AGB are redder than the stars in M5; this is because the metallicity of M5 \citep[{[Fe/H]}$=-1.29$;][]{Harris2010} is somewhat lower than that of the tracks.
\label{fig:schematicwithtracks}
}
\end{figure*}

Figure~\ref{fig:schematicwithtracks} illustrates how stars in post-ZAHB evolutionary stages populate the AHB region of the CMD\null. Here we repeat the M5 CMD data from Figure~\ref{fig:schematic}, and superpose theoretical evolutionary tracks from the grid computed recently by M+19.\footnote{We thank Marcelo Miller Bertolami for sending us detailed tables of his tracks with a finer time resolution than given in the M+19 paper.} The tracks plotted in Figure~\ref{fig:schematicwithtracks} are for a metallicity of $\rm[M/H]=-1.0$, and a ZAHB mass range of $0.53 < M/M_{\odot} < 0.70$, as indicated in the figure legend. To convert these tracks from the theoretical parameters of effective temperature, $\log \Teff$, and luminosity, $\log L/L_\odot$, to the observational $(B-V),M_V$ plane, we used the online PARSEC YBC web tool\footnote{\url{http://stev.oapd.inaf.it/YBC/}} \citep{Chen2019}.

To simplify Figure~\ref{fig:schematicwithtracks}, we edited the M+19 evolutionary tracks to remove rapid excursions caused by helium shell flashes within the star. Thus the plotted tracks should be regarded as semi-schematic, illustrating the main features of the stars' evolution, but omitting short-timescale departures from the tracks. (These excursions are so transitory that very few stars would be expected to be observed in such stages.) We did, however, retain a slow excursion experienced by the $0.53\,M_\odot$ model just as it departs from the AGB toward the blue, as well as two thermal pulses in the $0.55\,M_\odot$ model, one on the AGB and the second as it reaches the top of the WD sequence. The more massive models experience more frequent shell flashes, especially near the AGB tip. 

Figure~\ref{fig:schematicwithtracks} shows that the AHB stars lying within $\sim$1~mag of the ZAHB are likely to be objects that started on the BHB and EHB, and are currently evolving across the CMD toward the base of the AGB\null. The two luminous variable stars near the top of M5's CMD appear to be in the PAGB phase, caught during their blueward evolution in the CMD\null.  Not shown in the figure are tracks that could populate the PEAGB region lying between these two groups; such tracks would arise from objects at the hot end of the EHB, which is not populated in M5.

A key feature of post-ZAHB evolution is that the traversal from the BHB across the CMD to the base of the AGB is rapid, and the blueward evolution of PEAGB and PAGB stars is even faster. As a result, AHB stars are relatively rare in the old populations of GCs---essentially, they are in the Population~II analog of the Hertzsprung Gap that is seen in younger stellar systems. We also note that the tracks in Figure~\ref{fig:schematicwithtracks} are all for single stars. In theory, binary-star evolution can also produce stars in the AHB region of the CMD, e.g., by stripping away the stellar envelope and causing an early departure from the RGB or AGB, or through stellar mergers.  Similarly, \citet{Catelan2009} points out that the transition time from the RGB tip to the ZAHB resulting from the helium core flash may be as long as $\sim\!10^6$~yr, providing another method of populating the AHB region of the CMD\null.  Additionally, several authors, including \citet[][and references therein]{Brown2012}, have discussed scenarios in which extreme mass loss on the RGB can cause a star to evolve straight to the WD cooling track before the onset of the helium core flash; if this occurs, the star  will move through the AHB region of the CMD\null.  However, stars experiencing these alternative stellar-evolution scenarios should be quite rare. Most of the objects we detect in the AHB region of the CMD will have recently left the HB and are now evolving toward the AGB.

Several decades ago, evolutionary tracks calculated for post-HB stars ascending the AGB sometimes exhibited excursions to the blue \citep[see, for example,][and references therein]{Gingold1976,Gingold1985}.  These stars would move into the AHB region of the CMD, cross the instability strip, and then return to the AGB\null.  Discussions at the time \citep[e.g.,][]{Wallerstein1984,Wallerstein2002} suggested that these ``blue loops'' or ``Gingold noses'' were the production mechanism for the W~Vir variables seen in GCs. However, in the past two-plus decades, evolutionary studies using updated physics have failed to produce such pronounced blue loops; see the discussion in \S3 of \citet{Bono2020}. 

The most luminous stars in GCs and other old populations are the objects that have departed the top of the AGB and are evolving at nearly constant bolometric luminosity toward higher temperatures. Because of the temperature dependence of bolometric corrections, such objects are brightest at optical wavelengths as they pass through the temperature range corresponding to colors of $0\lesssim B-V\lesssim0.5$ (see Figure~\ref{fig:schematicwithtracks}).  We call these luminous objects ``yellow PAGB stars,'' hereafter yPAGB stars. They are the {\it visually brightest stars\/} in old stellar populations.

\subsection{Surveys for AHB Stars in Globular Clusters}
\label{subsec:ahb_surveys}

Luminous AHB stars that lie within the instability strip---the Type~II Cepheids---are relatively easy to discover via their variability. Thus the census of such stars in Galactic globular clusters is likely close to  complete, except possibly for objects in the crowded central cores of distant, condensed systems, or in relatively little-studied clusters.  These known variables, which have been cataloged by \citet[][hereafter C01]{Clement2001},\footnote{The Catalog of Variable Stars in Globular Clusters is maintained and updated by Christine Clement and is available online at \url{http://www.astro.utoronto.ca/~cclement/read.html}.} are generally subclassified according to their pulsation periods: (1)~Type~II Cepheids with periods of about 1 to 5~days are defined as BL~Herculis objects; (2)~Type~II Cepheids with $\sim$5- to $\sim$20-day periods are W~Virginis variables; and (3)~stars with periods greater than 20 days are classified as RV~Tauri objects. (The nomenclature for these variables and the exact period boundaries differ among the various authors of an extensive literature; see, for example, the reviews by \citealt{Wallerstein1984, Wallerstein2002}, and \citealt{Sandage2006}, and papers by \citealt{Sandage1994}, \citealt{Soszynski2008}, \citealt{Bono2020}, and references therein.)

Also conspicuous in GCs, especially at short wavelengths, are the hot BHB and EHB stars and their immediate more luminous descendants, including the AGB-manqu\'e objects. These stars stand out in space-based ultraviolet (UV) images \citep[for example,][]{Hill1992, Parise1994, Brown2010, Schiavon2012, Siegel2014, Prabhu2021}, and in deep optical data taken through a blue filter to define a color index, such as $U-V$ or $B-R$ \citep[e.g.,][]{Randall2016, Latour2018}. 

In contrast, at temperatures lower than $\sim$12,000~K, the identification of non-variable AHB stars in Galactic GCs is much less complete, and suffers from considerable contamination. The principal reason is that, in most photometric systems, these bright stars are difficult to distinguish from the general foreground stellar population; this is especially true redward of the main-sequence turnoff of Galactic-halo stars at $B-V\simeq0.45$. Moreover, most modern optical surveys, especially those from space, are aimed at reaching the faintest cluster members; in these images, the bright AHB stars are saturated.

The first large-scale survey aimed at identifying AHB stars in Galactic GCs was the classical photographic study by \citet[][hereafter ZNG]{Zinn1972}. The ZNG team blinked photographs of 27 GCs obtained in the $U$ and $V$ bands, and identified 156 ``UV-bright'' candidates that were the brightest non-variable objects in the $U$ band. Over a decade later, \citet{deBoer1987} added two more stars to this list by using $u$ and $V$ filters and a CCD camera to search for UV-bright stars in the cores of nine GCs.  It should be noted, however, that the term ``UV-bright'' is slightly misleading, since it suggests high effective temperatures.  While some of the ZNG objects, such as Barnard~29, von~Zeipel 1128, and M5 ZNG~1, are indeed luminous, hot PAGB stars, others are designated ``UV-bright'' simply because they are brighter than most cluster members in the $U$ band---often because they are unrelated foreground stars that happen to be superposed on the cluster.  In a follow-up study, \citet{Zinn1974} obtained radial velocities (RVs) for a sample of the ZNG candidates, and found that only about 40\% had RVs consistent with cluster membership.  Similarly, \citet{Harris1983} determined RVs for the ZNG stars in two GCs, and found that a significant fraction of the candidates were interlopers. \citet{Harris1983} also presented a catalog and a composite CMD for candidate AHB stars in 29 Galactic GCs, including the known Type~II Cepheids. Their paper noted that cluster membership remained uncertain for a significant fraction of the candidates, and concluded that ``although our [composite CMD] is an improvement over earlier diagrams, it is still seriously incomplete due to selection effects.''

With the recent availability of precise parallaxes and proper motions (PMs) from the \Gaia\/ Early Data Release~3 (EDR3; \citealt{Gaia2021}), it is now possible to apply stringent astrometric tests of cluster membership. In a recent analysis, one of us \citep{Bond2021} identified all of the ZNG stars (ZNG had only published finding charts, not celestial coordinates), and then used the EDR3 astrometry to test their membership. This study found that only 45\% of the ZNG candidates are likely to be cluster members.


To our knowledge, there has not been a comprehensive search for GC AHB stars lying between the AGB and an effective temperature of about 12,000~K which goes beyond the studies described above.  In this paper, we present the results of a ground-based survey aimed at discovering and verifying a large sample of these yellow AHB stars. We use two tools to identify the AHB stars and remove field contaminants: (1)~ground-based photometric observations in the ``\uBVI\/'' system, which is optimized to detect low-surface gravity cluster members and distinguish them from foreground stars; and (2)~\Gaia\/ EDR3 PMs and parallaxes. Objects that satisfy both the photometric and astrometric criteria are almost certainly AHB members of their host clusters.

\section{A \uBVI\/ Globular-Cluster Survey \label{sec:survey}}

AHB stars have low masses and high luminosities, and hence very large radii and low surface gravities. In our temperature range of interest, this means that their spectral-energy distributions (SEDs) are characterized by a very large drop in flux below the Balmer limit at $\sim$3650\,\AA\null. The data reported in this paper were obtained in the \uBVI\/ photometric system, which was developed for efficient measurement of this Balmer discontinuity.  This system combines the $u$ filter of \citet{Thuan1976}---whose bandpass lies almost entirely below the Balmer jump---with the classical broad-band {\it BVI\/} filters of Johnson-Kron-Cousins photometry. The design principles of the \uBVI\/ system can be found in \citet[][hereafter Paper~I\null]{Bond2005}. This paper showed that, for measuring the Balmer jump in a given exposure time, the Thuan-Gunn $u$ has the highest figure of merit of any standard ground-based bandpass, including the \citet{Stromgren1963} $u$, the Sloan Digital Sky Survey $u$ \citep{Fukugita1996}, and the Johnson $U$ \citep{Bessell1990} filters.  A network of standard stars for \uBVI\/ photometry was established by \citet[][hereafter Paper~II]{Siegel2005}, who list $u$ magnitudes for 103 stars in 14 equatorial fields.  The $B$, $V$, and $I$ magnitudes of these standards are given by \citet{Landolt1992}. Further details for the filter bandpasses, sensitivities to stellar parameters, and recommendations for observing practices and data reduction, are given in Papers~I and ~II\null. Note that our \uBVI\/ magnitudes are on the Vega zero-point system, except that Vega is defined to have $u=1.00$; this is the same convention used in the Str\"omgren system.

The GC \uBVI\/ observations discussed in this paper were obtained by H.E.B. with CCD cameras on the 0.9- and 1.5-m telescopes at Cerro Tololo Inter-American Observatory (CTIO), and the 0.9-m and Mayall 4-m telescopes at Kitt Peak National Observatory (KPNO) between 1994 December and 2001 March. Appendix~A gives details of these observing runs. Table~\ref{table:runs} lists the observing-run dates, telescope-detector combinations, plate scales, and fields of view.  Table~\ref{table:obs} presents an observing log detailing the observations of each GC\null. Note that due to its relatively low throughput, most of the integration time per cluster was through the $u$ filter. 

In addition to observing 100 Galactic GCs, we imaged nine of the ``Population~II'' GCs in the Magellanic Clouds \citep[][their Table 1]{Olszewski1996}. Observations of standard fields were obtained at regular intervals throughout the photometric nights, including some at both low and high airmasses to determine the atmospheric-extinction coefficients. In several cases, especially with the smaller field of view of the CTIO cameras, the target Galactic GCs were too large to be surveyed in a single pointing.  For these objects, $2\times2$ or occasionally $3\times3$ mosaics were used to cover the clusters. Exposure times were chosen so as to reach a signal-to-noise ratio (SNR) of at least 200--300 in all four filters at the anticipated absolute magnitude ($M_V\simeq-3.5$) of the brightest yPAGB stars.  In many clusters, this SNR was actually reached at a level $\sim$2~mag fainter, and in a few favorable cases, a SNR of about 200--300 was attained at the apparent magnitude of the HB\null. For some of the clusters, we added frames with very short exposures ($\sim$8--10~s or even less) so that the very brightest stars would not be saturated; shutter-time corrections were determined and applied, but were very small.

The CCD frames were bias-subtracted, trimmed, and flat-fielded using standard IRAF\footnote{IRAF was distributed by the National Optical Astronomy Observatories, operated by AURA, Inc., under cooperative agreement with the National Science Foundation.} tasks in the {\tt ccdproc} and {\tt quadproc} packages, before proceeding to the photometric reductions and calibrations described below.

The primary goal of the survey was to search for low-gravity yPAGB stars in the Galactic GC system, in order to test their utility as potential standard candles for measuring extragalactic distances. The basic theoretical and observational arguments that non-variable yPAGB stars may be excellent and easily detected ``Population~II'' candles were presented by \citet{Bond1997a, Bond1997b}. Our \uBVI\/ survey resulted in the discoveries of two new yPAGB stars, one in M79 \citep{Bond2016}, and one in M19 \citep{Bondetal2021}. Further discussion of yPAGB stars as standard candles will be given in separate papers. {\bf Our complete catalog of \uBVI\/ photometric measurements will be published in another separate publication. In the present paper we describe the full sample of intermediate-temperature AHB stars found in our survey.}

\section{Photometric Reductions and Calibration \label{sec:reductions}}

\subsection{Methods}
\label{subsec:methods}

Stellar photometry was performed on the \uBVI\/ survey frames using the point-spread-function (PSF)-fitting algorithms of \texttt{DAOPHOT}, \texttt{DAGROW}, and \texttt{ALLFRAME} \citep{Stetson1987, Stetson1990}.  The raw data were then transformed to \uBVI\  magnitudes using the standard stars of \citet{Landolt1992} and Paper~II, and matrix inversion of the photometric calibration equations, including terms for airmass, color, color-airmass, or color squared, as appropriate. 

As described above, some of our target clusters were observed as $2 \times 2$ or $3 \times 3$ mosaics, and some of these pointings were taken under non-photometric conditions.  In addition, some frames of the more distant GCs were repeated on non-photometric nights in order to increase the SNR of the measurements. In both cases, these data were incorporated into our analysis by applying the color, airmass, and higher-order terms determined on the photometric nights of the observing run, and then adjusting the zero-point offsets until the frame's photometry matched that of the photometric data in the regions of frame overlap.  Comparison of data from different observing runs generally showed consistency at the 1--2\% level.  Examination of the resultant catalogs then revealed that, by limiting the data to objects with {\tt DAOPHOT} goodness-of-fit parameters of \texttt{CHI} $<$ 3 and $-0.5<$ \texttt{SHARP} $<0.5$, we could remove most of the spurious detections associated with bright stars or regions of severe stellar overlap.  The latter was an issue in a few of the compact, mostly distant clusters, as discussed in more detail below.




In many cases, our data consist of single-epoch photometry, one frame each in the four \uBVI\/ filters. However, several of our clusters were observed multiple times on different dates. Moreover, for those clusters that were imaged using $2 \times 2$ or $3 \times 3$ mosiacs, the stars in the overlap regions were recorded multiple times. In such cases, we adopt the stars' error-weighted mean magnitudes taken over all the observations.

For variable stars, our photometry is for the epoch of our observation, or in the case of multiple observations, the mean over our handful of epochs. As a result these stars will generally not have their intensity-weighted mean magnitudes. This explains why, for example, the bright Cepheid in Figure~\ref{fig:schematic} appears to lie outside the instability strip. The situation is particularly problematic for the RRL variables, some of which happened to be observed near maximum light; these objects mimic true AHB stars. Similarly, the rapid variability of RRL variables around maximum light can occasionally result in an object appearing to have unusually blue or red colors, even when the \uBVI\/ exposures were taken in sequence. We elected to retain these spurious cases of mimicry, but we have flagged the known variable stars in our plots and data tables.   

\subsection{Photometric Completeness \label{sec:photometric_completeness}}

\begin{figure}
\includegraphics[width=3.35 in]{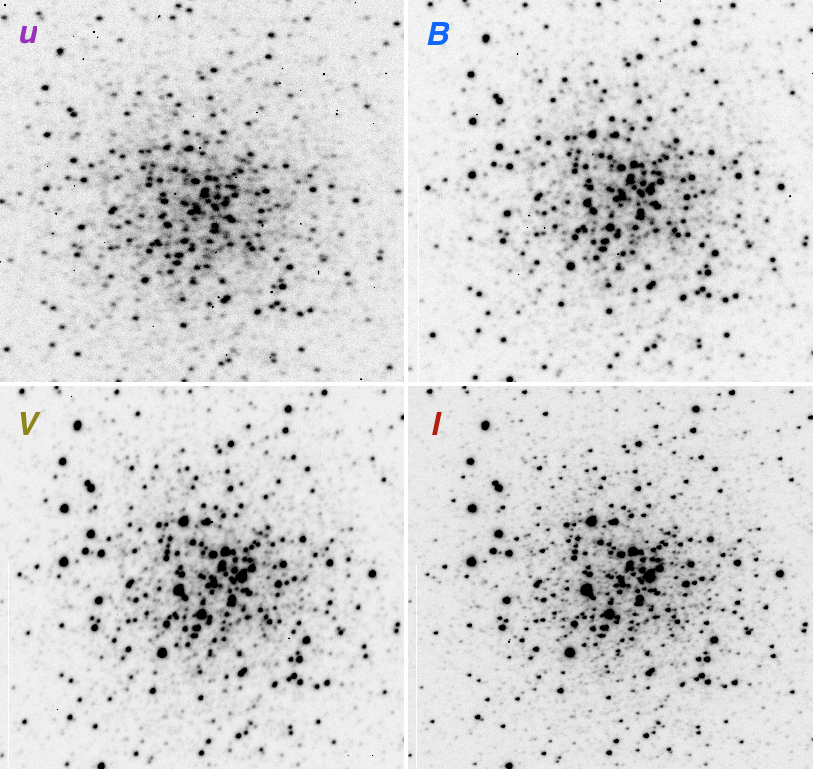}
\caption{Frames of M72 obtained with the CTIO 0.9-m telescope, in $u$ (upper left), $B$ (upper right), $V$ (lower left), and $I$ (lower right). The bright stars are well resolved into the cluster center. Exposure times were 600, 60, 45, and 60~s, respectively. Each frame is $160\arcsec$ high.
\label{fig:m72mosaic}
}
\end{figure}

Our goal is to detect AHB stars in each GC as completely as possible, even into the cluster centers. Figure~\ref{fig:m72mosaic} illustrates a set of \uBVI\/ images from our survey, for the typical GC M72.  It is clear from these frames that in clusters like this, stellar crowding is not severe, and AHB stars can be identified right into the cluster core.  

\begin{figure}
\includegraphics[width=3.35 in]{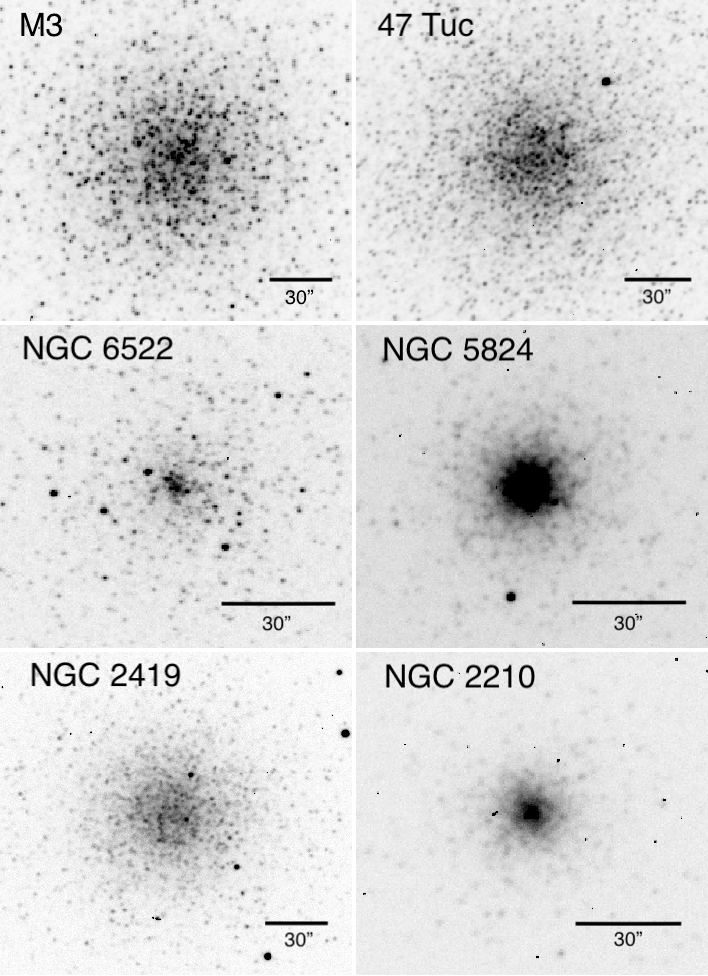}
\caption{
Frames of six globular clusters in the $u$ bandpass, illustrating the range of image qualities in our survey material. These frames were obtained with the 0.9-m telescopes of CTIO and KPNO, except for NGC\,2419, which is from the KPNO 4-m telescope.  The images on the top show two well-resolved clusters where AHB detections are possible into the very center.  The middle row displays NGC\,6522, which lies in Baade's Window, and the distant ($\sim$32\,kpc) cluster NGC\,5824.  In the former cluster, bright stars are well resolved to within few arcsec of the center, while detections in the latter are impossible within the central $10\arcsec$.  The bottom left panel  shows the extremely distant ($\sim$83\,kpc) system NGC\,2419, where bright-star photometry is complete into the cluster center. The bottom right panel displays a frame of the LMC cluster NGC\,2210, where severe stellar overlap prevents detections in the cluster core.
\label{fig:umosaic}
}
\end{figure}

While M72 is representative of most of our sample, there are systems where a combination of stellar density, cluster distance, and/or poor seeing quality makes AHB detections in the central regions problematic.  This is illustrated in Figure~\ref{fig:umosaic}, which displays the range of $u$-band image quality present in our survey material.  In clusters such as M3 and 47 Tuc, AHB detections are straightforward, even in the very center of the cluster, and this is the norm for most of our dataset.  However, in the more distant systems, such as NGC\,5824 and NGC\,2210, even the brightest individual stars are lost amidst the high surface brightness of the cluster centers.  

In order to make a numerical estimate of our completeness, we ran artificial-star tests on a subsample of clusters with different distances and concentrations. For each GC, we used the {\tt DAOPHOT} \texttt{ADDSTAR} program to place 10,000 artificial stars (25 at a time) onto the $u$, $B$, $V$, and $I$ images, giving them a uniform distribution over the magnitude range 14 to 24 and a radial distribution drawn from clusters' \citet{King1962} profiles, as defined by H10 and \citet{Lanzoni2019}.  We then processed the frames through our photometry pipeline to create a first estimate of the recovery fraction versus magnitude. These numbers were then modified by discounting the results of any object projected within one full-width-at-half-maximum seeing disk of a previously cataloged source; this accounted for the extreme crowding in the cluster centers.  We also discounted any star that failed the structural parameter cuts used in our final photometric catalogs.

\begin{figure}
\includegraphics[width=3.35 in]{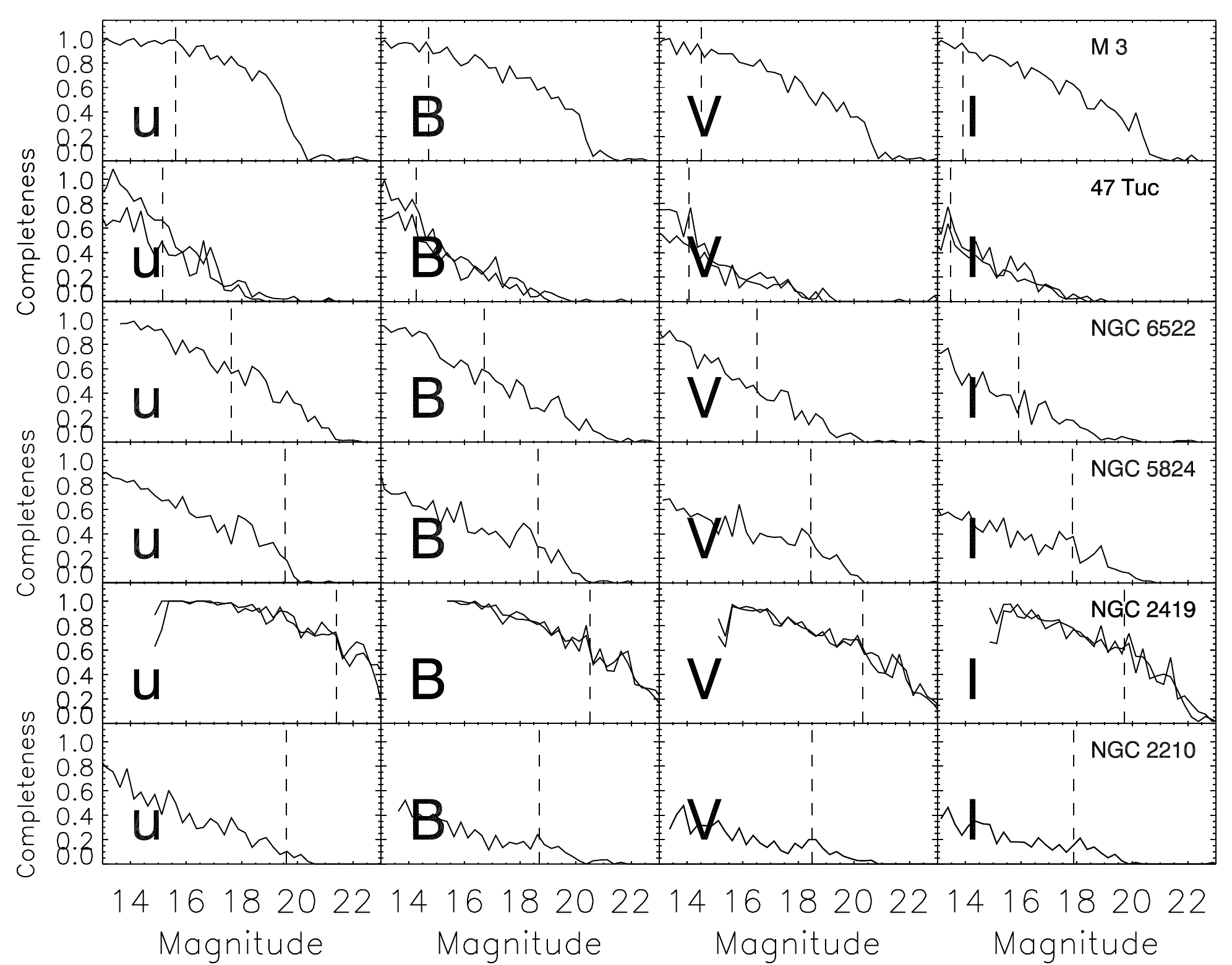}
\caption{Completeness functions for stars in the globular clusters M3, 47~Tuc, NGC\,6522, NGC\,5824, NGC\,2419 and NGC\,2210. The curves display the completeness fraction versus magnitude derived from artificial tests on individual CCD frames; panels with multiple curves indicate more than one frame was taken of the cluster.  The vertical lines mark the approximate magnitudes of the HB\null.  While our AHB surveys in clusters such as M3, 47~Tuc, and NGC\,6822 are virtually complete, a large fraction of the objects in the other systems are lost in the bright cluster centers.  This is especially true for observations of compact clusters in the $I$ filter.
}
\label{fig:allcomp}
\end{figure}

Figure~\ref{fig:allcomp} displays the completeness curves for the six globular clusters shown in Figure~\ref{fig:umosaic}. As can be seen, the completeness fraction falls as a function of magnitude and compactness.  For the well-observed cluster M3, we are nearly 100\% complete down to the HB\null. In contrast, incompleteness in the LMC cluster NGC~2210 is significant, due to extreme crowding.  In this case, most of the initial detections are excluded due to their poor \texttt{CHI} and \texttt{SHARP} values.  We can also see that the $I$ band tends to have the worst incompleteness, both due to increased crowding by red giants, and, in this case, a slightly poorer PSF\null.

Table~\ref{tab:exclusion} lists those GCs where the crowding was so severe that it limited our ability to detect AHB stars. The central regions of these clusters were excluded from our analysis; these regions are given in the table.  Also tabulated is the fraction of cluster light contained in the excluded regions, as derived from the systems' King parameters listed in H10 and \citet{Lanzoni2019}.  We note that in some cases, the brightest AHB objects projected onto the crowded regions may still be identifiable.  Nevertheless, due to their questionable photometry, we omitted these stars from our analysis.

\begin{deluxetable}{lcc}
\tablecaption{Excluded Regions Due to Crowding
\label{tab:exclusion}}
\tablehead {
\colhead{Cluster}
&\colhead{Radius [$\arcsec$]}
&\colhead{Fraction of Light Lost}
}
\startdata
M2        &20   &0.16 \\
M14       &25   &0.10 \\
M15       &38   &0.40 \\
M53       &40   &0.29 \\
M54       &30   &0.51 \\
M75       &20   &0.47 \\
M80       &30   &0.34 \\
NGC 2210  &20   &0.70 \\
NGC 2419  &20   &0.19 \\
NGC 2808  &40   &0.44 \\
NGC 5286  &15   &0.15 \\
NGC 5634  &15   &0.32 \\
NGC 5824  &10   &0.34 \\
NGC 6388  &35   &0.57 \\
NGC 6441  &15   &0.29 \\
\enddata
\end{deluxetable}

\section{Identification of AHB Stars}
\label{sec:ahb_identification}

We selected AHB stars in our target clusters---and eliminated field interlopers---using four criteria.  AHB stars must (1)~lie within the tidal radius of the host cluster; (2)~lie above the HB and blueward of the AGB in its cluster's CMD; (3)~have \uBVI\/ colors consistent with those expected for low-surface-gravity objects; and (4)~have \Gaia\/ EDR3 parallaxes and PMs in accord with cluster membership. The first criterion is not a tight constraint. To apply it, we calculated the clusters' tidal radii from information tabulated in H10 and \citet{Lanzoni2019}; in general these radii are large (and sometimes even larger than the field of view of our frames).

In the following subsections, we describe the use of \uBVI\/ photometry for selection of low-gravity stars, the creation of templates for defining the location of the HB in CMD and color-difference space, and the \Gaia\/ astrometric criteria. We then present the CMDs and color-difference diagrams for our target clusters and our catalog of AHB stars. 

In the discussion below, we adopt cluster distances and reddenings for Galactic GCs from H10.\footnote{\bf We note that, while our analysis was well underway, a valuable new compilation of distances and other data for Galactic GCs was published by \citet{Baumgardt2021}, accompanied by a useful website: \url{https://people.smp.uq.edu.au/HolgerBaumgardt/globular/}. Adoption of these improved distances would not significantly alter the main conclusions of the present paper.} For the LMC and SMC systems, we use distances of 50.1 \citep{Mackey2003a} and 60.3~kpc \citep{Mackey2003b}, respectively, and assume the reddenings tabulated by \citet{Pessev2008}. We adopt $R_V=3.1$ throughout. For $V-I$ reddening corrections as a function of $E(B-V)$, we use the formula of \citet{Dean1978}.

\subsection{Low-gravity Stars in the \uBVI\/ System}
\label{subsec:uBVI_system}

As discussed in Paper~I, we measure the Balmer jump of GC stars using the color difference $(u-B)-(B-V)$.  This difference is an analog of the $c_1$ Balmer-jump index in the classical four-color \citet{Stromgren1966} system, and we henceforth call it $c_2$. This $c_2$ index is better than $u-B$ for Balmer-jump measurements since it has much weaker sensitivities to both interstellar reddening and stellar metallicity.  Specifically, the calculations presented in Paper~I give a scaling of $E(c_2) =E[(u-B)-(B-V)]=-0.11\,E(B-V)$, for a standard interstellar reddening curve \citep{Cardelli1989}.   Moreover,  as Figures~4 and~5 of Paper~I illustrate, the use of $c_2$ greatly reduces a possible degeneracy between surface gravity and metallicity.  These plots, which display $u-B$ and $c_2$ as a function of $B-V$ and $V-I$, demonstrate the responses of these indices to stellar temperature, surface gravity, and metallicity using grids of model atmospheres with [Fe/H] = 0 and [Fe/H] = $-2$.  The plots show that $c_2$ is sensitive to surface gravity over a wide range of color, $0 \lesssim B-V \lesssim 1.0$, but only minimally dependent on metallicity. At hotter temperatures, the Balmer jump loses its sensitivity to $\log g$ and becomes dependent primarily on temperature\footnote{Actually, at effective temperatures hotter than $\sim$12,000~K, the sign of the dependence of the Balmer jump on $\log g$ reverses; see \url{https://www.stsci.edu/~bond/whereistheinfo.pdf}.}; at lower temperatures, the sensitivity to gravity becomes too weak to be useful. 

\subsection{Creating Template Cluster Diagrams}
\label{subsec:cluster_diagrams}

The first step in creating our catalog of AHB stars is the selection of objects that fall above the HB and blueward of the AGB in each cluster's CMD\null. To apply this criterion quantitatively, we define a specific region in the $M_V,(B-V)$ diagram, within which the stars must lie.  We do this with the aid of a template CMD, derived from a sample of lightly reddened GCs, in which the loci of HB, RGB, and AGB stars are well defined.




To create our template CMD, we combined data for nine high-latitude ($|b| \geq 30^\circ$) Galactic GCs, all having small reddenings [$E(B-V) \leq 0.03$], and spanning a wide range of metallicities ($-1.27 \leq {\rm [Fe/H]} \leq -2.29$) and HB morphologies. For the latter requirement, our choice of clusters took into account the horizontal-branch ratios (HBRs) tabulated by \citet[][hereafter LD94]{Lee1994}, defined by
\begin{equation}
    \textrm{HBR} = (B-R)/(B+R+V) \, ,
\end{equation}
where $B$, $R$, and $V$ represent the numbers of HB stars blueward of, redward of, and within the RRL instability strip, respectively. HBR generally reflects the color distribution of HB stars in the cluster: values above zero denote clusters with predominantly blue HBs, while negative numbers signify clusters with mostly red HB stars. We chose clusters with HBRs ranging from 0.25 to 0.90, thus providing a well-sampled HB over the full range of stellar colors.\footnote{By including clusters with a range of metallicities in our template, we have broadened both the HB and the RGB, due to the well-known dependence of HB luminosity and RGB color on [Fe/H] \citep[e.g.,][and references therein]{Sandage2006}.} 



\citet{Torelli2019} (and others) have pointed out that the HBR becomes insensitive to morphology when all of the HB stars are either redder than, or bluer than, the RRL instability strip. They defined an alternative HB morphology index, $\tauhb$, derived from cumulative number distributions along the HB in the $I$ magnitude and $V-I$ color, based on photometry from {\it Hubble Space Telescope\/} (\HST) images; it ranges from $\tauhb=0$ for an extremely red HB, to $\tauhb\simeq14$ for a very blue HB\null. This index has more sensitivity to the stellar distribution along the HB, but is available for fewer clusters, than the HBR\null. For our template clusters, $\tauhb$ ranges from 4.35 to 13.37.

\begin{figure}[hbt]
\includegraphics[width=3.35in]{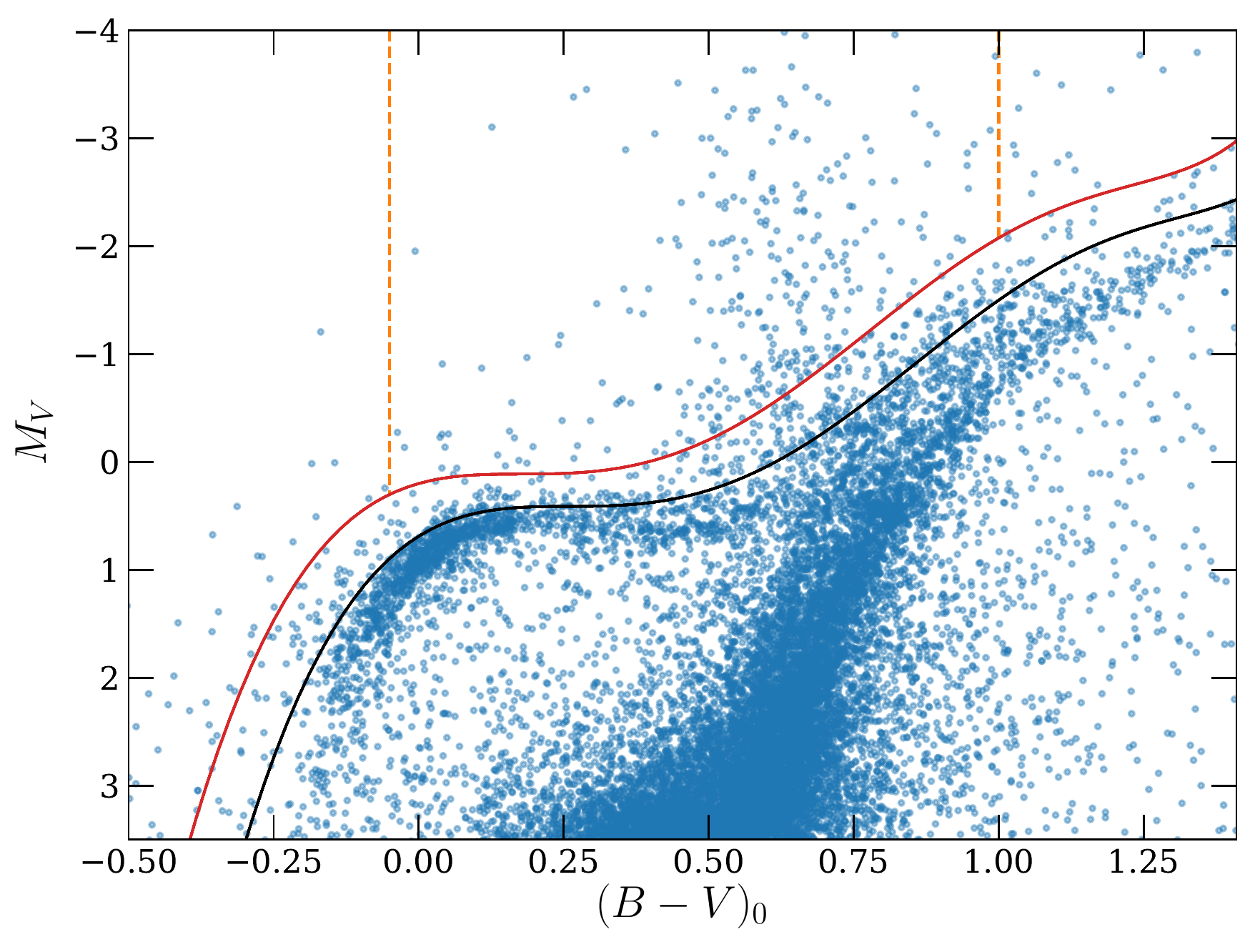}
\caption{Combined dereddened color-magnitude diagram for the nine Milky Way globular clusters listed in Table~\ref{tab:template}.  These clusters span a wide range of metallicities ($-1.27 \leq \textrm{[Fe/H]} \leq -2.29$) and HB morphologies ($0.25 \leq \textrm{HBR} \leq 0.90$) and have only a small amount of foreground reddening ($E[B-V] \leq 0.03$).  The upper envelope of our composite horizontal branch and AGB is shown as a black curve. The red curve illustrates the selection limit for our sample:  AHB stars must lie above the HB/AGB upper edge by 0.3~mag, and be bluer in color by 0.08~mag. The two vertical dashed lines mark the color limits for our AHB sample selection, $-0.05\leq(B-V)_0\leq1.00$. Field stars have not been removed from the figure.}
\label{fig:cmd_fit_definition}
\end{figure}

Table~\ref{tab:template} lists the clusters used to form our template CMD\null. The composite CMD of these systems is shown in Figure~\ref{fig:cmd_fit_definition}. Although
the figure includes all the stars detected on the CCD frames, the clusters are at high Galactic latitudes and suffer only minor field contamination. 

\begin{deluxetable}{lccccc}[b]
\tablecaption{CMD and Color-Difference Template Clusters\label{tab:template}}
\tablehead {
\colhead{Cluster}
&\colhead{$(m-M)_0$\tablenotemark{a}}
&\colhead{$E(B-V)$\tablenotemark{a}}
&\colhead{[Fe/H]\tablenotemark{a}}
&\colhead{HBR\tablenotemark{b}} 
&\colhead{$\tauhb$\tablenotemark{c}} 
}
\startdata
M5   &14.38  &0.03 &$-1.27$  &0.37 & 5.04 \\
M13  &14.26  &0.02 &$-1.54$  &0.97 & 13.37 \\
M30  &14.54  &0.03 &$-2.12$  &0.88 & 6.40 \\
M79  &15.55  &0.01 &$-1.57$  &0.89 & $\dots$ \\
NGC\,4147 &16.43 &0.02 &$-1.83$ &0.55 & 5.96 \\
NGC\,5053 &16.20 &0.03 &$-2.29$ &0.61 & 4.35 \\
NGC\,5466 &16.02 &0.00 &$-2.22$ &0.68 & 5.02 \\
NGC\,6229 &17.42 &0.01 &$-1.43$ &0.25 & $\dots$ \\
NGC\,7492 &17.10 &0.00 &$-1.51$ &0.90 & $\dots$ \\
\enddata
\tablenotetext{a}{Distance modulus, reddening, and metallicity from H10.}
\tablenotetext{b}{Horizontal-Branch Ratio from LD94.}
\tablenotetext{c}{Horizontal-branch index from \citet{Torelli2019}.}
\end{deluxetable}

Using this composite CMD, we traced the upper envelope of HB and AGB stars over the color range $-0.4 \leq (B-V)_0 \leq 1.5$, and fit this locus with a sixth-degree polynomial (represented by the black curve in Figure~\ref{fig:cmd_fit_definition}). We then shifted this relation by 0.3~mag in $M_V$ (brighter) and 0.08~mag in $B-V$ (bluer), to define the red curve shown in the figure.  This curve is given by
\begin{equation}
\begin{split}
M_V = -3.74 x^6 & +7.03 x^5 +6.83 x^4 -19.86 x^3 \\ &+8.98 x^2 -1.51 x +0.20 \, ,
\end{split}
\label{eq:cmd}
\end{equation}
where $x=(B-V)_0$.
We require all our AHB candidates to lie above this curve and in the color range $-0.05\leq(B-V)_0\leq1.00$, as shown by the two vertical lines in Figure~\ref{fig:cmd_fit_definition}. We eliminate bluer stars from consideration because, as discussed in \S\ref{subsec:uBVI_system}, and demonstrated below in \S\ref{sec:comparisons}, the $c_2$ index loses its sensitivity to $\log g$ at high temperatures.  (These stars can still be identified via their extremely blue colors, and we plan to present a catalog of the hot cluster members found in our survey in a separate paper.) Similarly, we do not include stars redder than $(B-V)_0=1.00$ in this study, again due to the $c_2$ index's lack of sensitivity to surface gravity at cooler temperatures.

The next step in selecting AHB stars is to eliminate foreground objects by requiring our candidates to have the low surface gravities characteristic of luminous low-mass stars. As discussed above (\S\ref{subsec:uBVI_system}) and in Paper~I, we employ a color-difference diagram (CDD)---dereddened $c_2$ versus $V-I$ color---and use it to identify cluster members with low $\log g$.  For this purpose, we use $V-I$ as our temperature index rather than $B-V$, due to its lower sensitivity to metallicity. 

\begin{figure}
\includegraphics[width=3.35in]{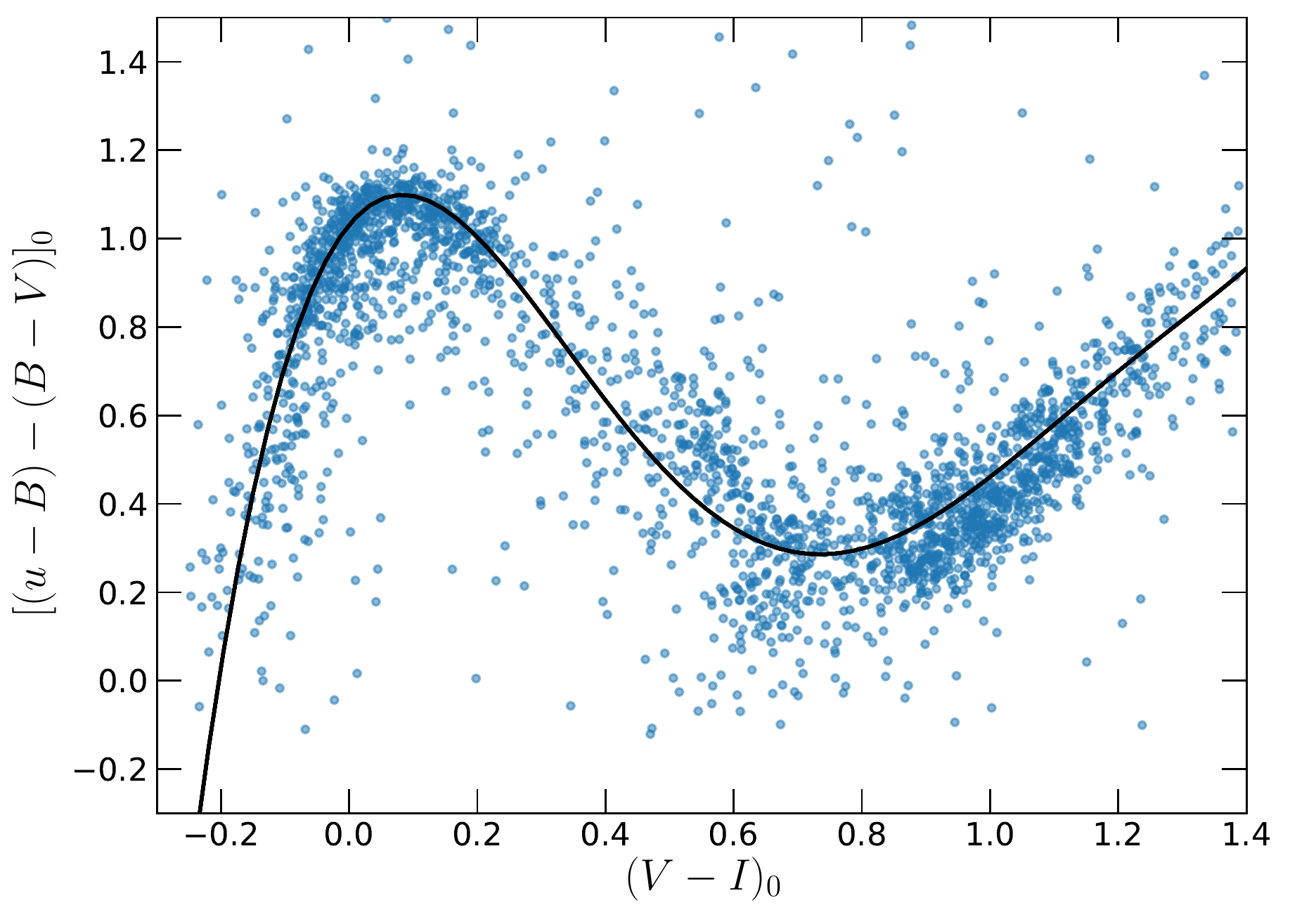}
\caption{Dereddened color-difference diagram for horizontal-branch and early AGB stars in the nine template clusters. The locus of the stars is displayed via the dark curve.  As shown by \citet{Bond2005}, the amplitude of the Balmer discontinuity in stars in this $(V-I)_0$ color range, is quite sensitive to surface gravity . As a result, objects lying above the curve (i.e., with fainter $u$ magnitudes) have lower surface gravities than ZAHB objects (or field main-sequence stars) and likely lie above the horizontal branch.}
\label{fig:cdd_fit_definition}
\end{figure}

Figure~\ref{fig:cdd_fit_definition} plots the CDD for the dereddened HB and AGB stars in the nine template clusters. To create this diagram, we selected only those stars with visual absolute magnitudes within $\pm 0.5$ mag of the black curve shown in Figure~\ref{fig:cmd_fit_definition} and fitted these data over the color range $-0.25 \leq (V-I)_0 \leq 1.5$ with a sixth-degree polynomial:
\begin{equation}
\begin{split}
(c_2)_0 = 0.98 y^6 -1.46 y^5 -6.28 y^4 +17.04 y^3 \\ -12.60 y^2 +1.75 y +1.03 \, ,
\end{split}
\label{eq:cdd}
\end{equation}
where $y=(V-I)_0$.
This fit to the template clusters' HB and AGB stars is represented by the black curve in Figure~\ref{fig:cdd_fit_definition}.  The shape of this curve reflects the effect of temperature on the amplitude of the Balmer discontinuity, superposed on the general trend of $c_2$ becoming progressively redder as $\Teff$ decreases.  We eliminated all AHB candidates from the CMD selection that fell below this relation; such objects have higher surface gravities than HB and AGB stars, and are likely to be foreground interlopers.

\subsection{Template Adjustments for Individual CMDs}
\label{subsec:adjustments}

In order to apply our CMD template to Galactic GCs, we adopted the cluster distances and reddenings given in H10. However, this assumption resulted in cases where the upper edge of a cluster's HB is inconsistent with our template. This discrepancy likely arises from an imperfect knowledge of the system's distance and/or reddening, the possibility of small errors in the zero-points of our photometry, and/or the dependence of HB luminosity upon metallicity.  Moreover,  for some clusters, the foreground reddening of the Milky Way varies across the field, causing the upper edge of the HB to be ill-defined.

To minimize the effect of these errors, we slid the fiducial curve defined in Equation~\ref{eq:cmd} vertically in the CMD and applied an edge-fitting algorithm to the resulting star counts.  Specifically, we began by moving the curve to a location 2~mag above the default HB/AGB given by Equation~\ref{eq:cmd}, and counted the number of stars below the curve. We then shifted the curve downward in steps of 0.1 mag until it reached a position 0.6~mag fainter than the default HB, and at each location we counted the number of stars below the curve. The step giving the largest change in counts was taken to indicate the location of the HB's upper edge, and the zero-point shift corresponding to this step was applied to the AHB selection curve for the cluster. In most cases, the adopted shifts were small, generally less than $\pm 0.2$~mag.

A few clusters did not have a discernible HB in our data, due to issues such as large photometric errors at the magnitude of the horizontal branch, severe differential extinction, or simply the lack of a substantial HB population.  In these systems, we retained the default AHB selection region using the H10 distance and reddening, or in a very few cases estimated the position of the HB by eye. 

\subsection{Astrometric Membership Criteria}
\label{subsec:astrometric_criteria}

At this stage we have selected a list of candidate AHB stars that fall within the cluster tidal radii and satisfy the photometric requirements of lying in the AHB region of the host cluster's CMD and CDD\null.  To remove any remaining field stars, we now further require that the candidates' astrometric properties be consistent with cluster membership. 

To make a statistically valid cluster/field-star separation, we use data from \Gaia\/ EDR3,\footnote{Our AHB candidates are all relatively bright, and nearly all are contained in EDR3. The few missing objects are likely due to source crowding near the cluster centers.} and apply a Gaussian-mixture model \citep[GMM; e.g.,][]{scikit-learn, Kuhn2017, McLachlan2019} to the stars' positions, parallaxes, and PMs. This analysis assumes that there are one or more kinematic populations of field stars, which are superposed on a single cluster population.  

\begin{figure*}
\includegraphics[width=\textwidth]{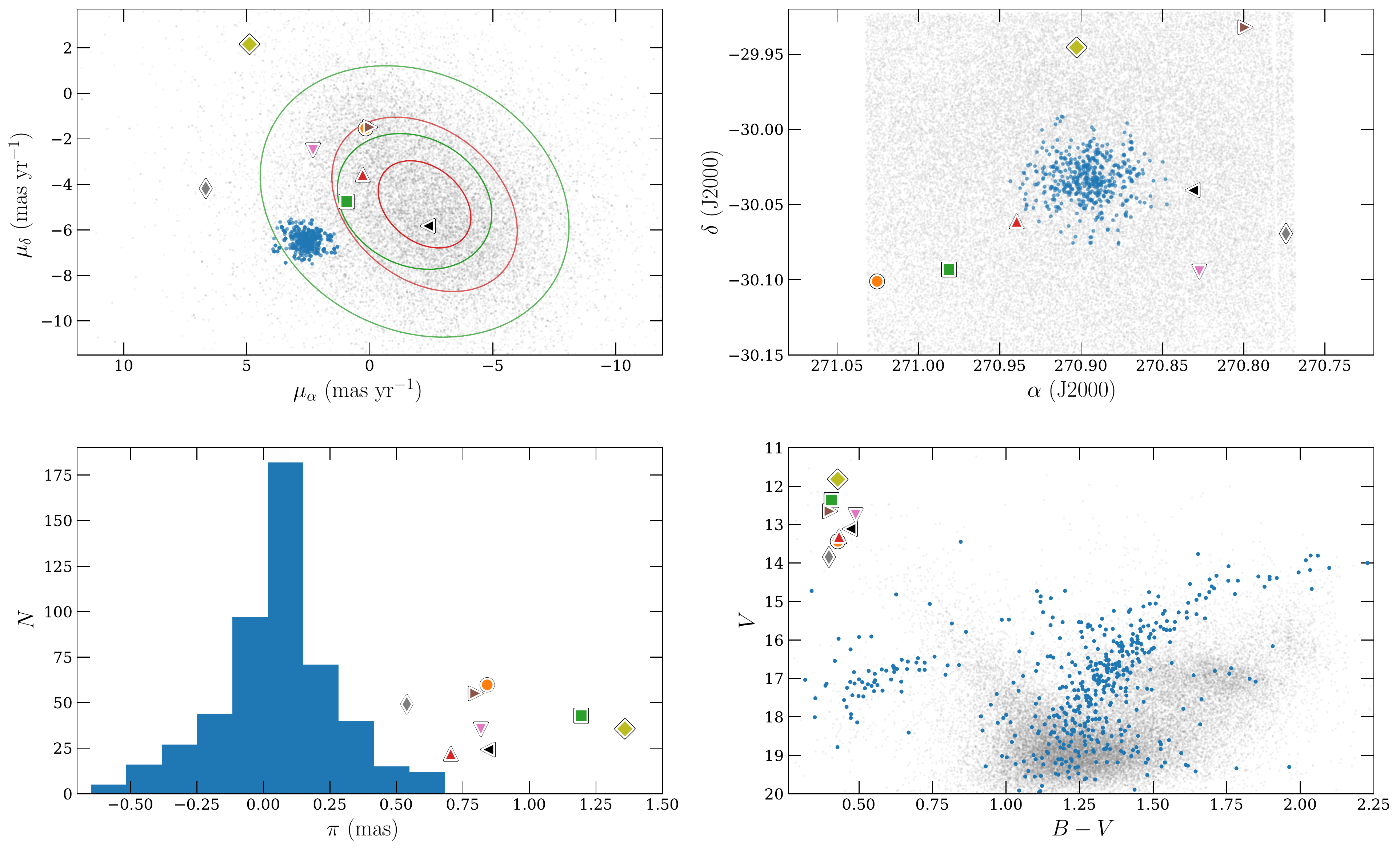}
\caption{Illustration of field-star rejection in NGC~6522---a Galactic-bulge cluster with heavy field contamination---based on our Gaussian-mixture model. Stars indicated to be cluster members are shown in blue, and field stars are displayed in gray. Eight bright objects that could be mistaken for cluster AHB stars are highlighted by special symbols; our analysis indicates that are all non-members. \textit{Top left:} Proper-motion diagram. The field stars are modeled by two Gaussians, and are represented by green and red ellipses drawn at 1 and 2$\sigma$ intervals. A single, much smaller, Gaussian represents the cluster members. \textit{Top right:} Positions on the sky of cluster members and field stars.  The cluster's tidal radius is larger than the size of our CCD\null.  \textit{Bottom left:} Histogram of the parallaxes of cluster members (blue bars). The parallaxes of the eight bright stars are marked with the same symbols as in the other panels; their heights above the $x$-axis are proportional to their angular distance from the cluster center. \textit{Bottom right:} Color-magnitude diagram for the cluster members and field stars. The highlighted bright stars are all non-members. The cluster member at $(B-V,V)=(0.84,13.45)$ is confirmed to have a large Balmer jump by our \uBVI\ photometry.}
\label{fig:cand_selection}
\end{figure*}

The inputs to the GMM are the stars' PMs ($\mu_{\alpha}$ and $\mu_{\delta}$), parallaxes, and angular separations from the cluster centers.  To reduce the affect of astrometric uncertainties, we limited our analysis to stars brighter than $M_V=+3.5$; this proved to be a good compromise between our desire for a large sample of stars, and our need to exclude objects with large measurement errors. The GMM then assumes that each of the four parameters has a Gaussian distribution within each population, and assigns to each star a probability of membership in each of the groups. Figure~\ref{fig:cand_selection} illustrates how the GMM works using data for the Baade's Window GC NGC~6522. Because of this system's low Galactic latitude ($b=-3\fdeg9$), contamination by field stars in this cluster is substantial, and forced us to model the field-star population using two separate components.   

In the figure, cluster stars are plotted in blue, and field objects are shown in gray. For illustrative purposes, the figure also displays eight field stars which, based on their apparent magnitude and color, could be mistaken for AHB stars belonging to the cluster.  Spatially, these eight bright stars lie outside the cluster core, but well within the system's $15\farcm 8$ tidal radius, which is larger than our CCD's field of view (top-right panel). These stars also have slightly larger parallaxes than the stars in the cluster, although in some cases the numbers are still consistent with cluster membership (bottom-left panel).  However, as is illustrated in the figure, the PMs of the cluster stars are tightly clumped, while those of field objects have a broad distribution, which is elongated parallel to the Galactic equator (top-left panel).   The separation is not perfect, and there is a small amount of kinematic, spatial, and parallax overlap between the cluster and field populations.  Nevertheless, by requiring that our AHB candidates have a GMM probability of cluster membership greater than 0.8, \textit{and\/} a probability of membership in any of the field populations less than 0.2, \textit{and\/} have a large Balmer jump as measured by our \uBVI\ photometry, we can be reasonably certain that our list of AHB candidates is uncontaminated.

We do note that the GMM's assignment of cluster membership is not completely unique, as the procedure requires a choice of the number of distinct kinematic groups present among the field stars.  For most clusters, the assumption of a single field population is sufficient to produce a clean cluster CMD with little sign of field contamination.  However, in some low-Galactic-latitude systems such as NGC\,6522, two to four field populations are necessary to remove obvious interlopers from the sample.  In these cases the optimum number of field components was determined by visual inspection of the CMD, and the number of components that produced the greatest number of cluster stars with the least amount of contamination was chosen for our analysis.

For several distant and/or sparsely populated clusters, the number of HB stars detected with sufficient SNR was too small for a GMM analysis. In these cases, we vetted the AHB candidates found via our \uBVI\/ photometry using solely the \Gaia\/ EDR3 parallaxes and proper motions. (None of the candidates passed these tests.) These clusters were: AM 4, E3, Eridanus, Pyxis, Hodge~11, and Pal~1, 3, 4, 5, 11, 12, 13, 14, and~15.

Lastly, there were five clusters for which we obtained data, but the field contamination was so severe, and/or there was such substantial and/or variable foreground extinction, that a reliable search for AHB member stars was not possible. These clusters were: NGC\,1916, NGC\,2019, NGC\,6517, NGC\,6528, and Pal~2.

In summary, we obtained \uBVI\/ observations of 109 clusters (100 Galactic, one in the Small Magellanic Cloud, and eight in the Large Magellanic Cloud). However, as just described, we excluded three Galactic and two Large Magellanic Cloud targets from our AHB search and analysis.

\subsection{CMDs, CDDs, and Catalog of AHB Stars\label{subsec:catalogs}}

In Appendix~B we present plots of the CMDs and CDDs from our \uBVI\/ data for all of the clusters in our survey.  The list of AHB candidates that satisfy all of our selection criteria is given in Table~\ref{tab:meta_catalog}.  Included in this table are the stars' \Gaia\/ EDR3 J2000 coordinates, PMs, and parallaxes, our own \uBVI\ photometry, the values of $M_V$ and $(B-V)_0$ (derived principally using the distance moduli and reddenings of H10), and the cluster-membership probabilities. The final two columns give a classification, described below in \S\ref{subsec:classification}, and previous identifications of the stars.

Our catalog contains a total of 438 candidate AHB stars. They are found in 64 out of the 104 clusters that we searched. However, we note that 59 of the objects are known RRL variables, which should not be considered true AHB candidates; in most or all of these cases, we happened to make our observations when the variables were near maximum light. We note that the $(B-V)_0$ colors of these objects are preferentially blue, consistent with them being RRL variables near maximum.

\begin{deluxetable*}{llll}
\tablecaption{Contents of AHB Catalog\label{tab:meta_catalog}}
\tablehead{
\colhead{Number}
&\colhead{Units}
&\colhead{Label}
&\colhead{Explanation}
}
\startdata
1  &\dots   &Cluster Name  &Name of Globular Cluster \\
2  &\dots   &Star ID       &Identification Number \\
3  &deg     &RA            &RA, J2000, decimal degrees  \\
4  &deg     &DEC           &DEC, J2000, decimal degrees  \\
5  &mas/yr  &pmRA          &RA Proper Motion (milliarcsec~yr$^{-1}$) \\
6  &mas/yr  &pmDEC         &DEC Proper Motion (milliarcsec~yr$^{-1}$) \\
7  &arcsec  &Radius        &Distance from Cluster Center (arcsec) \\
8  &mag     &$M_V$         &Absolute $V$-band magnitude \\
9  &mag     &$(B-V)_0$     &Intrinsic $(B-V)$ color \\
10 &mag     &$(c_2)_0$     &Intrinsic $(u-B) - (B-V)$ \\
11 &mag     &$u$           &Observed $u$-band magnitude \\
12 &mag     &$e_u$         &Uncertainty in $u$ \\
13 &mag     &$B$           &Observed $B$-band magnitude \\
14 &mag     &$e_B$         &Uncertainty in $B$ \\
15 &mag     &$V$           &Observed $V$-band magnitude \\
16 &mag     &$e_V$         &Uncertainty in $V$ \\
17 &mag     &$I$           &Observed $I$-band magnitude \\
18 &mag     &$e_I$         &Uncertainty in $I$ \\
19 &mas     &plx           &Parallax (milliarcsec) \\
20 &\dots   &$P$           &Membership probability \\
21 &\dots   &Class         &Classification \\
22 &\dots   &Alt-ID        &Previous identifications\tablenotemark{a} \\
\enddata
\tablenotetext{\textrm{a}}{V = Variable Star number from \citet{Clement2001};
ZNG = ID from \citet{Zinn1972}}
\tablecomments{This table is available in its entirety in machine-readable and Virtual
Observatory (VO) forms.}
\end{deluxetable*}

\section{Comparisons with Other Surveys \label{sec:comparisons}}

As shown in \S\ref{sec:photometric_completeness}, with the exception of the central regions of a handful of distant and/or compact clusters, our photometric catalogs extend down to or below the level of the HB\null.  Moreover, since nearly all of our AHB candidates are contained in \Gaia\/ EDR3, our AHB identification procedures should have produced a catalog that is close to complete. To test this expectation, we compared our dataset to three collections of known AHB stars: Type~II Cepheids identified via their variability, UV-bright objects found in the ZNG survey, and AHB stars in the rich cluster \oCen. 

\subsection{Recovery of Known Variable Stars}
\label{subsec:variables}

The first test of the effectiveness of our detection strategy is to determine whether our list of AHB stars contains the known Type~II Cepheid variables in the target clusters. Such a comparison will be imperfect, as we have only single- or few-epoch data for the clusters, and during a pulsation cycle, a lower-luminosity variable star may cross into and out of the AHB regions of the color-magnitude and color-difference diagrams. Nevertheless, since the instability strip does lie in the temperature range where the Balmer jump's sensitivity to surface gravity is greatest, our technique should recover the vast majority of RV Tauri and W Virginis stars listed in the C01 catalog, and many of the BL~Her variables.

\begin{figure*}
\centering
\includegraphics[width=236pt]{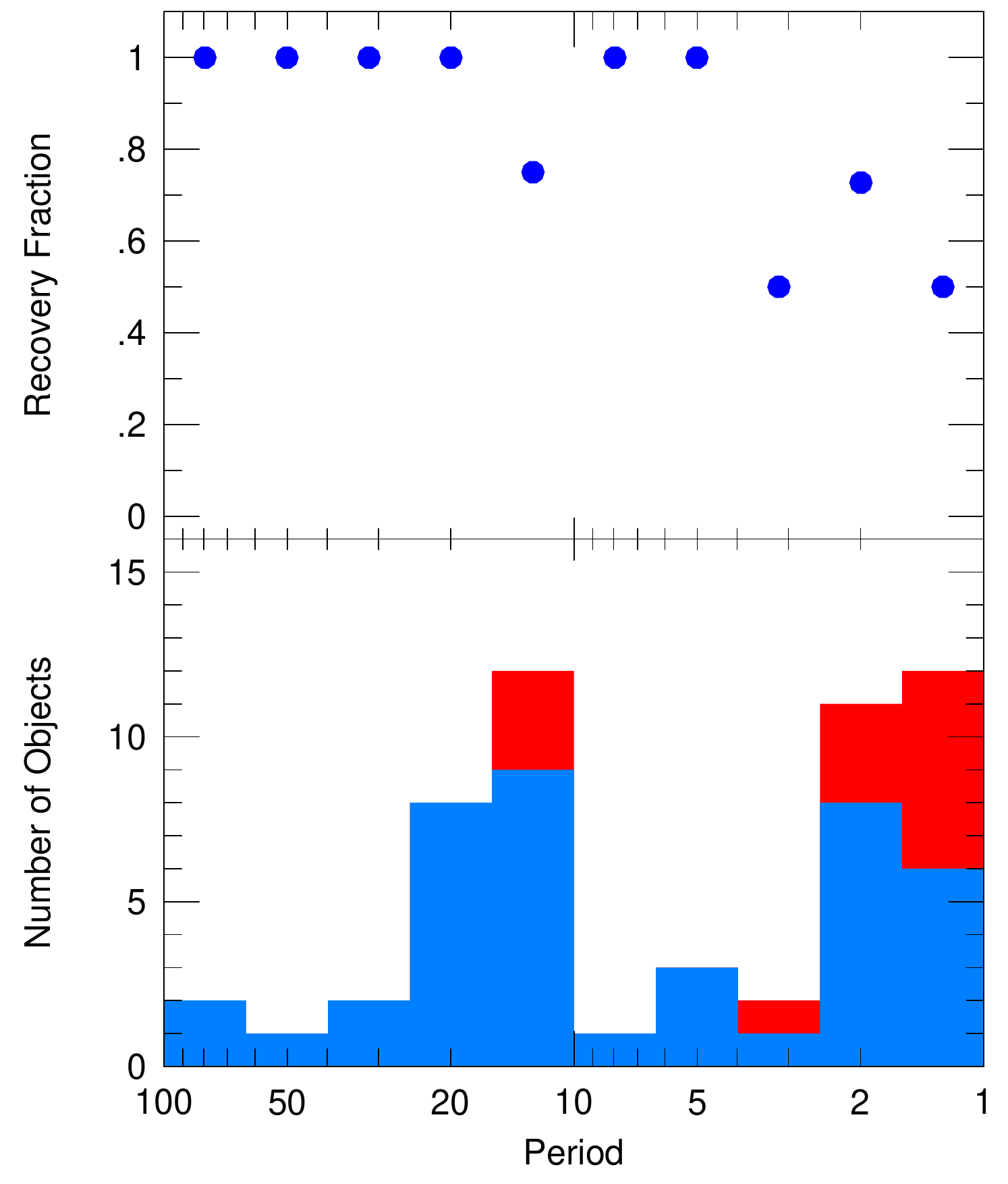}
\hfil
\includegraphics[width=236pt]{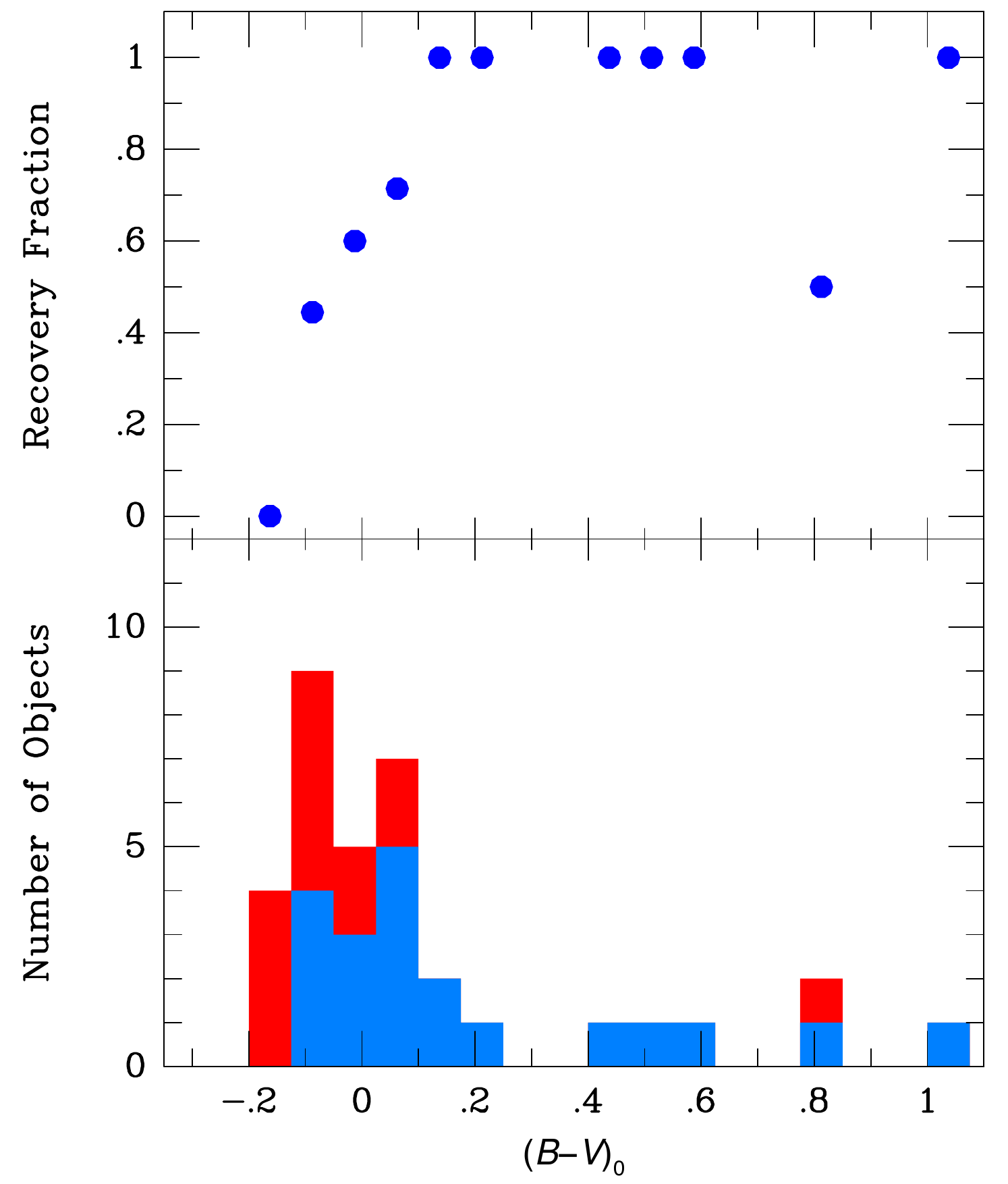}
\caption{Our recovery fraction of previously cataloged AHB and PAGB stars in Milky Way globular glusters.  The left-hand panel displays the results for BL~Herculis, W Virginis, and RV Tauri variables as a function of period; the right-hand panel shows the recovery of the ZNG stars confirmed by \citet{Bond2021} as a function of color.  The top panels give our fraction of recoveries; the bottom panels shows the number of total (red) and recovered (blue) stars within each bin.  The plots demonstrate that our sample is highly complete for ``yellow'' AHB stars, but its effectiveness  begins to decline at $(B-V)_0 \lesssim 0.1$, due to the decreased sensitivity of the Balmer jump to surface gravity.}
\label{fig:ahb_comparison}
\end{figure*}

To perform this test, we identified all objects on our \uBVI\/ frames within $3\arcsec$ of a cataloged Type~II Cepheid or RV Tauri variable and having an apparent brightness consistent with that expected from the star's listed intensity-weighted mean magnitude and pulsation amplitude.  For the vast majority of stars, these simple criteria produced an unambiguous match, with the closest possible star having a median astrometric offset of $0\farcs 4$ from the quoted position, and the next closest object generally being $\gtrsim 2\farcs 0$ away.  We then examined the fraction of recoveries as a function of surface gravity, using the stars' pulsation periods as a proxy for the latter.  The result is shown in the left-hand panel of Figure~\ref{fig:ahb_comparison}.

The panel shows that, for the longer-period variables, our survey technique is very effective.  The only long-period ($P > 4$~day) Cepheids not recovered in our survey are objects that \texttt{DAOPHOT} flagged as having bad \texttt{CHI} and \texttt{SHARP} values due to image crowding.  More specifically, only one well-measured Cepheid (a BL~ Her object with a 4.15 day period) did not satisfy our CMD and CDD criteria.  Thus, where accurate photometry is possible, the Balmer-jump criterion is extremely successful in picking out low-surface-gravity stars in the instability strip.

The left-hand panel of Figure~\ref{fig:ahb_comparison} also demonstrates that, at shorter periods, our recovery fraction of Population~II Cepheids is reduced.  This is simply due to the fact that, if a short-period variable is observed near minimum, its luminosity may be too low to satisfy our AHB brightness criterion.  Conversely, when we repeat our variable-star identification procedure with objects listed in the C01 catalog as RRL stars, we find that $\sim$15\% of the objects that we identify as AHB objects are actually variables with periods of less than one day.  Clearly, these RRL stars were caught near maximum, when their high luminosities and low surface gravities caused them to satisfy our CMD and CDD criteria.  We retained these contaminants in our catalog, but they are noted as RRL variables.

\subsection{Recovery of Zinn et al.\ UV-Bright Stars}
\label{subsec:zinn}

A second dataset for comparison is the UV-bright stars identified by ZNG\null.  As described in \S\ref{subsec:ahb_surveys}, these are an inhomogeneous set of non-variable objects selected solely on the basis of their apparent brightness in the $U$-band filter.  Although a large fraction of these stars are foreground interlopers, \citet{Bond2021} recently used the objects' \Gaia\/ PMs, parallaxes, and colors to determine their membership status and to classify each as a HB, RGB, AHB, AGB, or PAGB star.  We can use these data to quantify the effectiveness of our Balmer-jump criterion as a function of color and to test the predictions of Paper~I, which are based on grids of model atmospheres.

The right-hand panel of Figure~\ref{fig:ahb_comparison} displays the fraction of ZNG stars classified by \citet{Bond2021} as AHB or PAGB cluster members that lie above the CDD threshold curve of Figure~\ref{fig:cdd_fit_definition}.  For purposes of this comparison, we consider all stars with colors in the range $-0.3 < (B-V)_0 < 1.25$, and not just the objects within the color limits of our survey. Two properties stand out.  The first is the distribution of colors:  very few of the ZNG AHB and PAGB stars are redder than $(B-V)_0 = 0.2$.  This is to be expected: AHB stars that are redward of the instability strip are difficult to distinguish from RGB and AGB objects. Moreover, since ZNG selected their stars via their apparent brightness in the $U$~band, one might expect the distribution of stellar colors to be skewed towards the blue.  Thus, the ratio of blue to red objects shown by the figure's histogram is somewhat expected.

The second and more important feature of the figure is the fraction of post-HB ZNG recoveries as a function of color.  According to the model-atmosphere analysis in Paper~I, the strength of the Balmer break, and therefore the value of the $c_2$ index, is most sensitive to stellar surface gravity in the color range $0.1 < (V-I) < 0.9$. If we translate this result to $B-V$ using the color-temperature relation for low-gravity stars \citep{Worthey2011}, then we should expect the fraction of ZNG recoveries to be greatest in the range $0.1 < (B-V) < 0.7$, and decline rapidly towards the blue and more slowly in the red.  This is exactly what is seen.  Every ZNG star classified by \citet{Bond2021} as an AHB or PAGB object is recovered within the color interval $0.05 < (B-V)_0 < 0.7$. Blueward of this range, the survey's effectiveness drops precipitously, so that by $(B-V)_0 < -0.1$, AHB stars are no longer detectable via their Balmer jump. There are far less data for the red side of the distribution, but of the three ZNG stars with colors $(B-V)_0 > 0.6$, two were recovered by the technique. This recovery fraction justifies our use of $-0.05 < (B-V)_0 < 1.0$ as the color limits of our survey catalog.

\subsection{AHB Stars in \oCen}

\citet{McDonald2009} presented a CMD of bright stars in \oCen\ that are likely cluster members on the basis of their ground-based PMs \citep{vanLeeuwen2000} and, in some cases, RVs \citep{vanLoon2007}.  Sixteen of their objects lie above the HB and blueward of the AGB (their Figure~3 and Table~6). We searched for these stars in our Table~\ref{tab:meta_catalog} catalog, and found that we had independently identified seven as AHB objects with highly probable cluster memberships. Six of our recoveries have RVs given by \citet{vanLoon2007} which, along with the \Gaia\/ astrometry, are consistent with membership in the cluster.  The seventh is the Type~II Cepheid V48. 

Of the remaining nine stars, we found the following: (1)~one object is redder than our catalog's color cutoff; (2)~two have \Gaia\/ EDR3 parallaxes and\slash or PMs (and in one case an RV in EDR3) inconsistent with membership; (3)~three are blended with nearby stars, as indicated by large values of the {\tt RUWE} parameter in EDR3 and by our inspection of \HST\/ images; and (4)~two do not have a parallax and PM listed in EDR3, and thus we could not verify their membership.

This leaves one object that is a genuine AHB star based on its CMD position, its \Gaia\/ parallax and PM, and its RV (from both \citealt{Reijns2006} and EDR3), which nevertheless did not qualify for inclusion in our Table~\ref{tab:meta_catalog}. This is star number 37295 in \citet{vanLeeuwen2000}, at J2000 position $(\alpha,\delta)=(201\fdeg82400,-47\fdeg43161)$. We measure its magnitudes to be $(u,B,V,I)=(14.125,13.162,12.244,11.669)$.  If we correct these for a reddening of $E(B-V)=0.12$ (H10), then the location of this star in our CDD  (Figure~\ref{fig:cdd_fit_definition}) is $(V-I)_0=0.427$, $(c_2)_0=0.068$. This implies that the star does not have a large Balmer jump---in fact, its value of $(c_2)_0$ is even lower than those of HB stars, in spite of its high luminosity ($M_V=-1.69$). Our data were obtained during two different observing runs, so an artifact affecting our photometry appears unlikely. We note that a low value of the Balmer-jump index in luminous stars is a characteristic of carbon- and {\it s}-process-rich objects (\citealt{Bond2019}; the effect is due to enhanced CH absorption in the $B$ band). Thus we speculate that this star is chemically peculiar. A spectroscopic investigation of this anomalous star would be of interest.

Setting aside this one object, we again find strong support for a high level of completeness for our AHB survey.


\section{Binaries and Blends as AHB Impostors}
\label{sec:interlopers}

\begin{figure}[bt]
\centering
\includegraphics[width=3.35in]{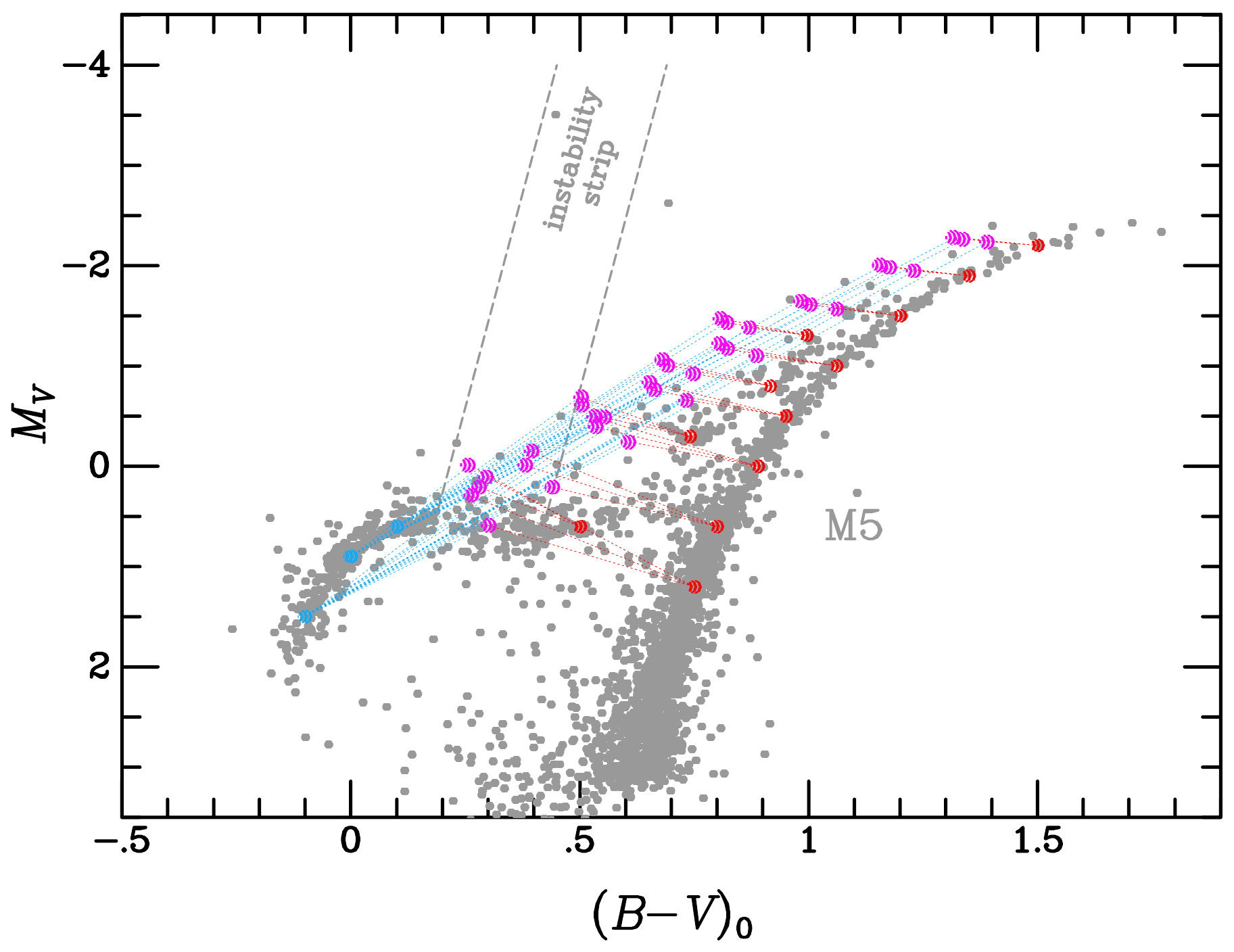}
\caption{
Positions in the CMD of unresolved binaries and blends, superposed on the CMD of the typical cluster M5. We combined the light of pairs of representative BHB/EHB stars (blue filled circles) and RGB/AGB stars (red filled circles) to produce the colors and magnitudes of the blends (pink filled circles). The dotted blue lines connect the individual components to the combined light.  Such combinations can explain the existence of non-variable objects that appear to lie inside the instability strip, and within about 1~mag of the HB\null.  Brighter non-variables in the instability strip can only be produced by blends of multiple objects. Blends can also produce brighter objects appearing to lie slightly blueward of the AGB.
}
\label{fig:blended_pairs}
\end{figure}

A physical binary star, or simply an unresolved pair of overlapping stars, can produce an object that appears to lie in the AHB region of a cluster's CMD\null. This effect is particularly important in the crowded central regions of clusters, especially in the more distant ones, where blending or binarity can result in a non-variable ``star'' that appears to lie in the pulsational instability strip. This is illustrated in Figure~\ref{fig:blended_pairs}. Here we have plotted the CMD of M5 (from Figures~\ref{fig:schematic} and~\ref{fig:schematicwithtracks}), and selected the locations of representative BHB/EHB (filled blue circles) and RGB/AGB stars (filled red circles).   We then calculated the colors and magnitudes of binaries consisting of one of the blue stars and one of the red stars (filled pink circles). The dotted lines connect the positions of the two components to the locations of their combined light.

The figure shows that blended objects can indeed populate the AHB region of the CMD, and place pairs of non-variable stars within the instability strip.  However, these objects cannot be more luminous than $M_V\simeq-0.5$ (apart from higher-order multiples), unless the blue component is itself an AHB star.  We will discuss a few examples of such interlopers in the next section.  We do note that, in most cases, blended objects with extreme temperature differences can be identified via their anomalous locations in a $B-V$ versus $V-I$ color-color diagram.  It is also possible to recognize some of the blends by examining images taken with \HST, if such data exist. 

\section{Comparisons with Post-HB Evolutionary Theory}
\label{sec:theory-comp}

In this section we compare the AHB populations that we see in a set of representative GCs with predictions based on the recent theoretical post-HB evolutionary tracks of M+19. These examples illustrate the usefulness of our AHB catalog for investigating late stages of low-mass stellar evolution and testing theoretical calculations.

In Figures~\ref{fig:ngc6362_with_tracks} through \ref{fig:m15_with_tracks}, and the associated discussion, we present our CMDs for four Galactic GCs chosen to cover a range of metallicities and HB morphologies. Figures~\ref{fig:m14_with_tracks} and~\ref{fig:m10_with_tracks} show the CMDs of two additional clusters, M14 and M10, both of which are remarkably rich in AHB stars. To assure pure stellar samples in our CMDs, we require each star to have a cluster membership probability greater than 0.8, and a probability of belonging to any of the field populations less than 0.2, as calculated from the \Gaia\/ astrometry via the procedure described in \S\ref{subsec:astrometric_criteria}. For each cluster, we adopt the metallicity, distance, and reddening from H10, as indicated in the figure legends, and plot member stars as gray filled circles, with known RRL and Type~II Cepheid variables encircled in red. We then superpose evolutionary tracks from M+19, having metallicities consistent with the [Fe/H] of each GC, and with an appropriate range of ZAHB masses chosen to match the temperature range of HB stars seen in the cluster. As in \S\ref{subsec:post-hb_evolution}, the theoretical quantities ($\Teff$, $\log L/L_\odot$) of the tracks have been converted to observational $B-V$ and $M_V$, using the PARSEC YBC web tool \citep{Chen2019}. Also, as explained in \S\ref{subsec:post-hb_evolution}, for the sake of clarity we have edited the track data to remove rapid excursions, including those due to TPs. We indicate the evolutionary time steps by placing special symbols on each track. These steps are marked at intervals that start at 10~Myr, then decrease to 1~Myr, and finally to 0.1 or 0.01~Myr, as the evolution accelerates. The time-step plotting symbols are encoded as shown in the figure legends. 

\subsection{NGC 6362}
\label{subsec:6362}

\begin{figure}
\centering
\includegraphics[width=3.35in]{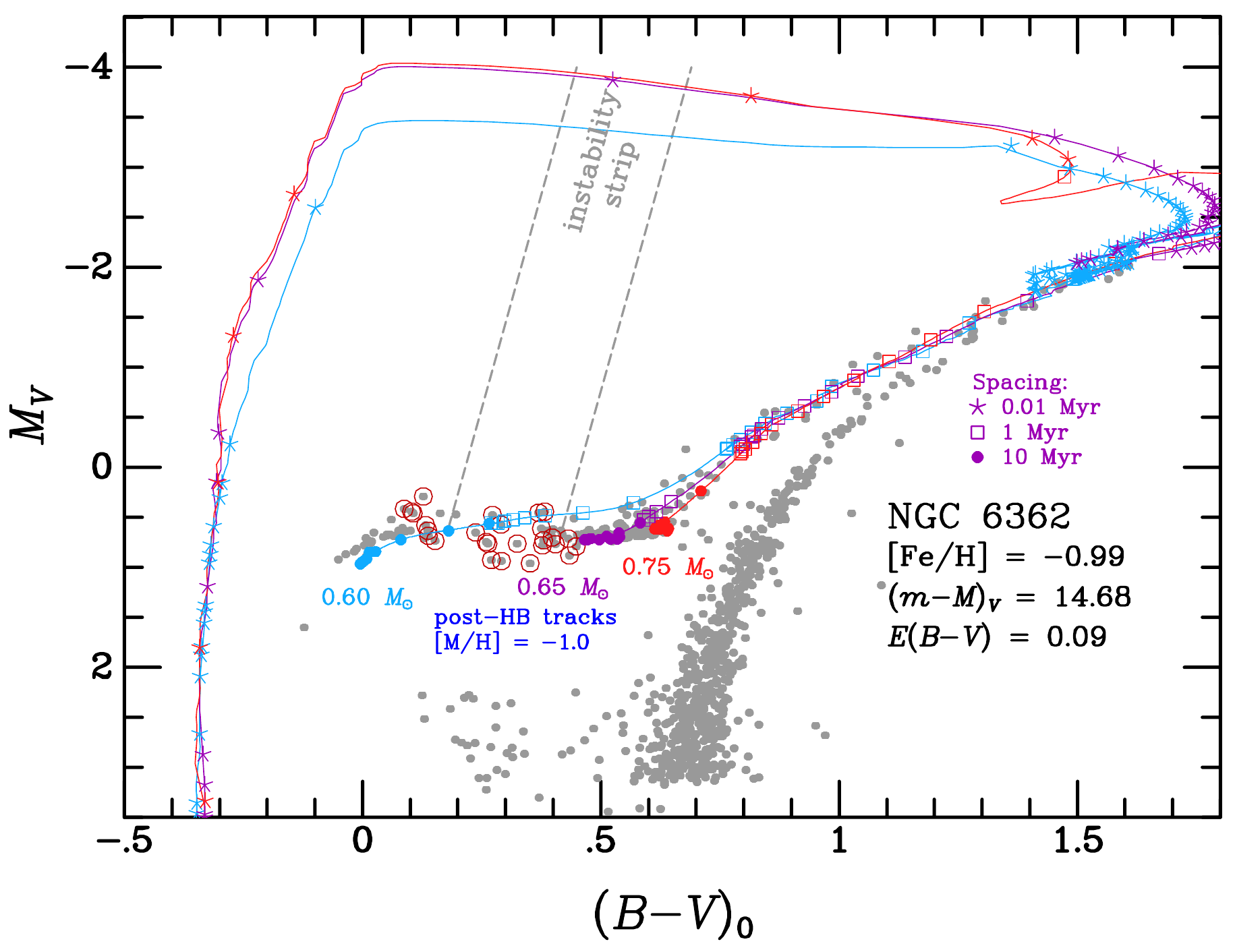}
\caption{ 
Color-magnitude diagram for members of NGC\,6362 (gray points). Known RR~Lyrae variables are encircled in red; for the reasons discussed in \S\ref{subsec:methods}, some appear to lie outside the instability strip. Superposed are post-HB evolutionary tracks for the three ZAHB masses labelled in the figure (see text for details). NGC\,6362 lacks any stars above the HB, due to the absence of very blue stars on its ZAHB, and to the evolution of more-massive ZAHB stars being along the HB rather than appreciably above it. Evolutionary time steps are marked at intervals given by the symbols in the legend. The time steps show that post-AGB evolution in this cluster is expected to be extremely rapid. This, along with the relative scarcity of ``feeder'' stars on the AGB, explains the observed lack of luminous, yellow stars in the cluster. 
}
\label{fig:ngc6362_with_tracks}
\end{figure}

Figure~\ref{fig:ngc6362_with_tracks} shows the CMD of the lightly reddened GC NGC\,6362. This cluster is relatively metal-rich ($\rm[Fe/H]=-0.99$); accordingly it has a predominantly ``red'' HB, with an HBR index of $-0.58$ (LD94) and $\tauhb=2.24$ \citep{Torelli2019}, and it contains about three dozen cataloged RRL variables (C01). Superposed on the CMD are post-HB tracks with a metallicity of $\rm[M/H]=-1.0$. 

Since the post-HB evolution of stars of these masses and metallicity is essentially along the ZAHB, no AHB stars are expected to be seen, and indeed, none are present in our data. Similarly, the tracks predict that the cluster will not contain any objects in the PEAGB region of the CMD, due to the lack of progenitors on the extremely blue HB\null. While NGC\,6362 could in principle, contain very luminous PAGB stars, the time steps marked in the figure show that the evolution rate during this phase is extremely rapid; thus the likelihood of finding such stars is small. Additionally, there are relatively few luminous AGB stars in this cluster to act as ``feeders'' for the PAGB population.  Since the late, high-temperature post-AGB evolution slows down around $M_V=0$, we might have expected to see some stars in this pre-WD phase of stellar evolution.  However, none were found---again, probably due to the small number of progenitors.  Overall, our data are consistent with theoretical expectations: we detected no PEAGB or PAGB stars, and in fact no AHB stars at all that are sufficiently bright for inclusion in our catalog. 





\subsection{M79 (NGC 1904) and NGC 5986} 
\label{subsec:m79}

\begin{figure*}
\centering
\includegraphics[width=3.4in]{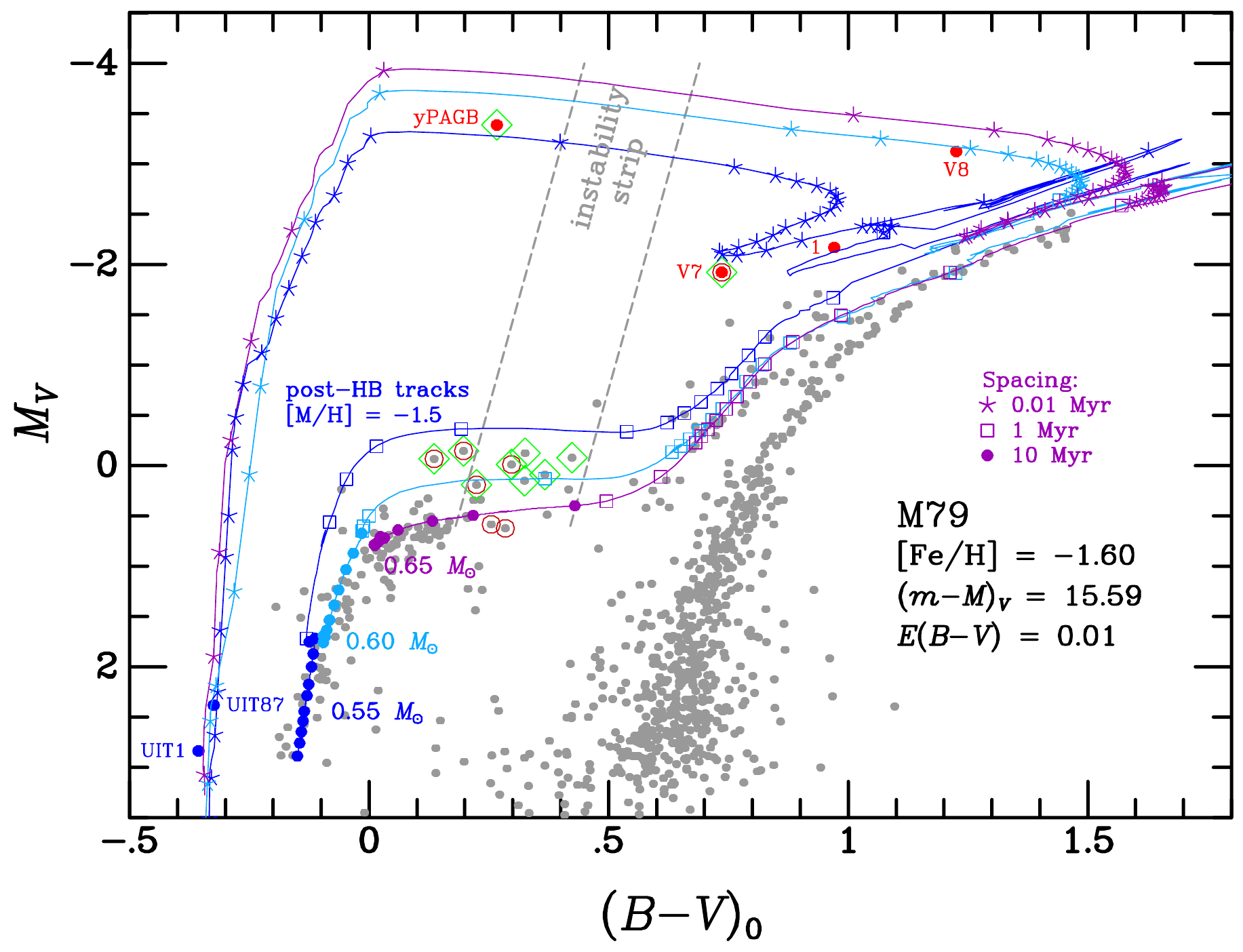}
\hfill 
\includegraphics[width=3.4in]{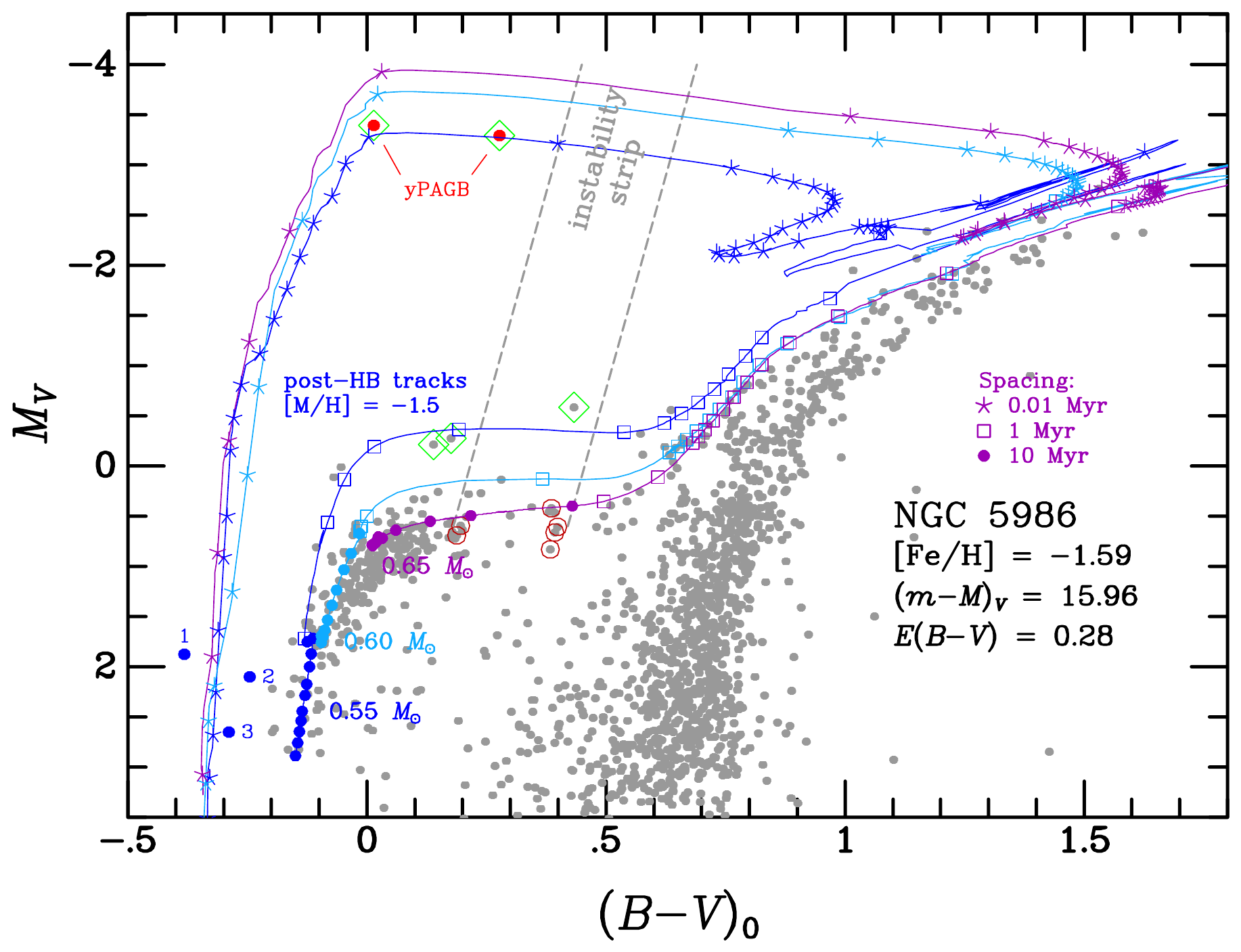}
\caption{ 
Color-magnitude diagram for M79 (left panel) and NGC\,5986 (right panel). Member stars are plotted as gray filled circles. For emphasis, we use filled red circles to mark four yellow and red post-AGB stars in M79, and two yellow post-AGB stars in NGC\,5986. Hot post-AGB stars are marked with filled blue circles at the lower left in both diagrams. Known RR~Lyrae and Cepheid variable stars are encircled in red; some of the variables, including the bright Cepheid V7 in M79, appear to lie outside the instability strip, for the reasons discussed in \S\ref{subsec:methods}. Stars confirmed to have low $\log g$ from our \uBVI\/ photometry are enclosed in green diamonds.  Superposed are post-HB evolutionary tracks for the three ZAHB masses labelled in the panels. Evolutionary time steps are marked at intervals given by the symbols in the legends. Both M79 and NGC\,5986 have much bluer ZAHBs than NGC\,6362 in the previous figure. The post-AGB evolutionary sequences are at lower luminosities than they are in NGC\,6362, and their timescales are longer, accounting for the presence of the six post-AGB stars. The apparently non-variable stars in the instability strip are discussed in the text.
}
\label{fig:m79_with_tracks}
\end{figure*}

The two panels in Figure~\ref{fig:m79_with_tracks} show our CMDs for the nearly unreddened GC M79 and the moderately reddened system NGC\,5986. These two clusters have intermediate metallicities of $\rm[Fe/H]=-1.60$ and $-1.59$, respectively, and nearly identical HB morphologies. As a consequence of their lower metallicities, the clusters' HB stars are systematically bluer than those in NGC\,6362; the LD94 HBRs for M79 and NGC\,5986 are 0.89 and 0.95, respectively, and the \citet{Torelli2019} $\tauhb$ value for NGC\,5986 is 7.85. M79 contains only 11 known or suspected RRL variables, and NGC\,5986 only ten (C01). In the figures, the lone W~Vir Cepheid in M79, along with the RRL variables, are encircled in red,\footnote{The 40-day variable V13 in NGC\,5986 is called a Cepheid by C01, but in our photometry it lies near the AGB tip at $([B-V]_0,M_V) = (1.27, -2.31)$. We suggest it be considered a semiregular (SR) variable. The cluster also contains a second SR variable, V4, which our photometry places at the tip of the AGB at $([B-V]_0,M_V) = (1.53,-2.76)$.} while stars that qualify for inclusion in our AHB catalog (Table~\ref{tab:meta_catalog}) are enclosed in green diamonds. 

The theoretical tracks shown in Figure~\ref{fig:m79_with_tracks} appear to account for the principal features of the AHB populations of the two clusters. Initially the post-HB stars evolve slowly; the evolution then accelerates and the stars move quickly across the CMD to the base of the AGB, creating an analog of the Hertzsprung Gap seen in Population~I systems (as discussed in \S\ref{subsec:post-hb_evolution}). Both clusters contain clumps of about a half-dozen stars around $(B-V)_0\simeq 0.6, M_V\simeq+0.5$, which at first might seem to be RHB stars. However, the evolutionary tracks indicate that these objects are actually post-ZAHB stars with masses of about $0.65\,M_\odot$. In contrast, the lowest-mass ZAHB stars evolve above the HB on their way to the AGB base, and account for the presence of several AHB stars lying up to almost 1~mag above the ZAHB over the color range $0.0 \lesssim (B-V)_0 \lesssim 0.5$. Based on these single-star evolutionary tracks, we do not expect there to be any AHB stars lying in the space between $\sim$1~mag above the ZAHB and the luminous post-AGB tracks departing from near the top of the AGB---and none are seen in our data, with the possible exception of the M79 variable V7. 

At the end of their ascent of the AGB, post-HB stars evolve to higher effective temperatures. In M79 and NGC\,5986, PAGB evolution is slower than in NGC\,6362, increasing the probability of finding objects in this phase. Indeed, as marked by the filled red circles in the left panel of Figure~\ref{fig:m79_with_tracks}, M79 contains four luminous yellow stars. One of them is the yPAGB star discovered by B16; the other three include the 14.0-day Cepheid V7,\footnote{V7 was observed in our data at a relatively cool pulsation phase. Although we interpret it here as a candidate post-AGB or PEAGB star, we note that M79 contains a sequence of optically faint and very hot EHB stars, not visible in our relatively shallow photometry, but seen in deep images obtained by \HST\/ and the {\it Galaxy Evolution Explorer\/} (\GALEX) \citep[e.g.,][]{Lanzoni2007}. Below, in \S\ref{subsec:m14}, we explore the possibility that these EHB stars may be the progenitors of W~Vir Cepheids like V7. If so, V7 is not a post-AGB star, but is crossing the instability strip for the first time.} and two redder objects (which could be AGB stars undergoing TPs). Since the latter two stars did not meet the criteria for inclusion in our AHB star catalog, we give their positions and our dereddened $BV$ photometry in Table~\ref{table:red_stars_in_m79}. The brighter star is the semiregular variable V8, for which a light curve is given by \citet{Bond2016}. The fainter object, lying near the center of M79, has not, to our knowledge, been previously cataloged.

Remarkably, as shown by the two filled red circles in the right panel of Figure~\ref{fig:m79_with_tracks}, NGC\,5986 contains two luminous yPAGB stars, discovered by \citet{Bond1977}, and discussed by \citet{Alves2001}. The tracks plotted in both panels of Figure~\ref{fig:m79_with_tracks} indicate that these visually bright and conspicuous post-AGB objects are likely to be descendants of BHB stars with ZAHB masses of about 0.55--$0.60\,M_\odot$. All three yPAGB stars in M79 and NGC\,5986 lie very close to the PAGB track for a star with a ZAHB mass of $0.55\,M_\odot$. This is perhaps not surprising, since the post-AGB evolutionary timescales increase rapidly with increasing mass.

\begin{deluxetable}{lCCCC}
\tablecaption{Two Red Post-AGB Stars in M79\label{table:red_stars_in_m79}}
\tablehead{
\colhead{ID}
&\colhead{RA [J2000]} 
&\colhead{Dec [J2000]}
&\colhead{$(B-V)_0$}
&\colhead{$M_V$}\\
\colhead{}
&\colhead{[deg]} 
&\colhead{[deg]}
&\colhead{}
&\colhead{}
}
\startdata
V8 & 81.04812 & -24.52731 & 1.226 & -3.122 \\
1 & 81.04596 & -24.52492 & 0.971 & -2.166 \\
\enddata
\end{deluxetable}


The post-HB tracks in Figure~\ref{fig:m79_with_tracks} also show that, after the PAGB stars have reached high temperatures and moved to the top of the WD cooling sequence, their evolution slows down.  Consistent with this deceleration, we find two very hot PAGB stars in M79, and three in NGC\,5986.  These are emphasized by the filled blue circles in the lower-left corners of both panels. The two hot PAGB stars labeled in M79 were recognized in the space-ultraviolet study of the cluster with the {\it Astro-I\/} Spacelab Ultraviolet Imaging Telescope \citep{Hill1992, Hill1996}, and are cataloged as UIT\,1 and UIT\,87. To our knowledge, the three hot PAGB stars in NGC\,5986 have not been recognized previously. We give the positions and our $BV$ photometry of these faint, hot stars in Table~\ref{table:hot_stars_m79+5986}.  Note also that both clusters contain a significant population of visually fainter, but very hot, PAGB, AGB-manqu\'e, and EHB stars, which have been detected in space-based ultraviolet images \citep[e.g.,][]{Altner1993,Hill1996,Lanzoni2007,Schiavon2012,Siegel2014}.


Our data may imply a mismatch between the timescales of the M+19 evolutionary tracks and the locations of the pre-WDs shown in the CMDs of M79 and NGC\,5986.  According to the models, the rate of post-AGB evolution begins to slow at high luminosity (around $M_V \simeq -3$), suggesting that hot stars, such UIT 1 and 87, should be present to much brighter $V$-band magnitudes than seen here.  The fact that all five of the hot stars found in the clusters have much fainter visual absolute magnitudes (between +2 and +3) suggests that the evolutionary rates for stars in these stages of post-AGB evolution may need some re-examination.

\begin{deluxetable}{lCCCC}
\tablecaption{Blue Stars in M79 and NGC 5986\label{table:hot_stars_m79+5986}}
\tablehead{
\colhead{ID}
&\colhead{RA [J2000]} 
&\colhead{Dec [J2000]}
&\colhead{$(B-V)_0$}
&\colhead{$M_V$}\\
\colhead{}
&\colhead{[deg]} 
&\colhead{[deg]}
&\colhead{}
&\colhead{}
}
\decimals
\startdata
\noalign{\smallskip}
\multispan5{\hfil M79 (NGC 1904) \hfil}\\
\noalign{\smallskip}
UIT 87 &  81.05638 & -24.53331 & -0.324 & 2.379 \\
UIT 1  &  81.04971 & -24.53892 & -0.356 & 2.837 \\
\noalign{\smallskip}
\multispan5{\hfil NGC 5986 \hfil}\\
\noalign{\smallskip}
1      & 236.51725 & -37.77986 & -0.382 & 1.875 \\
2      & 236.50388 & -37.78881 & -0.245 & 2.100 \\ 
3      & 236.52100 & -37.73750 & -0.289 & 2.647 \\  
\noalign{\smallskip}
\enddata
\end{deluxetable}

Another curiosity of Figure~\ref{fig:m79_with_tracks} is the presence of several stars in both clusters that appear to lie within the pulsational instability strip, but are not known RRL variables. To investigate this phenomenon, we examined high-resolution \HST\/ frames of the clusters available in the Hubble Legacy Archive (HLA)\footnote{\url{https://hla.stsci.edu/hlaview.html}} from M79 programs GO-6095 (PI: G.~Djorgovski) and GO-6607 (PI: F.~Ferraro) and NGC\,5986 programs GO-10775 (PI: A.~Sarajedini) and GO-13297 (PI: G.~Piotto). We find that nearly all of these stars are close pairs of blue and red objects, with separations ranging from barely resolved to about $1\farcs3$, and the pairs often fall near the crowded centers of the clusters. They are thus examples of the stellar blends discussed in \S\ref{sec:interlopers}, which at ground-based resolution can produce apparently non-variable interlopers within the instability strip. It should be noted, however, that some of these objects nevertheless have large Balmer jumps in our \uBVI\/ photometry, indicating that the bluer components may still be AHB objects.


\subsection{M15 (NGC 7078)}
\label{subsec:m15}

For our next comparison, we selected the extremely metal-poor system M15 ($\rm[Fe/H]=-2.37$). This cluster's HB covers a wide range of colors; consistent with its low metallicity, M15 contains numerous very blue HB stars (the LD94 HBR is 0.72, and $\tauhb=6.63$), but it also hosts more than 165 known RRL variables (C01).

\begin{figure}
\centering
\includegraphics[width=3.35in]{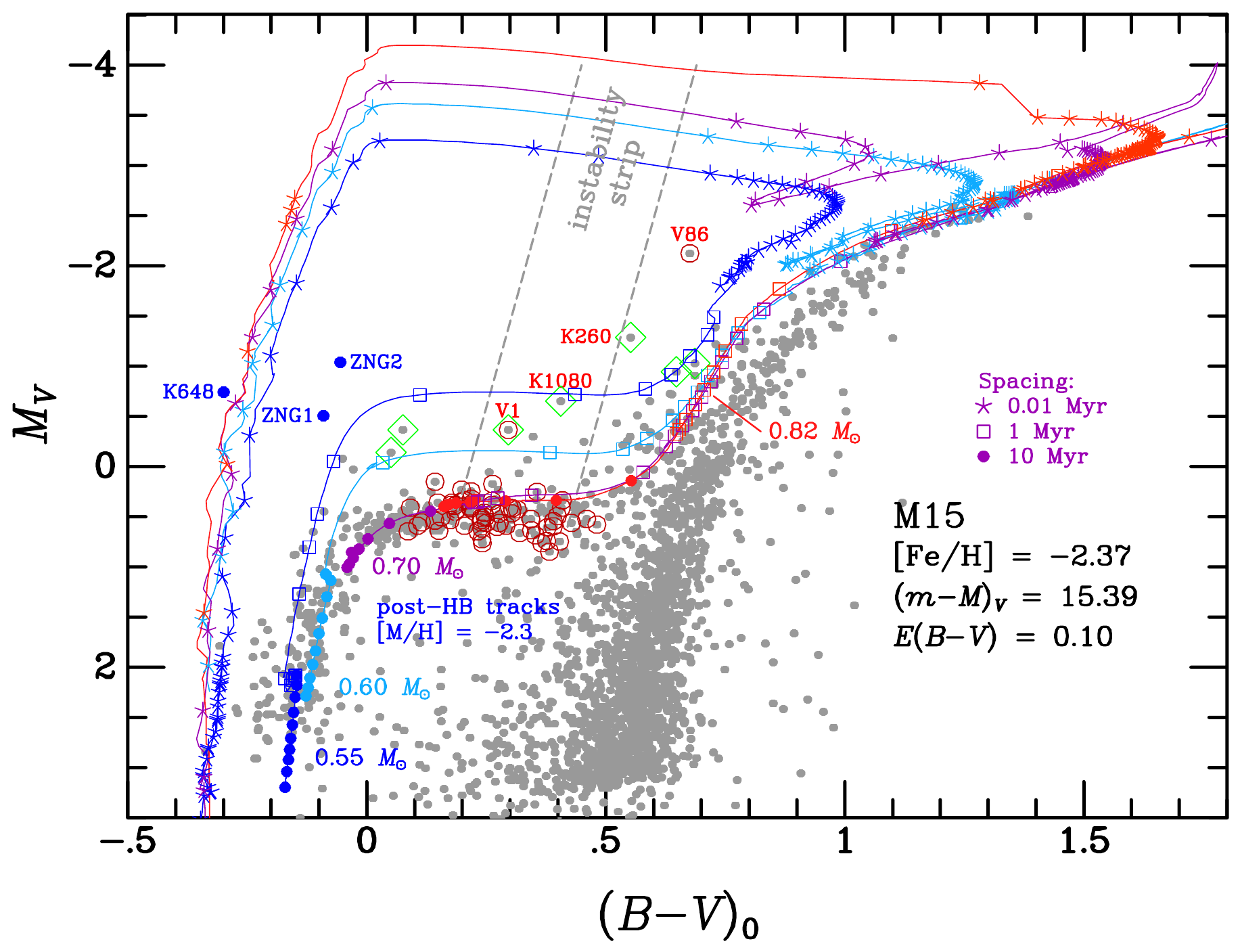}
\caption{
Color-magnitude diagram for members of M15 lying more than $38''$ from the cluster center (gray points). Several individual stars of interest are labelled and discussed in the text; these include the two luminous hot stars K\,648 (central star of the planetary nebula Ps~1) and ZNG\,1, plotted as filled blue circles. Known RR~Lyrae and Cepheid variable stars are encircled in red; as in the previous two figures, some of the variables, including the bright Cepheid V86, appear to lie outside the instability strip, for the reasons discussed in \S\ref{subsec:methods}. Stars confirmed to have low $\log g$ from our \uBVI\/ photometry are enclosed in green diamonds. Superposed are post-HB evolutionary tracks for the four ZAHB masses labelled in the figure. Evolutionary time steps are marked at intervals given by the symbols in the legend. 
}
\label{fig:m15_with_tracks}
\end{figure}

Our CMD for the cluster is plotted in Figure~\ref{fig:m15_with_tracks}. Because of extreme stellar crowding in the GC's inner regions, the plot only includes member stars lying more than $38''$ from the center. As in the previous CMD figures, we encircle the known RRL and Type~II Cepheids stars in red.  M15 contains three of the latter: V1, V86, and V142, with pulsation periods of 1.44, 16.83, and 1.23~days, respectively (C01). However, V142 lies too close to the cluster center for reliable photometry in our groundbased \uBVI\/ survey.  V86 also lies close to ($14\arcsec$) the cluster center, but the star is luminous enough to have usable photometry, and it is retained in the figure. (The fact that V86 appears to lie outside the instability strip is due to our limited phase coverage. The object's mean magnitude, as measured by \citealt{FusiPecci1980}, places it comfortably within the instability strip.)

M15 contains several luminous hot stars, of which the best known is K\,648, central star of the planetary nebula Ps\,1 (see \S\ref{subsec:ahb_intro}). Our photometry of K\,648 is affected by its surrounding nebula, so its location in Figure~\ref{fig:m15_with_tracks} is based on the stellar parameters given by \citet{Rauch2002} and \citet{Otsuka2015} and the YBC web tool described in \S\ref{subsec:post-hb_evolution}. Also plotted are two hot UV-bright stars ZNG\,1 and ZNG\,2. Our photometry of ZNG\,1 is affected by a barely resolved nearby red giant, so we instead employed stellar parameters from \citet{Mooney2004} and the YBC tool to obtain its estimated color and absolute magnitude. Table~\ref{table:hot_stars_m15} gives the positions and dereddened $BV$ photometry for these three blue stars.

\begin{deluxetable}{lCCCC}[hb]
\tablecaption{Blue Stars in M15\label{table:hot_stars_m15}}
\tablehead{
\colhead{ID}
&\colhead{RA [J2000]} 
&\colhead{Dec [J2000]}
&\colhead{$(B-V)_0$}
&\colhead{$M_V$}\\
\colhead{}
&\colhead{[deg]} 
&\colhead{[deg]}
&\colhead{}
&\colhead{}
}
\decimals
\startdata
ZNG 2 & 322.53099 & 12.18547 & -0.055 & -1.036 \\
ZNG 1\tablenotemark{a} & 322.49247 & 12.19514 & -0.09  & -0.51 \\
K 648\tablenotemark{a} & 322.49748 & 12.17395 & -0.30  & -0.74 \\
\enddata
\tablenotetext{a}{Photometry of ZNG\,1 and K\,648 estimated from spectroscopic
stellar parameters; see text.}
\end{deluxetable}






As shown in Figure~\ref{fig:m15_with_tracks}, the post-HB evolution of metal-poor stars with ZAHB masses of $\gtrsim\!0.70\,M_\odot$ is essentially along the ZAHB\null.  We therefore expect no AHB production from these relatively high-mass objects.  (In fact, some of the RRL variables and apparent RHB objects may actually be post-HB stars, rather than on the ZAHB\null.) However, BHB stars with ZAHB masses of about $0.55\,M_\odot$ evolve to cooler temperatures at luminosities as much as $\sim$1.3 mag above those of ZAHB objects. Consistent with this expectation, M15 contains several such AHB stars; these are enclosed in green diamonds to indicate their large Balmer jumps and low surface gravities. The 1.44-day Cepheid V1 lies in this AHB region of the CMD, as does V142 \citep{Tuairisg2003}.

An apparent anomaly is the object K\,1080, whose low $\log g$ is verified by our \uBVI\/ photometry. This star, whose membership is also confirmed by RV measurements \citep[e.g.,][]{Gebhardt1997}, seems to lie within the instability strip, but is not a known variable. We inspected \HST\/ images from several programs (including GO-12604, PI G.~Piotto; and GO-13295, PI S.~Larsen), and in these frames it appears that the star is marginally resolved. Thus K\,1080 is likely another case of a blend of a blue and red non-variable object (see \S\ref{sec:interlopers}). However, based on the object's luminosity and large Balmer jump, one of the system's components must be a true AHB star, and lie blueward of the instability strip.  The other non-variable objects that appear to lie within the instability strip do not have large Balmer jumps; these are likely blended red/blue pairs. 

M15 contains no red or yellow PAGB stars. This is consistent with the rapid evolutionary timescales at the top of its CMD, along with the scarcity of AGB feeders. However, K\,648 is at a location in the CMD consistent with it being a hot PAGB object whose  evolutionary rate is slowing down. On the other hand, several authors \citep[e.g.,][]{Alves2001} have argued that K\,648 is more luminous than expected for single-star stellar evolution, and must be the descendant of a merged binary.

Perhaps the most puzzling feature of the M15 CMD is the sequence of AHB stars at intermediate luminosities, which includes ZNG\,1 and~2, K\,260, and the long-period W~Vir Cepheid V86. K\,260 in particular has been the focus of several studies \citep[e.g.,][and references therein]{Jasniewicz2004,Masseron2019} and is definitely a cluster member. These objects might conventionally be considered PEAGB stars, evolving blueward following an early departure from the AGB\null. However, the single-star evolutionary tracks of M+19 appear unable to explain these stars, since they only depart the AGB toward higher temperatures at considerably brighter levels. 

Alternatively, these anomalous stars may be descended from hot and visually faint EHB stars, with ZAHB masses less than $0.55\,M_\odot$. In this case, the objects would be evolving {\it redward\/} through the AHB region of the CMD\null. Although not shown well in our relatively shallow photometry, deep CMDs of M15 reveal that the cluster does contains a sparse population of very hot EHB stars \citep[e.g.,][and references therein]{BondM15_2020}. These objects could be the progenitors of the intermediate luminosity AHB sequence seen in M15. We will explore this possibility further in the next subsection, and find support for it.

\smallbreak

\subsection{Two AHB-Rich Clusters: M14 (NGC 6402) and M10 (NGC 6254)}
\label{subsec:m14}

\begin{figure*}
\centering
\includegraphics[width=3.4in]{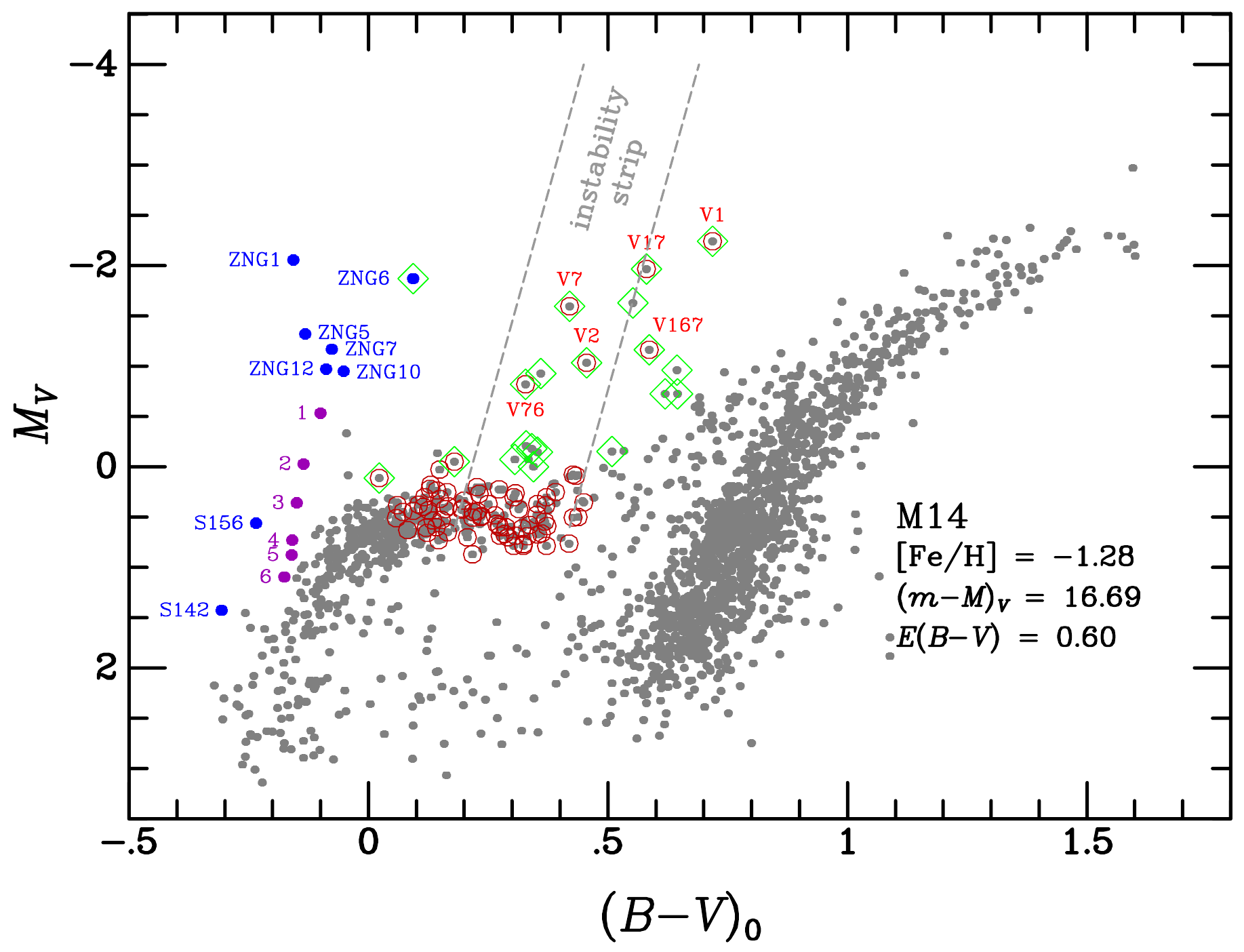}
\hfill
\includegraphics[width=3.4in]{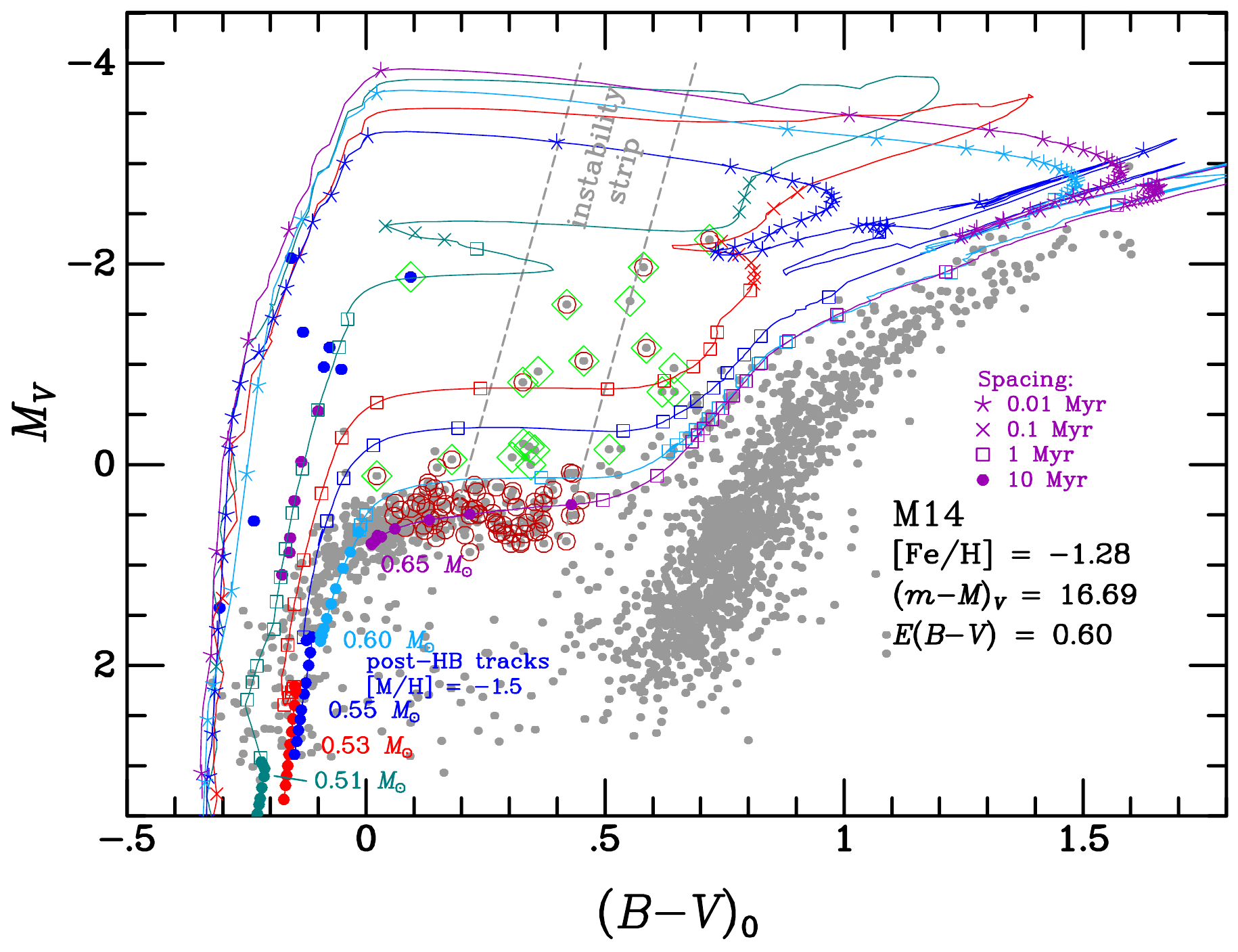}
\caption{ 
Color-magnitude diagram for members of M14 lying more than $25''$ from the cluster center (gray points). In the left panel, several individual hot blue stars of interest are labelled and are discussed in the text. The numerous RR~Lyrae variables and six Cepheids are encircled in red, and the Cepheids are labelled with their designations. Stars confirmed to have low $\log g$ from our \uBVI\/ photometry are enclosed in green diamonds. In the right panel, we superpose post-HB evolutionary tracks for the five ZAHB masses labelled in the figure. Evolutionary time steps are marked at intervals given by the symbols in the legend. 
}
\label{fig:m14_with_tracks}
\end{figure*}

Finally we examine two GCs that are unusually rich in AHB stars. The first is M14, an intermediate-metallicity system ($\rm[Fe/H]=-1.28$) that is among the dozen most luminous GCs in the Milky Way---but which unfortunately suffers moderate and spatially variable extinction of about $E(B-V)=0.60$.  Consequently, the cluster has been explored less extensively than other better-known GCs. M14's extraordinary population of AHB stars is displayed in the CMDs plotted in both panels of Figure~\ref{fig:m14_with_tracks}, with the objects having large Balmer jumps in our \uBVI\/ photometry enclosed in green diamonds. Due to crowding, we only consider member stars lying more than $25''$ from the cluster center, as indicated in Table~\ref{tab:exclusion}. M14's HB contains predominantly blue stars: LD94 give an HBR of $>$0.65 based on older data; however, more recent deep photometry, both ground-based \citep{ContrerasPena2013} and with \HST\/ \citep{Piotto2002}, shows that M14 also has a long ``blue tail'' of visually faint, hot EHB stars  that are not visible in our relatively shallow photometry. The HB also extends redward into the instability strip, and M14 hosts over 100 RRL variables (C01). 

For clarity, the left panel of Figure~\ref{fig:m14_with_tracks} shows only the CMD, with several stars of interest labelled. In the right panel, the CMD is overlain with post-HB evolutionary tracks from M+19 for the five ZAHB masses indicated in the labels. The two lowest-mass tracks, 0.51 and $0.53\,M_\odot$, are included because of the presence of the blue-tail EHB stars.

As is the case for the clusters considered in the previous subsections, ZAHB stars with masses around $\sim$0.55 to $0.60\,M_\odot$ initially evolve upward in the CMD toward higher luminosities, and then turn onto nearly horizontal paths toward the base of the AGB\null. Over the range from $(B-V)_0\simeq0$ to 0.5 these tracks are at visual brightnesses of up to $\sim\!1$~mag above the ZAHB\null. Thus we expect to see a sequence of M14 stars around $M_V \simeq -0.5$. Indeed, there are a few such objects in our CMD, although we caution that some of points displayed in the figure  (in particular, the non-variable objects  in the instability strip at $M_V\simeq0$) may actually be blends, as described in \S\ref{sec:interlopers}. The available \HST\/ imaging of M14 has only limited spatial coverage; however, with the aid of Wide Field Planetary Camera~2 frames taken by G.~Piotto (GO-8118) and F.~Ferraro (GO-11975), we did verify that several of objects located above the HB are indeed close pairs.

M14 contains six Type~II Cepheids, listed by C01 and \citet{ContrerasPena2018}, including two BL~Her variables (periods of 1.89 and 2.79~days) and four W~Virginis Cepheids ($P = 6.20$, 12.09, 13.60, and 18.76~days). Only three other Galactic GCs are known to contain more Cepheids (\citealt{Bono2020}, Table~A.1). All six of these variables are labelled in the left panel of Figure~\ref{fig:m14_with_tracks}. The post-HB tracks in the right panel of the figure suggest that the two BL~Her variables are descended from stars of ZAHB masses of about $0.55\,M_\odot$. 

In addition to the Cepheids, M14 contains a remarkable number of hot AHB stars.  One of these stars, ZNG\,6, is identified as an AHB object in Table~\ref{tab:meta_catalog}; the \Gaia\/ positions and our dereddened photometry for the remaining objects are given in Table~\ref{table:hot_stars_m14+m10}. The six brightest of these stars were recognized by ZNG and are plotted as filled blue circles in both panels, and are labelled in the left panel of the figure. The very brightest and hottest of these stars is ZNG\,1. Two optically fainter, extremely hot stars, which were cataloged by \citet{Schiavon2012} based on {\it GALEX\/} ultraviolet photometry, are also labelled in the figure. ZNG\,1 and the Schiavon et al.\ objects appear to be on their final descent toward the WD cooling track; unfortunately, their initial masses are uncertain, since all of the tracks are nearly superposed on each other. 

Interestingly, a group of five cooler ZNG stars lies redward of the post-AGB tracks. The brightest of this group, ZNG\,6, shows a large Balmer jump in our \uBVI\/ photometry and, as noted above, is included our AHB catalog (Table~\ref{tab:meta_catalog}); the remaining ZNG stars are bluer than our catalog cutoff at $(B-V)_0=-0.05$. These objects appear to fall at the top of a separate sequence traced by half a dozen fainter and previously uncataloged hot stars. We emphasize these objects by plotting them with filled purple circles in Figure~\ref{fig:m14_with_tracks}.  These stars are likely evolving {\it upward\/} in the CMD, and were once blue-tail EHB stars with initial ZAHB masses in the range $0.51 M_{\odot} \lesssim M \lesssim 0.53 M_{\odot}$.  The evolutionary tracks of these objects turn toward cooler temperatures and cross the instability strip at a luminosity well above the HB\null.  The exact luminosity where this happens is very sensitive to ZAHB mass; however, since the evolution at this stage is relatively slow, as shown by symbols spacing of 1~Myr, it should be possible to find stars in this phase of evolution and determine their status. The brightest of these stars should eventually turn back toward higher temperatures before reaching the AGB, and re-cross the instability strip, but on much shorter timescales. The lowest-mass track shown in the figure is for $0.51\,M_\odot$.  Stars with this mass only briefly cross through the instability strip near the end of their evolution. ZAHB stars of even lower masses become AGB-manqu\'e objects, and never produce AHB objects or Cepheids.

\begin{deluxetable}{lCCCC}
\tablecaption{Blue Stars in M14 and M10\label{table:hot_stars_m14+m10}}
\tablehead{
\colhead{ID}
&\colhead{RA [J2000]} 
&\colhead{Dec [J2000]}
&\colhead{$(B-V)_0$}
&\colhead{$M_V$}\\
\colhead{}
&\colhead{[deg]} 
&\colhead{[deg]}
&\colhead{}
&\colhead{}
}
\decimals
\startdata
\noalign{\smallskip}
\multispan5{\hfil M14 (NGC 6402) \hfil}\\
\noalign{\smallskip}
ZNG 1   & 264.38821   &	-3.24772  & -0.156  & -2.056   \\
ZNG 5	& 264.37908   &	-3.25108  & -0.131  & -1.319   \\
ZNG 7	& 264.38279   & -3.24794  & -0.076  & -1.167   \\
ZNG 12  & 264.42592   & -3.24986  & -0.088  & -0.969   \\
ZNG 10	& 264.39900   &	-3.27500  & -0.051  & -0.950   \\
1	& 264.41812   &	-3.25728  & -0.100  & -0.531   \\
2	& 264.42775   &	-3.23200  & -0.136  & -0.031   \\
3	& 264.44512   &	-3.26456  & -0.150  &  0.358   \\
S 156	& 264.37875   &	-3.29439  & -0.234  &  0.562   \\
4	& 264.39233   &	-3.13619  & -0.159  &  0.729   \\
5	& 264.39517   &	-3.19219  & -0.160  &  0.873   \\
6	& 264.35929   &	-3.17403  & -0.176  &  1.096   \\
S 142	& 264.35858   &	-3.24872  & -0.307  &  1.425   \\
\noalign{\smallskip}
\multispan5{\hfil M10 (NGC 6254) \hfil}\\
\noalign{\smallskip}
ZNG 1	& 254.28871   &	-4.07339  & -0.305  & -0.567   \\
ZNG 2	& 254.29117   &	-4.07453  & -0.103  & -0.307   \\
ZNG 5	& 254.27337   &	-4.11444  & -0.073  & -0.043   \\
ZNG 8	& 254.24025   & -4.12103  & -0.088  &  0.215   \\
1	& 254.30162   &	-4.09675  & -0.046  &  0.216   \\
2	& 254.27933   &	-4.10208  & -0.099  &  0.265   \\
3	& 254.23421   &	-3.92800  & -0.273  &  1.212   \\
S 241	& 254.31158   &	-4.08417  & -0.227  &  1.398   \\
\enddata
\end{deluxetable}

\begin{figure*}
\centering
\includegraphics[width=3.4in]{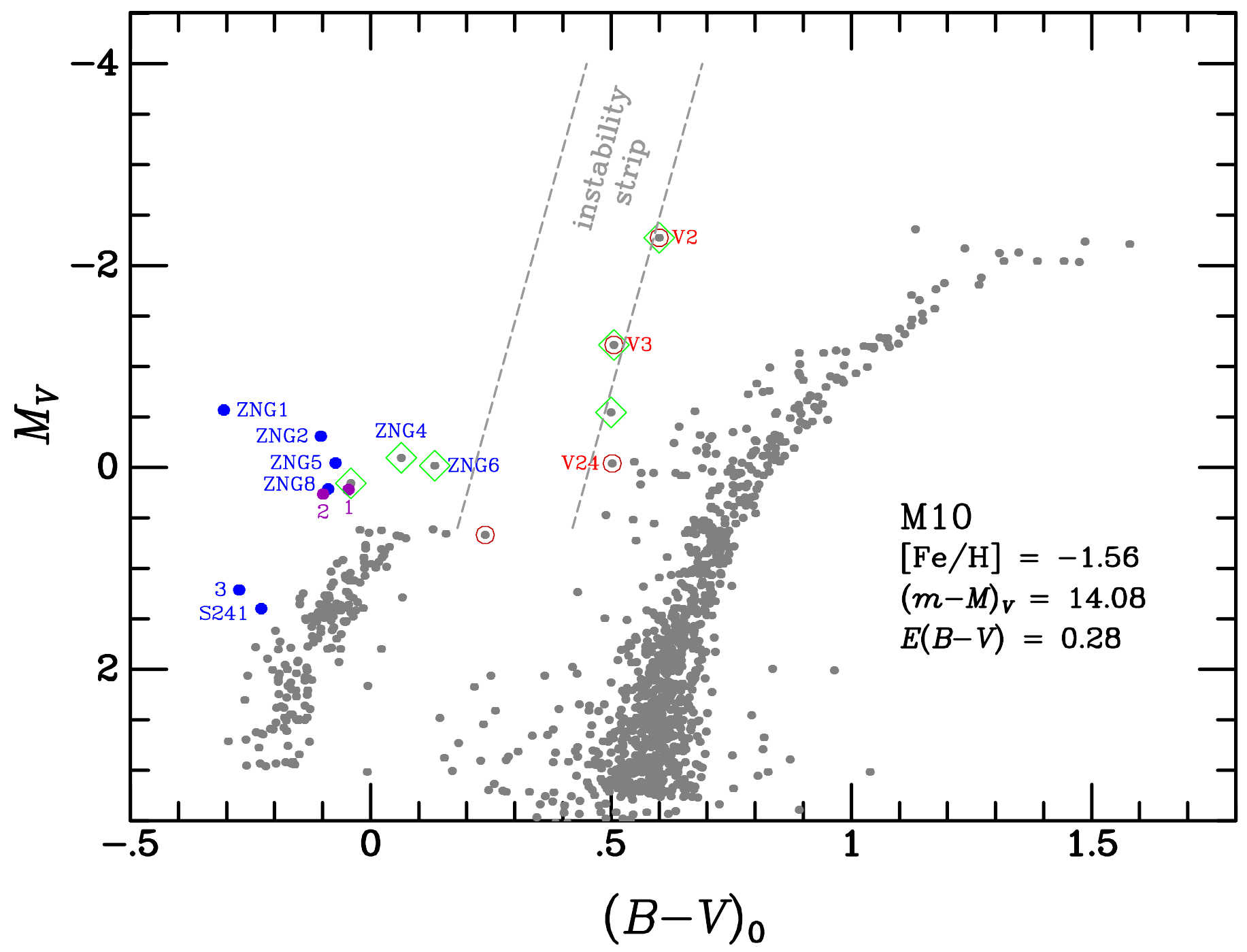}
\hfill
\includegraphics[width=3.4in]{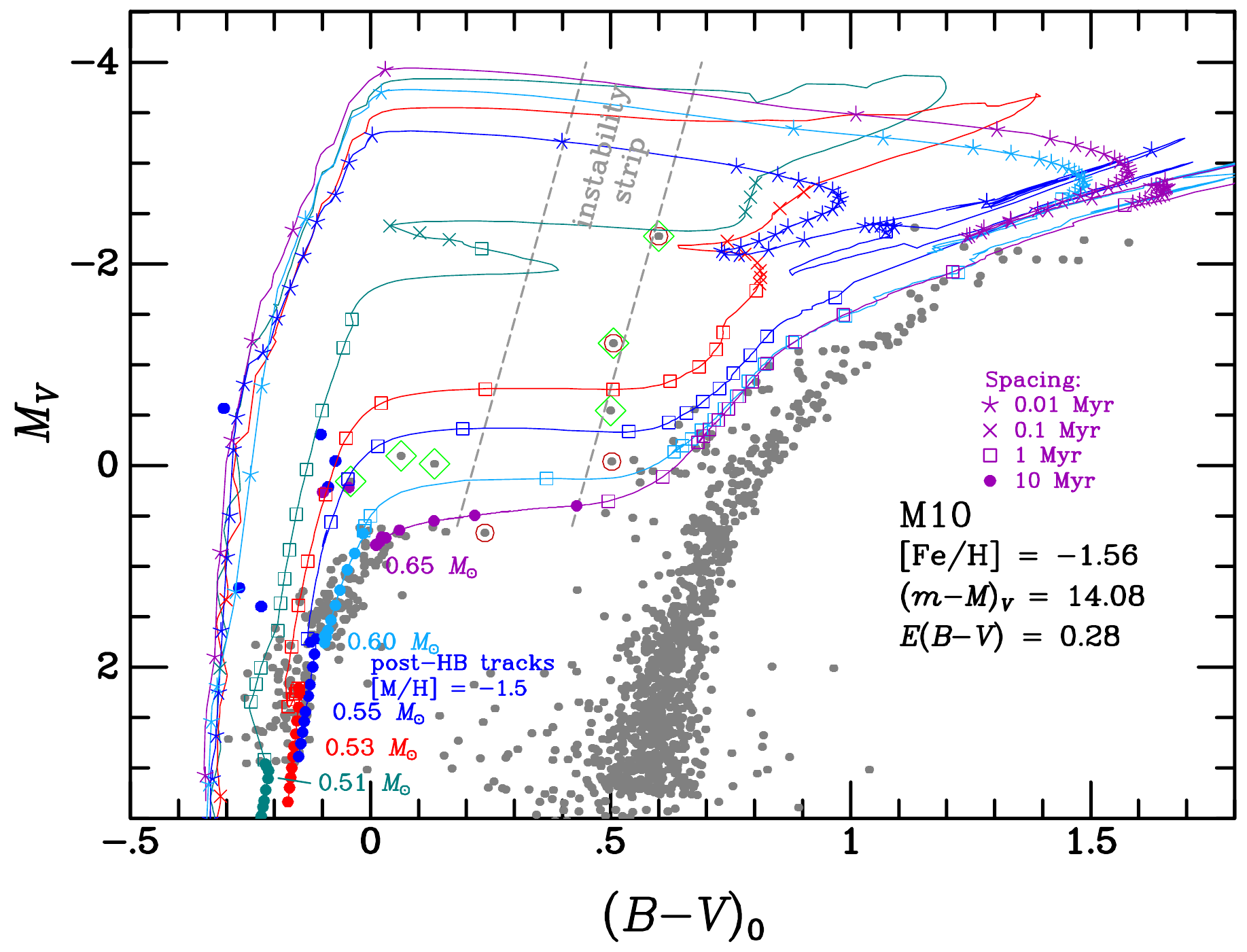}
\caption{ 
Color-magnitude diagram for M10 (gray points). In the left panel, several individual hot blue stars of interest are labelled; they are discussed in the text. The single RR~Lyrae variable and three Cepheids are encircled in red, and the Cepheids are labelled in the left panel with their designations. Stars confirmed to have low $\log g$ from our \uBVI\/ photometry are enclosed in green diamonds. In the right panel, we superpose post-HB evolutionary tracks for the five ZAHB masses labelled in the figure. Evolutionary time steps are marked at intervals given by the symbols in the legend. 
}
\label{fig:m10_with_tracks}
\end{figure*}

We now turn to M10, a GC that is slightly more metal deficient than M14 ($\rm[Fe/H]=-1.56$). Although it is less populous, with an absolute visual luminosity that is only 22\% that of M14, it nevertheless joins M14 in hosting a remarkable population of AHB stars. We show our CMD of M10 in both panels of Figure~\ref{fig:m10_with_tracks}. In the left panel, several stars of interest are labelled; in the right panel, we superpose five evolutionary tracks from M+19 for the ZAHB masses indicated in the figure. M10's HB is extremely blue: the cluster hosts only one known RRL variable, its LD94 HBR is 0.94, and the \citet{Torelli2019} HB index is $\tauhb=11.03$. M10, like M14, contains a sparse blue tail of very hot, optically faint EHB stars, which are seen in deep CMDs obtained from the ground \citep[e.g.,][]{vonBraun2002, Pollard2005, ArellanoFerro2020} and with \HST\/ \citep{Piotto1999}.

M10 hosts three Type~II Cepheids. Two of them, V2 and V3, belong to the W~Vir class, with periods of 7.83 and 18.82~days respectively (see C01); the third is a shorter-period (2.31~days) BL~Her variable discovered recently by \citet{ArellanoFerro2020} and designated V24. All three are labelled in the CMD in the left panel of Figure~\ref{fig:m10_with_tracks}. As in previous figures, our cataloged stars with large Balmer jumps are enclosed in green diamonds. One of these objects lies between V24 and V3 in the CMD, just redward of the instability strip, and is not a known variable.

M10---like M14---contains half a dozen bright blue stars that were first identified by ZNG\null. We plot them as filled blue circles in the CMD diagrams, label them in the left panel, and, for the objects not already in our AHB catalog, give their positions and $BV$ photometry in Table~\ref{table:hot_stars_m14+m10},  To these objects, we add three slightly fainter blue stars found in our data; these are displayed either as gray points enclosed in a green diamond (for the star which is listed in our AHB catalog) or as filled purple circles (for the two that are slightly bluer than our catalog cutoff).  The brightest and hottest of these objects, ZNG\,1, lies on the superposed post-AGB tracks shown in the right-hand panel of Figure~\ref{fig:m10_with_tracks}. A fainter, uncataloged hot star, plotted with a filled blue circle and labelled ``3,'', also appears to lie on the post-AGB tracks.

As in the case of M14, the remaining ZNG stars lie redward of the post-AGB sequence---as does a fainter hot blue star identified in \GALEX\/ images by \citet{Schiavon2012}, labelled as S241 in the left-hand panel. These objects lie at locations consistent with them being post-ZAHB stars that are evolving {\it upward\/} in the CMD\null. ZNG\,4\footnote{ZNG\,4 was misidentified with a nearby red giant in \citet{Bond2021}. The correct identification is given here in Table~\ref{tab:meta_catalog}.} and ZNG\,6 appear to be stars with ZAHB masses of about 0.55 to $0.60\,M_\odot$, and are the immediate progenitors of BL~Her Cepheids like V24. Presumably, these objects will reach the base of the AGB and evolve up towards higher luminosities. Stars with slightly lower ZAHB masses of about 0.51 to $0.55\,M_\odot$, including ZNG\,2, 5, and 8, and the stars marked with purple circles and labeled ``1'' and ``2,'' are plausibly destined to become W~Vir Cepheids like V2 and V3.

The evolutionary status of Type~II Cepheids has been discussed recently by \citet{Bono2020} (see also the extensive references therein). We concur with Bono et al.\ that the short-period (BL~Her) Cepheids are in post-HB evolutionary states, on their approach to the base of the AGB\null. Bono et al.\ argue that the longer-period W~Vir variables are a mixture of PEAGB and PAGB stars evolving toward higher temperature. However, in the context of the M+19 models considered here, we interpret the W~Vir variables as being descended from low-mass EHB stars. In this picture, they are predominantly on their {\it first\/} crossings of the instability strip, evolving toward {\it lower\/} temperatures. At least in the cases of M14 and M10, the sequence of their immediate warmer progenitors appears to be detected. Specifically, we see a group of hot post-EHB stars increasing their luminosity on a fairly slow timescale, and forming a group of blue stars, the brightest of which were identified by ZNG\null. From there, the stars' evolution quickens, as they move redward in the CMD, passing through the instability strip. Few, if any, of the W~Vir stars can be in PEAGB or PAGB states on a second crossing toward higher temperature, as a consequence of the very rapid evolutionary timescales at those stages. Of course, this scenario needs to be tested with further calculations of post-HB evolution, using a range of parameters including masses and chemical composition. 

Measuring period changes in the Type~II Cepheids provides another empirical test of the direction of their evolution. \citet{Wehlau1982} studied period changes of 12 GC Cepheids with periods of 1.13 to 7.90~days, and found increasing periods in nine of them, and no case of a decreasing period. More recently, \citet{Osborn2019} listed period changes measured for 18 BL~Her variables (six in the field, the rest in GCs), with periods of 1.11 to 5.11~days; all but two show increasing periods, and again none have decreasing periods. These findings are consistent with evolution toward lower temperatures, for which there is a general consensus in the extensive literature on this subject. The situation with longer-period W~Vir Cepheids is less clear, as reviewed by \citet{Neilson2016}. Here there are again predominantly positive period changes, indicating evolution to the red as we have suggested. However, several W~Vir stars show decreasing periods, especially at longer pulsation periods. This could imply that at least some of these objects are either low-mass stars of $\sim\!0.52\,M_\odot$ that are turning back toward higher temperature, stars that are undergoing TPs, or objects on their final post-AGB evolution to high temperature.  This last possibility is almost certainly the case for long-period RV~Tau variables. However, at long periods, especially among objects classified as RV~Tau stars, the pulsations can become erratic, making it difficult to measure changes due to secular evolution. 

Lastly, we point out that in the foregoing discussion we only considered evolutionary tracks for single stars. It is likely that scenarios involving binary interactions may be capable of, for example, stripping the envelopes of stars ascending the RGB or AGB, and sending them into the AHB region of the CMD. 

\section{The AHB Populations in Globular Clusters}
\label{sec:ahb_pop}

In this section we discuss a few general properties of the population of AHB stars in GCs. We start with a classification scheme, and then examine correlations with the metallicities and HB morphologies of the host clusters. Finally we give a brief discussion of the luminous PAGB stars.

\subsection{AHB Classification Scheme \label{subsec:classification}}

\begin{figure}[hbt]
\centering
\includegraphics[width=3.35in]{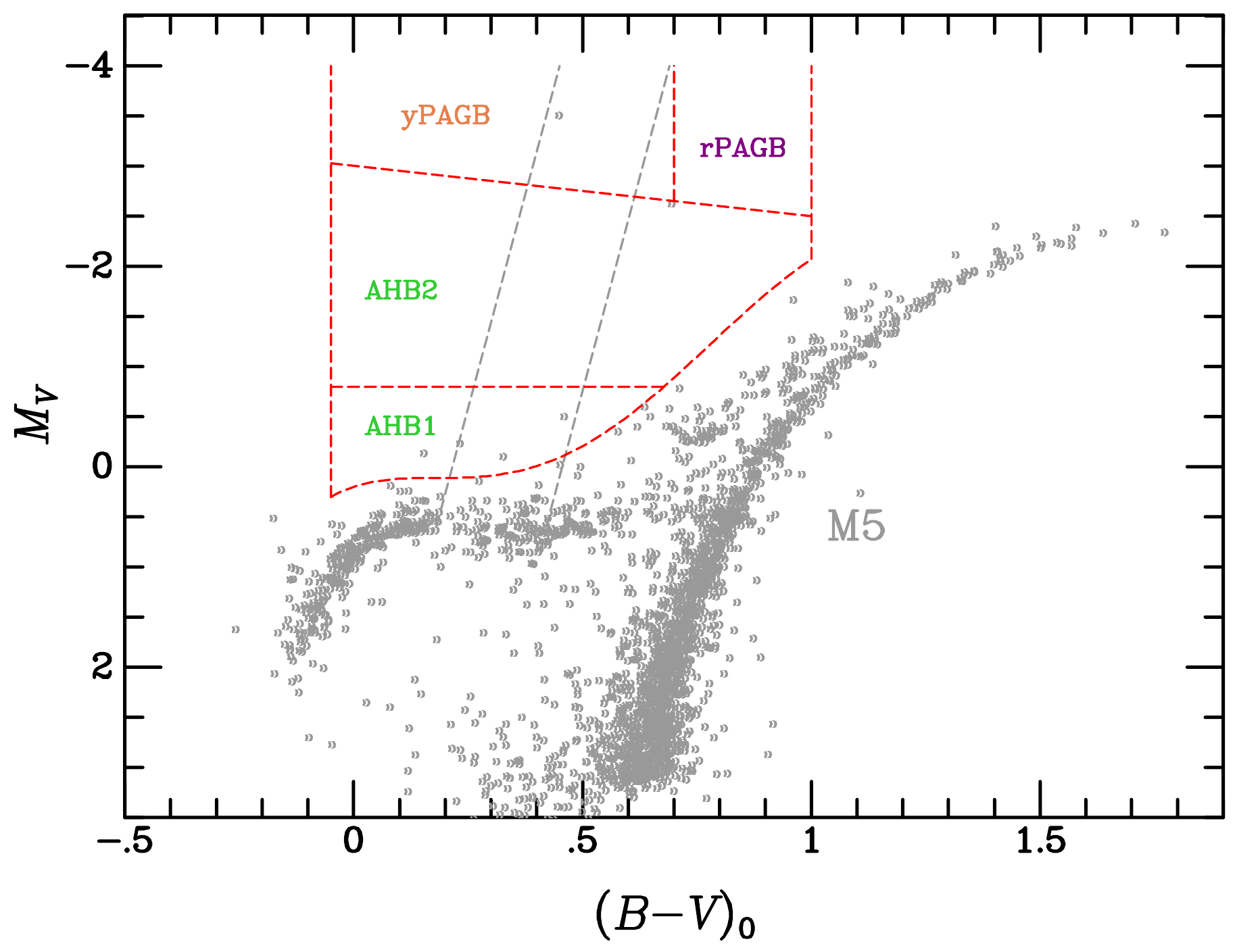}
\caption{ 
Classification scheme for AHB stars in globular clusters. AHB1: stars fainter than $M_V=-0.8$, evolving from the blue horizontal branch toward the base of the AGB. AHB2: mixture of stars departing from the early AGB toward higher temperatures, and low-mass extreme horizontal-branch stars evolving toward the AGB for the first time. yPAGB and rPAGB: yellow and red post-AGB stars. See \S\ref{subsec:classification} for details. 
}
\label{fig:classification}
\end{figure}

We adopt a simple classification scheme for our list of low surface-gravity AHB stars, based on their locations in the dereddened $M_V$ versus $(B-V)_0$ CMDs of their host clusters.  This scheme is a modified version of that presented by \citet{Bond2021}, which was based on photometry in the \Gaia\/ $M_G$ versus $BP-RP$ system. In the present paper, we only consider AHB stars in the color range $-0.05 \leq (B-V)_0 \leq 1.0$, within which our \uBVI\/ photometry is sensitive to the size of the Balmer discontinuity. Figure~\ref{fig:classification} shows our four classification boxes superposed on the CMD of M5. Figure~\ref{fig:schematicwithtracks} in \S\ref{subsec:post-hb_evolution} illustrates a sample of M+19 post-ZAHB evolutionary tracks superposed on this same CMD.

The classification boxes are as follows:

1. AHB1: objects lying between the AHB selection cutoff defined by Equation~\ref{eq:cmd}, and an absolute magnitude of $M_V=-0.8$.  The ``AHB1'' terminology is adopted from \citet[][and references therein]{Sandage2006}, and the upper limit for the brightness of the class is roughly the same as that used by \citet{Sandage2006}; it is the luminosity of a Type~II Cepheid with a period of about 3~days. As indicated by the evolutionary tracks shown in the previous section, AHB1 stars were initially on the BHB and EHB, and are now evolving across the CMD, above the level of the ZAHB, on their way to the base of the AGB.

2. AHB2: stars with absolute magnitudes brighter than $M_V=-0.8$ [for $-0.05\le(B-V)_0\le0.678$] or brighter than the limit given by Equation~\ref{eq:cmd} [for $0.678<(B-V)_0\le1.0$], but fainter than the lower limit of post-AGB stars defined below. These objects include both Type~II Cepheids with periods longer than $\sim$3~days, and non-variable stars in this luminosity and color range. We caution that the term ``AHB2'' has a different meaning in \citet{Sandage2006} and papers cited therein, and should not be confused with our usage here.  Objects in this region of the CMD are conventionally considered to be PEAGB stars, which have reached the AGB and are now evolving back toward higher temperatures. However, our analysis of M15, M14, and M10 in the previous section suggests that many AHB2 objects may actually be low-mass post-EHB stars evolving to cooler temperatures for the first time. Additionally, as mentioned at the end of \S7, it is likely that binary interactions can produce AHB2 stars.

3. Yellow and red post-AGB stars: these rare objects are the visually brightest stars in GCs; they have likely departed from near the top of the AGB and are now evolving rapidly toward higher temperatures. The faint limit of the post-AGB region runs from $M_V=-3.025$ at $(B-V)_0=-0.05$ to $M_V=-2.5$ at $(B-V)_0=1$; this line corresponds to an approximately constant bolometric luminosity. We subdivide the post-AGB stars into yellow (yPAGB) and red (rPAGB) objects, with the division at $(B-V)_0=0.7$. 

Column~21 in our catalog of AHB stars (Table~\ref{tab:meta_catalog}) gives the stellar classifications according to this scheme. 

\null

\medbreak

\subsection{Metallicity Correlation\label{subsec:metallicity}}

\begin{figure}
\centering
\includegraphics[width=3.35in]{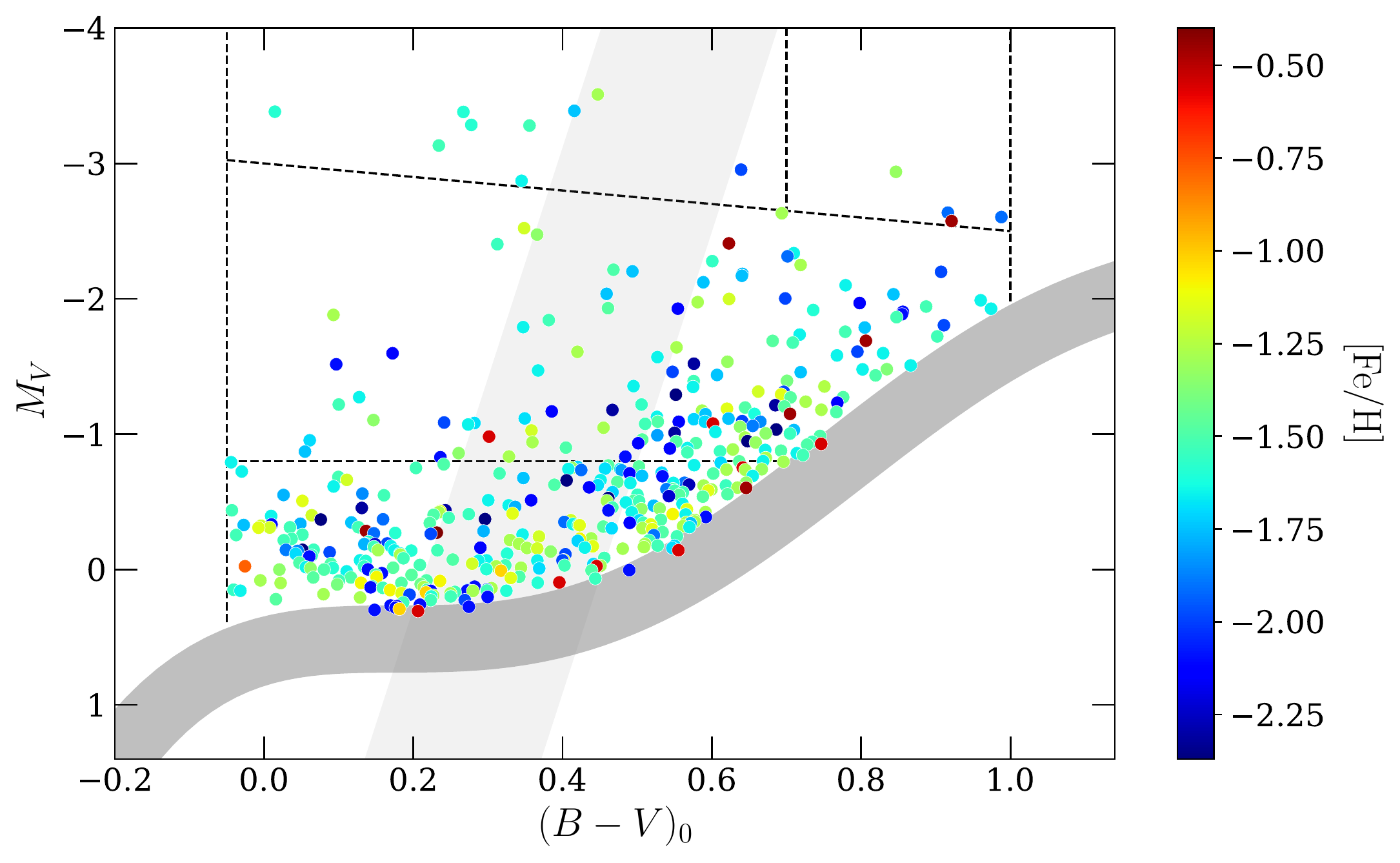}
\caption{
Color--absolute-magnitude diagram of our entire catalog of AHB stars.  The stars are color-coded by the metallicities of their host clusters, as indicated by the [Fe/H] color bar on the right. The nominal location of the Cepheid instability strip is represented in this figure and in Figure~\ref{fig:all_ahb_stars2} by the light-gray region, and the horizontal branch by the dark-gray region.}
\label{fig:all_ahb_stars}
\end{figure}

In Figure~\ref{fig:all_ahb_stars} we plot the CMD of our entire catalog of 438 AHB stars in 61 GCs, with the classification boxes from \S\ref{subsec:classification} superposed. The points are colored according to the metallicities of their host clusters, as indicated by the [Fe/H] color bar on the right.

Based on the comparison with the theoretical single-star evolutionary tracks shown in \S\ref{sec:theory-comp}, we have a strong expectation that AHB stars, especially those classified as AHB1, should preferentially be found in clusters of low metallicities. This is because AHB1 stars are likely descended from BHB and EHB objects, which in turn are found primarily in clusters with the lowest [Fe/H] values (the ``first parameter''). Moreover, while PAGB stars can in principle be present in clusters over the full range of metallicities, we expect most to associated with metal-poor systems.  As we point out in \S\ref{subsec:6362}, at the high-[Fe/H] end of the distribution, PAGB evolutionary timescales are relatively rapid; thus low-metallicity clusters are more favorable for hosting these objects.

Figure~\ref{fig:all_ahb_stars} generally agrees with these expectations. There is a strong preference, especially in the AHB2, yPAGB, and rPAGB classification boxes, for the stars to be associated with clusters with relatively low metallicities. Note also that there are more AHB1 stars than AHB2 objects: 221 AHB1 (after subtracting RRL interlopers caught at maximum; see \S\ref{subsec:catalogs}) versus 145 AHB2. This is qualitatively consistent with the slower evolutionary timescales for AHB1 stars evolving toward the base of the AGB\null.  In contrast, AHB2 stars are in more advanced evolutionary stages (as shown by the time-steps encoded into the figures of \S\ref{sec:theory-comp}). For the very rapidly evolving and luminous PAGB stars, we find only 9 yPAGB objects, and only 4 rPAGB stars, in our entire sample.

One apparent discrepancy between expectations and observations is posed by the 15 red points in Figure~\ref{fig:all_ahb_stars}, which represent AHB stars within the most metal-rich GCs in our survey. However, all but one of these stars belong to either NGC\,6388 and NGC\,6441---both of which are anomalous systems. The metallicities of NGC\,6388 and NGC\,6441 are $\rm[Fe/H]=-0.55$ and $-0.46$, respectively (H10), and, consistent with their high metal content, the HBs of both GCs are dominated by large numbers of RHB stars. LD94 do not give HBRs for these clusters, but the \citet{Torelli2019} $\tauhb$ values are near the ``red'' end of the distribution: 1.88 and 1.55. Nevertheless, and unusually, the HBs of both clusters extend to the blue, and even contain a few very hot and optically faint ``blue-tail'' EHB stars \citep[e.g,][]{Rich1997,Pritzl2003,BondM15_2020}. In addition, both clusters are remarkably rich in RRL variables and Type~II Cepheids \citep[e.g.,][]{Corwin2006}; in fact, these two GCs have the largest known populations of Cepheids in the entire Milky Way GC system: 12 and 8 stars, respectively. Thus, as in M14 and M10 (see \S\ref{subsec:m14}), we find an association of AHB stars and W~Vir Cepheids with the presence of extremely hot HB stars.\footnote{We believe the connection of bright nonvariable AHB2 stars with clusters containing hot HB stars is a new result, but the association of W~Vir Cepheids with such clusters has been known since the study of \citet{Wallerstein1970}.}

\subsection{Correlation with Horizontal-Branch Morphology\label{subsec:morphology}}

There is no one-to-one relation between a GC's metallicity and the color distribution of its HB stars (the ``second-parameter'' phenomenon).  However, a robust prediction of the M+19 stellar-evolution tracks is that the population of a cluster's AHB stars should be well-correlated with its HB morphology. GCs with predominantly red HBs should have few AHB stars, especially in the AHB1 and AHB2 categories, though they may produce a few luminous PAGB objects. Conversely, clusters with significant populations of BHB stars are expected to produce AHB1 objects, and systems with a sequence of faint, hot EHB stars should generate AHB2 objects, as discussed in \S\ref{sec:theory-comp} and at the end of the previous subsection.  These BHB systems should also contain relatively more PAGB stars.

\begin{figure}
\centering
\includegraphics[width=3.35in]{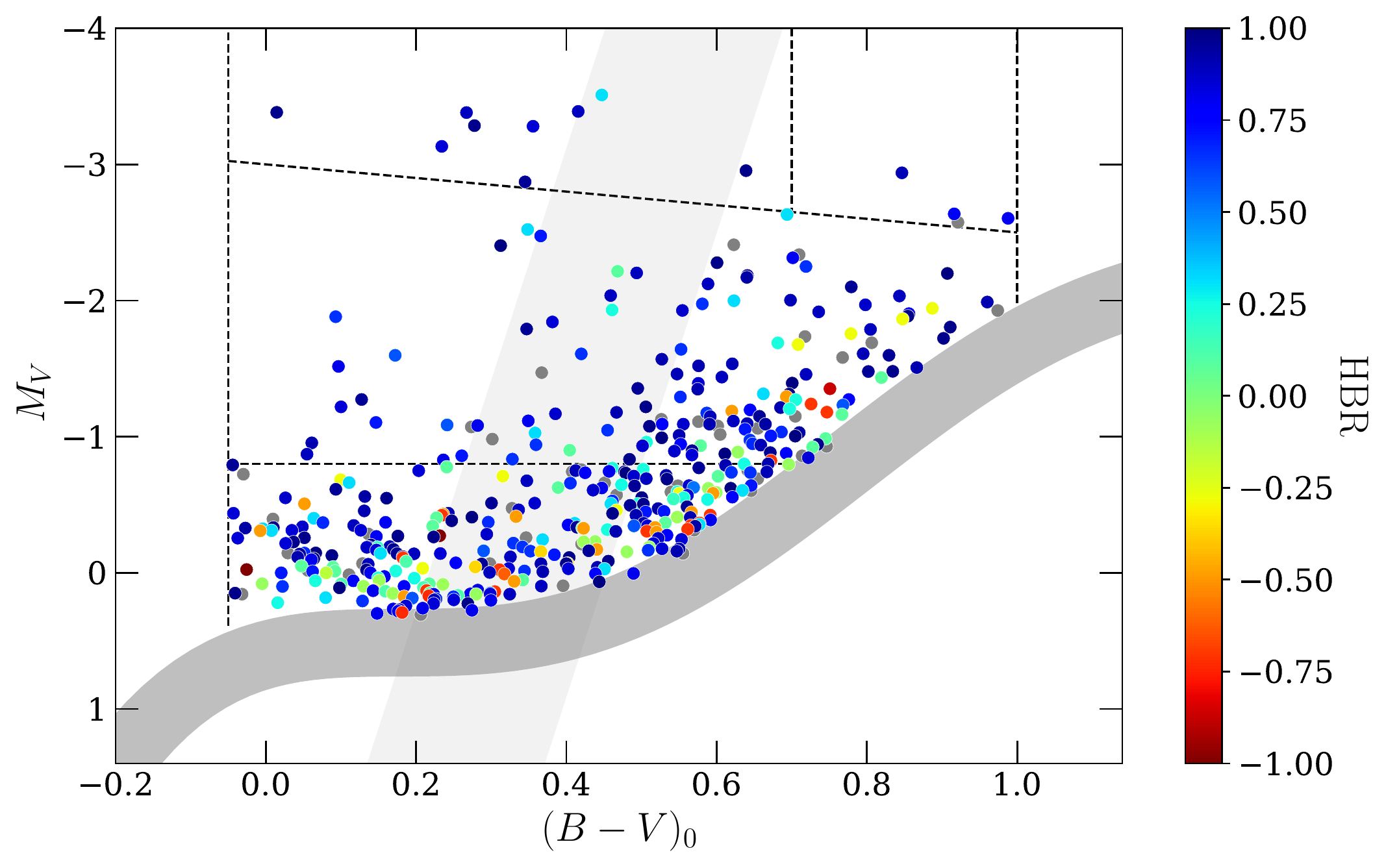}
\caption{ 
Color--absolute-magnitude diagram of our entire catalog of AHB stars (Table~\ref{tab:meta_catalog}), color-coded by the LD94 horizontal-branch ratio of their host clusters, as indicated by the color bar. Stars from clusters with no published HBR are shown in grey.  The approximate location of the instability strip and the location of the horizontal branch are shown in light gray and dark gray, respectively.
}
\label{fig:all_ahb_stars2}
\end{figure}

Figure~\ref{fig:all_ahb_stars2} repeats the previous CMD for our AHB catalog, but color-codes the points according to their clusters' LD94 HDR values. Now we see that the AHB stars are indeed strongly correlated with clusters having blue HBs. Most AHB host clusters have HBR values greater than 0.5; this is not surprising since it appears well established that AHB objects are the direct descendants of hot ZAHB stars. But the brighter AHB2 stars also appear to form primarily in clusters with blue HBs. This supports our view that AHB2 stars, including W~Vir Cepheids, are mostly objects that were on the BHB and EHB and are now evolving toward the AGB, not away from it.

Again there is an apparent anomaly in the form of a few AHB stars being associated with clusters with red horizontal branches. These are the objects discussed in the previous subsection, which come primarily from the two anomalous GCs, NGC\,6388 and 6441. These metal-rich clusters have extremely red HBs, with values of HBR approaching the limit of $-1$, but they also possess a population of hot BHB and EHB stars.

As a further illustration of the strong dependence of the AHB population upon HB morphology, we selected a set of 31 relatively nearby, and generally lightly reddened clusters. These systems all contain well-populated HBs, and have high-quality \uBVI\ data. We sorted these clusters into two groups: those with ``red'' HBs (11 GCs totalling $1.5 \times 10^6 L_{\odot}$ of surveyed $V$-band luminosity) and those with ``blue'' HBs (20 GCs with $L_V \simeq 2.1 \times 10^6 L_{\odot}$).  Table~\ref{tab:red_vs_blue} lists these clusters, along with their LD94 HBRs, $\tauhb$ indices \citep{Torelli2019}, reddenings, and metallicities (H10). All of the red-HB clusters have negative HBR ratios, and more than half of them having the smallest possible value of $-1.00$. Conversely, the blue-HB clusters mostly have HBR values greater than 0.85. 


\def\d{\dots}

\begin{deluxetable}{lCCCC}
\tablecaption{Clusters with Red and Blue Horizontal Branches\label{tab:red_vs_blue}}
\tablehead{
\colhead{Cluster}
&\colhead{HBR\tablenotemark{a}} 
&\colhead{$\tauhb$\tablenotemark{b}} 
&\colhead{$E(B-V)$\tablenotemark{c}}
&\colhead{[Fe/H]\tablenotemark{c}}
}
\decimals
\startdata
\noalign{\smallskip}
\multispan5{\hfil ``Red'' Horizontal Branches \hfil} \\
\noalign{\smallskip}
M69      & -1.00 & \d   & 0.18 & -0.64 \\
M107     & -0.76 & 0.28 & 0.33 & -1.02 \\
NGC 362  & -0.87 & 3.24 & 0.05 & -1.26 \\
NGC 1261 & -0.70 & 2.42 & 0.01 & -1.27 \\
NGC 6352 & -1.00 & 1.44 & 0.22 & -0.64 \\
NGC 6356 & -1.00 & \d   & 0.28 & -0.40 \\
NGC 6362 & -0.58 & 2.24 & 0.09 & -0.99 \\
NGC 6496 & -1.00 & 0.11 & 0.15 & -0.46 \\
NGC 6652 & -1.00 & 0.64 & 0.09 & -0.81 \\
NGC 6723 & -0.08 & 3.38 & 0.05 & -1.10 \\
47 Tuc   & -1.00 & 1.58 & 0.04 & -0.72 \\
\noalign{\smallskip}
\multispan5{\hfil ``Blue'' Horizontal Branches \hfil} \\
\noalign{\smallskip}
M10 	 & 0.94 & 11.03 & 0.28 & -1.56 \\
M12 	 & 0.92 &  9.05 & 0.19 & -1.37 \\
M22 	 & 0.94 &  6.53 & 0.34 & -1.70 \\
M30 	 & 0.88 &  6.40 & 0.03 & -2.27 \\
M53 	 & 0.76 &  6.67 & 0.02 & -2.10 \\
M55 	 & 0.91 &  6.59 & 0.08 & -1.94 \\
M79 	 & 0.89 & \d    & 0.01 & -1.60 \\
M80 	 & 0.92 &  7.86 & 0.18 & -1.75 \\
M92 	 & 0.88 &  8.95 & 0.02 & -2.31 \\
NGC 288  & 0.95 &  9.39 & 0.03 & -1.32 \\
NGC 4372 & 1.00 & \d    & 0.39 & -2.17 \\
NGC 5466 & 0.68 &  5.02 & 0.00 & -1.98 \\
NGC 5897 & 0.91 & \d    & 0.09 & -1.90 \\
NGC 5986 & 0.95 &  7.85 & 0.28 & -1.59 \\
NGC 6101 & 0.84 &  5.43 & 0.05 & -1.98 \\
NGC 6144 & 1.00 &  4.98 & 0.36 & -1.76 \\
NGC 6397 & 0.93 &  8.29 & 0.18 & -2.02 \\
NGC 6541 & 1.00 & 10.26 & 0.14 & -1.81 \\
NGC 6752 & 1.00 & 13.94 & 0.04 & -1.54 \\
NGC 7492 & 0.90 & \d    & 0.00 & -1.78 \\
\enddata
\tablenotetext{a}{Horizontal-Branch Ratio from LD94.}
\tablenotetext{b}{Horizontal-branch index from \citet{Torelli2019}.}
\tablenotetext{c}{From H10.}
\end{deluxetable}

\begin{figure}
\includegraphics[width=3.35in]{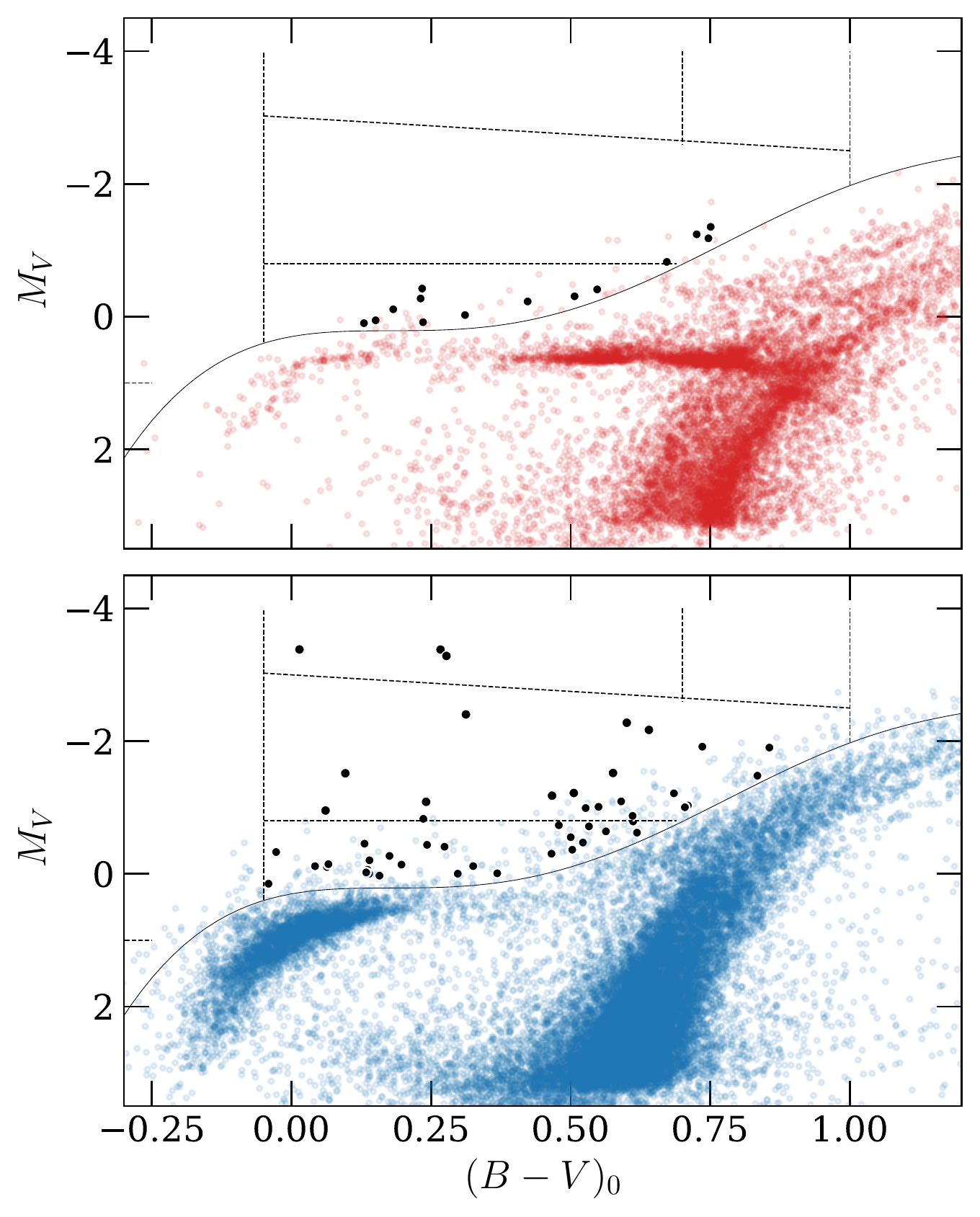}
\caption{Combined color-magnitude diagrams of clusters with very red (top) and very blue (bottom) HBs, as listed in Table~\ref{tab:red_vs_blue}.  Only those stars with membership probabilities greater than 0.8 are plotted. AHB objects from our catalog are displayed as filled black circles. The classification boxes from Figure~\ref{fig:classification} are superposed.  Clusters with red HBs contain no PAGB stars, and only four AHB2 stars, all near the red limit of our survey.  In contrast, clusters with blue HBs have several PAGB stars, and numerous AHB2 stars. Both groups contain AHB1 stars, but they are more numerous in the blue-HB clusters. These findings are generally consistent with AHB1 stars originating from objects on the blue HB, and AHB2 stars being the descendants of more extreme BHB or EHB objects.
}
\label{fig:red_vs_blue_cmds}
\end{figure}

The two panels in Figure~\ref{fig:red_vs_blue_cmds} show the combined CMDs for all of the cluster stars (membership probability greater than 0.8) in these two groups. The filled black circles mark the AHB stars that are listed in our catalog in Table~\ref{tab:meta_catalog}. Remarkably, the clusters with red HBs contain no PAGB stars, and only four AHB2 stars, all of which are at the red limit of our survey and barely qualified for inclusion. In contrast, the blue group hosts nearly two dozen AHB2 stars, and three PAGB objects. The red group does contain about 10 AHB1 stars, but these are plausibly descended from the weak sequence of BHB stars in their host clusters. The blue group hosts about two dozen AHB1 objects, which is consistent with this group's considerably richer population of BHB and EHB stars.

\subsection{Post-AGB Stars}

\begin{deluxetable*}{llCcCCCCl}[hbt]
\tablecaption{Post-AGB Stars in Galactic Globular Clusters\label{tab:post-AGB}}
\tablehead{
\colhead{Cluster}
&\colhead{Desig.} 
&\colhead{$M_V$} 
&\colhead{Classif.\tablenotemark{a}}
&\colhead{RA [J2000]} 
&\colhead{Dec [J2000]} 
&\colhead{HBR\tablenotemark{b}}
&\colhead{$\tau_{\rm HB}$\tablenotemark{c}}
&\colhead{Notes} \\
\colhead{}
&\colhead{} 
&\colhead{} 
&\colhead{}
&\colhead{[deg]} 
&\colhead{[deg]} 
&\colhead{}
&\colhead{}
&\colhead{} 
}
\decimals
\startdata
NGC 1904/M79           & PAGB      & -3.38 & yPAGB  &      081.04318 & -24.48910  &  0.89  &  \d   &	   	\\
NGC 5139/$\omega$ Cen  & V1        & -3.28 & yPAGB  &      201.52153 & -47.39518  &  0.89  &  \d   &	 RV Tau \\
\quad\quad$''$         & HD 116745 & -3.13 & yPAGB  &      201.60964 & -47.27429  &  $''$  &  \d   &	    		\\
NGC 5824               & PAGB 1    & -2.64 & rPAGB  &      225.99058 & -33.06770  &  0.82  &  \d   &	    		\\
\quad\quad$''$         & PAGB 2    & -2.60 & rPAGB  &      225.99838 & -33.06708  &  $''$  &  \d   &	    		\\
NGC 5904/M5            & V84       & -3.51 & yPAGB  &      229.65062 & +02.07117  &  0.37  &  5.04 &	 RV Tau 	\\
NGC 5986               & PAGB 1    & -3.38 & yPAGB  &      236.51396 & -37.77896  &  0.95  &  7.85 &			\\
\quad\quad$''$         & PAGB 2    & -3.28 & yPAGB  &      236.52057 & -37.78403  &  $''$  &  $''$ &			\\
NGC 6273/M19           & ZNG 4     & -3.39 & yPAGB  &      255.64659 & -26.25670  &  0.86  &  \d   &			\\
NGC 6441               & V127      & -2.57 & rPAGB\rlap{\tablenotemark{d}} & 267.55020 & -37.05339  & -0.73  & 1.55 & W Vir \\
NGC 6626/M28           & V17, ZNG\,5 & -2.94 & rPAGB  &      276.14933 & -24.88778  &  0.88  &  \d   & RV Tau	\\
NGC 6779/M56           & V6        & -2.95 & yPAGB  &      289.14905 & +30.19413  &  0.98  &  7.38 &	 RV Tau 	\\
NGC 7089/M2            & V11       & -2.87 & yPAGB  &      323.38507 & -00.81828  &  0.96  &  8.23 &	 RV Tau 	\\
\enddata
\tablenotetext{a}{Classification from Table~\ref{tab:meta_catalog}: yPAGB = yellow post-AGB; rPAGB = red post-AGB.}
\tablenotetext{b}{Host-cluster Horizontal-Branch Ratio from \citet{Borkova2000} for \oCen, \citet{Catelan2009} for NGC\,6441, our own data for M19, and LD94 for the rest.}
\tablenotetext{c}{Host-cluster horizontal-branch index from \citet{Torelli2019}.}
\tablenotetext{d}{The Cepheid V127 was observed by us at maximum light and should likely be classified AHB2 based on its mean magnitudes; see text.}
\end{deluxetable*}

Thirteen stars in our AHB catalog are classified as yellow or red PAGB stars: nine yPAGB objects, and four rPAGB stars. Details of these stars and their host clusters are listed in Table~\ref{tab:post-AGB}, including the stars' visual absolute magnitudes, $M_V$, and their hosts' HBRs and/or horizontal-branch indices, $\tauhb$. 

The thirteen PAGB stars are contained in ten clusters. Six of the PAGB stars are known variables, five of which belong to the RV~Tau class. Note that the absolute magnitudes of the variable stars given in Table~\ref{tab:post-AGB} are from our observations and do not reflect the stars' mean magnitudes averaged over their pulsation periods.\footnote{The sixth variable, V127 in NGC\,6441, is a 19.77-day W~Vir Cepheid. Based on its light curve \citep{Pritzl2003}, it appears that our observations were made at maximum light. If we had used its mean magnitudes, we would have classified it as AHB2.} The remaining seven non-variables consist of two red PAGB stars, and five yellow PAGB stars. 

As discussed above (\S\ref{sec:survey}), a search for yPAGB stars was the original motivation for our \uBVI\/ cluster survey. In addition to the previously known non-variable yPAGB star in \oCen\/ (see \S\ref{sec:intro}) and two stars in NGC\,5986 \citep{Alves2001}, the newly discovered members of this class are in M79 \citep{Bond2016} and M19 \citep{Bondetal2021}. As noted in \S\ref{sec:survey}, in a separate paper we will discuss the potential of using extragalactic analogs of these luminous but rare yPAGB stars as ``Population~II'' standard candles.  Here we simply point out that the five non-variable yPAGB stars have a mean absolute magnitude of $M_V=-3.31\pm0.05$, with a standard deviation of only 0.11~mag. 

Remarkably, all of the PAGB stars belong to clusters with blue HBs. Except for M5 and the anomalous NGC\,6441 (which does contain BHB stars, as discussed in \S\ref{subsec:morphology}), the host clusters have HBR values above 0.8, and except for NGC\,6441 they all have $\tauhb>5$. (M5 does contain a rich population of BHB stars, as shown in Figure~\ref{fig:schematic}; but it also hosts an appreciable number of RRL variables and RHB stars, which reduces its HBR value. There is, to our knowledge, no published HBR or $\tauhb$ for M19, but based on our own data we find $\rm HBR = 0.86$.)

These findings strongly suggest that the PAGB stars observed in GCs are the descendants of objects that had low envelope masses when they arrived on the ZAHB\null. These stars, when they make their final departure from the AGB toward higher temperatures, have relatively long evolutionary timescales, as we discussed in \S\ref{subsec:m79}. In contrast, PAGB stars in GCs with red HBs evolve considerably more rapidly; see, for example, Figure~\ref{fig:ngc6362_with_tracks}. These short timescales likely account for the observed absence of PAGB stars in red-HB clusters that lack the BHB and EHB progenitors.

Another factor to consider is whether circumstellar dust could dim or obscure a GC's PAGB stars at optical wavelengths. Infrared and millimeter-wave studies of luminous pulsating AGB stars in, for example, NGC\,362 \citep{Boyer2009}, 47~Tuc \citep{McDonald2011,McDonald2019}, and \oCen\ \citep{McDonald2011}, show that these stars are producing dust in their stellar winds. The mechanism is considered to be that, at the low surface gravities near the AGB tip, pulsations are able to levitate material to regions cool enough for dust formation (primarily metallic iron in these low-mass oxygen-rich stars); then radiation pressure on the dust drives the wind (see, for example, \citealt{McDonald2018}, and references therein). Once the star leaves the AGB, the pulsations end (until it reaches the Cepheid instability strip), and with it dust formation. As shown in \S\ref{sec:theory-comp}, PAGB evolutionary timescales at the low stellar masses in GCs---although ``rapid'' in the context discussed above---are still long enough (several to many tens of thousands of years) for the dust to have dissipated by the time the star enters the PAGB region of the CMD\null. 

This expectation is borne out by infrared observations that find no evidence of circumstellar dust around the yPAGB stars in \oCen\ \citep[HD\,116745;][]{McDonald2011}, M79 \citep{Bond2016}, and M19 \citep{Bondetal2021}. The only exception appears to be some of the RV~Tau variables in GCs. These PAGB objects, which have evolved into the top of the Cepheid instability strip, are so luminous and have such low surface gravities, that they may be able to resume dust formation through the pulsational mechanism described above. Circumstellar dust has indeed been detected in the RV~Tau variable V1 in \oCen\ \citep{McDonald2011}; however, its visual absolute magnitude is still so bright (see Figure~\ref{fig:CMDs}) that the optical depth of the dust must be very low. Moreover, \citet{Gezer2015} examined mid-infrared photometry from the {\it Wide-field Infrared Survey Explorer\/} of three of the RV~Tau variables in our Table~\ref{tab:post-AGB}---M5 V84, M28 V17, and M56 V6---and found no evidence for an infrared excess. We conclude that dust obscuration is not a significant issue for our visual survey for PAGB stars in GCs.


\section{Summary}

We have conducted a search for evolved stars lying above the horizontal branch and blueward of the AGB in the color-magnitude diagrams of 104 globular clusters---97 in the Milky Way, and seven of the ``Population~II'' clusters in the Magellanic Clouds. We performed CCD photometry of these clusters in the \uBVI\/ system, which is optimized for the detection of low-surface-gravity stars with large Balmer discontinuities in their spectral-energy distributions. The candidates selected from our \uBVI\/ photometry lie in the color range $-0.05\le(B-V)_0\le 1.0$, within which our Balmer-jump index, $c_2=(u-B)-(B-V)$, is most sensitive to $\log g$. Our candidates were then further tested for cluster membership via parallaxes and proper motions from \Gaia\/ EDR3, and a Gaussian-mixture model. Our final catalog of AHB stars (Table~\ref{tab:meta_catalog}) contains 438 objects, belonging to 64 clusters, and having membership probabilities greater than 0.8.  Because our survey was aimed at the brightest cluster members, we were able to perform our search all the way into the cluster centers, except for 15 clusters in which the stellar crowding was too severe; in these cases, we excluded the central regions at radii less than $10''$ to $40''$ (see Table~\ref{tab:exclusion}).

We confirmed a high level of survey completeness by comparing our catalog with previous searches for AHB stars in a small number of clusters studied by ZNG, in \oCen, and with catalogs of Type~II Cepheids in globular clusters. However, there are several caveats. Our data are based on a small number of observation epochs (often only one), and thus variable stars can lie away from their mean locations in our CMDs. In particular, 59 RR~Lyrae variables, observed near maximum light, are included in our catalog. Our catalog also contains a few interlopers that are actually blends of red and blue stars, or are physical binaries, which mimic true AHB stars; some of these can be recognized because they fall within the Cepheid instability strip, but are not known variable stars. However, these impostors generally do not lie more than about one magnitude above the horizontal branch.

We apply a simple classification scheme to AHB stars, based upon their locations in the CMD (see Figure~\ref{fig:classification}) and a comparison with theoretical post-HB evolutionary tracks. (1)~AHB1 stars are brighter than the HB and fainter than $M_V=-0.8$; they are the descendants of BHB stars, evolving across the CMD to lower temperatures on tracks parallel to the zero-age HB\null. As AHB1 stars pass through the Cepheid instability strip, they become BL~Her variables.  (2)~Post-AGB stars are the brightest AHB objects. They have departed the AGB and are evolving rapidly to higher temperatures. We subdivide them at $(B-V)_0=0.7$ into yellow (yPAGB) and red (rPAGB) post-AGB stars. RV~Tau variables are yPAGB stars that fall within the instability strip. (3)~AHB2 stars lie between the AHB1 and PAGB groups. They are primarily evolved from the hot EHB: they first ascend to high luminosities, and then cross the CMD toward lower temperatures. As they enter the instability strip, they become W~Vir Cepheids. The AHB2 category also includes objects that have departed the AGB and are evolving back toward higher temperatures---the post-early-AGB (PEAGB) objects. However, these are relatively rare, because of their more rapid evolutionary timescales.

We give a few illustrations of the astrophysical applications of our survey. We compared our results in several typical clusters with predictions from theoretical post-HB evolutionary tracks \citep{Moehler2019}, and we also investigated correlations of the AHB populations with the metallicities and horizontal-branch morphologies of the host clusters. In general, the theoretical tracks account for the main features of the AHB populations. They indicate that the AHB2 region of the CMD is populated primarily by descendants of stars that arrived on the ZAHB with very low envelope masses. In support of this expectation, we find that clusters containing AHB2 objects have blue horizontal branches. These clusters generally have intermediate to low metallicities---the ``first parameter.'' Conversely, AHB stars are rare or absent in metal-rich clusters that contain only red HB stars. However, there are exceptional clusters with relatively high metallicities, which still contain blue horizontal-branch stars and their AHB descendants. Moreover, we caution that we have only considered single-star evolution; binary interactions are also capable of populating the AHB region of the CMD.

We point out two clusters---M10 and M14---that are especially rich in AHB2 stars, including numerous W~Vir Cepheids. We find that both clusters contain a number of hotter, non-variable AHB2 stars, and a population of extremely hot ZAHB objects. This suggests an evolutionary sequence, in which BHB and EHB stars are the progenitors of the warm AHB2 objects, which then evolve into the instability strip and become W~Vir Cepheids. Thus the Cepheids are predominantly stars that are evolving toward the AGB, not away from it.

The visually brightest stars in globular clusters are the yPAGB objects. Our catalog contains nine of these objects (Table~\ref{tab:post-AGB}), of which five are non-variable and four are RV~Tau variables. All of the host clusters of these objects have relatively low metallicities and blue horizontal branches, indicating that ZAHB stars with low envelope masses are their progenitors. Metal-rich clusters can in principle also produce luminous PAGB stars, but the evolutionary timescales of these objects are so rapid that they are extremely rare. Non-variable yPAGB stars have a very narrow luminosity function, and we argue that their analogs in external galaxies are potential Population~II standard candles for distance measurement.

\medbreak

\acknowledgments

B.D.D. thanks the Zaccheus Daniel Trust for a grant that helped make this publication possible. He also thanks Eric Feigelson for guidance on statistical methods involving Gaussian-mixture models.

H.E.B. thanks his three co-authors for encouraging the analysis and publication of these extensive data, and for the enormous amount of labor they contributed---under pandemic conditions that prevented us from meeting in person. Thanks are also due to Zoom Video Communications, Inc., for making this collaboration possible.  

H.E.B.'s \uBVI\/ observations, and M.H.S.'s photometric reductions, were partially supported by NASA grant NAG 5-6821 under the ``UV, Visible, and Gravitational Astrophysics Research and Analysis'' program, and by the Director's Discretionary Research Fund at the Space Telescope Science Institute (STScI)\null. H.E.B. also thanks the staffs at Cerro Tololo and Kitt Peak for their support over many years.  

This research has made use of the SIMBAD database, operated at CDS, Strasbourg, France. This research has also made use of the VizieR catalogue access tool at CDS (DOI: 10.26093/cds/vizier). The original description of the VizieR service was published by \citet{Ochsenbein2000}. This research has made use of NASA's Astrophysics Data System Bibliographic Services.




This work has made use of data from the European Space Agency (ESA) mission {\it Gaia\/} (\url{https://www.cosmos.esa.int/gaia}), processed by the {\it Gaia\/} Data Processing and Analysis Consortium (DPAC, \url{https://www.cosmos.esa.int/web/gaia/dpac/consortium}). Funding for the DPAC has been provided by national institutions, in particular the institutions participating in the {\it Gaia\/} Multilateral Agreement.  

Based in part on observations made with the NASA\slash ESA {\it Hubble Space Telescope}, and obtained from the Hubble Legacy Archive, which is a collaboration between the Space Telescope Science Institute (STScI/NASA), the Space Telescope European Coordinating Facility (ST-ECF/ESA), and the Canadian Astronomy Data Centre (CADC/NRC/CSA). STScI is operated by the Association of Universities for Research in Astronomy, Inc.\ under NASA contract NAS 5-26555.

The Institute for Gravitation and the Cosmos is supported by the Eberly College of Science and the Office of the Senior Vice President for Research at Pennsylvania State University.

\facilities{KPNO: 4m, 0.9m, CTIO: 1.5m, 0.9m, Gaia, HST (WFPC2, WFC3, ACS)}

\clearpage

\onecolumngrid

\appendix

\section{Details of \uBVI\/ Observations}

Table~\ref{table:runs} gives some details of the observing runs at KPNO and CTIO which produced the \uBVI\/ photometry analyzed in this paper. There were 18 observing runs in all, from 1994 to 2001, using four different telescopes and four different (but similar) Tektronix CCD detectors. With the exception of three of the Mayall 4-m runs, the same $4\times4$-inch Gunn-Thuan $u$ filter was used throughout. We thank Ed Carder (NOAO) for assistance in constructing and characterizing this filter.

\begin{deluxetable*}{lllcc}[hbt]
\tablewidth{0 pt}
\tablecaption{Observing Runs 
\label{table:runs}
}  
\tablehead{
\colhead{Civil} & 
\colhead{Telescope} &
\colhead{CCD} &
\colhead{Plate Scale} &
\colhead{Field of View} \\
\colhead{Dates} & 
\colhead{} &
\colhead{} &
\colhead{[$''\,\rm pixel^{-1}$]} &
\colhead{[arcmin]}
}
\startdata
1994 Dec 1--3        & KPNO 4-m   & T2KB   & 0.470 &  16$\times$16               \\
1995 Jan 27--31      & CTIO 1.5-m & Tek4   & 0.440 &  15$\times$15     \\
1995 Oct 13--20      & CTIO 1.5-m & Tek3   & 0.440 &  15$\times$15                 \\
1996 Mar 11--13      & KPNO 4-m   & T2KB   & 0.470 &  16$\times$16                \\
1996 Sep 18--24      & KPNO 0.9-m & T2KA   & 0.688 &  23$\times$23                \\
1997 May 7--9        & KPNO 0.9-m & T2KA   & 0.688 &  23$\times$23                 \\
1997 May 27--Jun 1   & CTIO 0.9-m & Tek3   & 0.396 &  13$\times$13                \\
1997 Aug 3--10       & CTIO 0.9-m & Tek3   & 0.396 &  13$\times$13              \\
1997 Sep 17--22      & KPNO 0.9-m & T2KA   & 0.688 &  23$\times$23                \\
1997 Oct 3--5        & KPNO 4-m   & T2KB   & 0.420 &  14$\times$14  \\
1997 Nov 6--11       & CTIO 0.9-m & Tek3   & 0.396 &  13$\times$13                \\
1998 Mar 17--22      & KPNO 0.9-m & T2KA   & 0.688 &  23$\times$23              \\
1998 Apr 15--21      & CTIO 0.9-m & Tek3   & 0.396 &  13$\times$13                \\
1998 Aug 18--26      & CTIO 0.9-m & Tek3   & 0.396 &  13$\times$13                 \\
1999 Mar 12--15      & KPNO 0.9-m & T2KA   & 0.688 &  23$\times$23              \\
1999 Jun 10--15      & CTIO 0.9-m & Tek3   & 0.396 &  13$\times$13             \\
1999 Aug 24--28      & CTIO 0.9-m & Tek3   & 0.396 &  13$\times$13                \\
2001 Mar 22--27      & CTIO 0.9-m & Tek3   & 0.396 &  13$\times$13                 \\
\enddata
\end{deluxetable*}

\onecolumngrid

Table~\ref{table:obs} details the \uBVI\/ observations of each cluster.  Included in the table are the cluster names, fields observed (single pointing at the cluster center, or $2\times2$ or $3\times3$ mosaics), adopted distance modulus and reddening (mostly from H10), date, telescope, and \uBVI\/ exposure times.

\afterpage{
\def\t{$\times$}
\def\d{$\dots$}
\startlongtable
\begin{deluxetable*}{lcccllcccc}
\tablewidth{0 pt}
\tabletypesize{\footnotesize}
\tablecaption{{\it uBVI\/} Observations \\
\label{table:obs}}
\tablehead{
\colhead{} &
\colhead{} &
\colhead{} &
\colhead{} &
\colhead{} &
\colhead{} &
\multicolumn{4}{c}{------ Exposure Time [s] ------} \\
\colhead{Cluster} & 
\colhead{Field\tablenotemark{a}} &
\colhead{$(m-M)_V$\tablenotemark{b}} &
\colhead{$E(B-V)$\tablenotemark{b}} &
\colhead{UT Date} &
\colhead{Telescope\tablenotemark{c}} &
\colhead{$u$} &
\colhead{$B$} &
\colhead{$V$} &
\colhead{$I$} 
}
\startdata
%
%
\noalign{\vskip0.1in}
\multicolumn{9}{c}{Galactic Globular Clusters} \\
\noalign{\vskip0.1in}
NGC 104 (47 Tuc)  &NE  &13.37  &0.04 & 1997 Aug 6  & CT36  & 200 & 12 & 12 & 10  \\
                  &NE  &&     & 1997 Aug 6  & CT36  & 200 & 10 & 5  & 5   \\
                  &NE  &&     & 1997 Aug 6  & CT36  & \d  & 10 & 15 & 15  \\
                  &NW  &&     & 1997 Aug 6  & CT36  & 200 & 20 & 15 & 15  \\
                  &SW  &&     & 1997 Aug 6  & CT36  & 200 & 20 & 15 & 15  \\
                  &SE  &&     & 1997 Aug 7  & CT36  & 200 & 20 & 15 & 15  \\
                  &CTR &&     & 1997 Aug 11 & CT36  & 200 & 20 & 15 & 15  \\
                  &NE  &&     & 1997 Nov 9  & CT36  & 200 & 20 & 15 & 15  \\
                  &E   &&     & 1997 Nov 9  & CT36  & 200 & 20 & 15 & 15  \\
                  &SE  &&     & 1997 Nov 9  & CT36  & 200 & 20 & 15 & 15  \\
                  &N   &&     & 1997 Nov 9  & CT36  & 200 & 20 & 15 & 15  \\
                  &SW  &&     & 1997 Nov 9  & CT36  & 200 & 20 & 15 & 15  \\
                  &W   &&     & 1997 Nov 10 & CT36  & 200 & 20 & 15 & 15  \\
                  &NW  &&     & 1997 Nov 10 & CT36  & 200 & 20 & 15 & 15  \\
                  &CTR &&     & 1997 Nov 10 & CT36  & 200 & 20 & 15 & 15  \\
                  &E   &&     & 1997 Nov 10 & CT36  & 200 & 20 & 15 & 15  \\
NGC 288           &CTR &14.84 &0.03 & 1997 Aug 4  & CT36  & 400 & 30 & 20 & 30  \\
                  &NE  &&     & 1997 Aug 7  & CT36  & 400 & 30 & 20 & 30  \\
                  &NW  &&     & 1997 Aug 7  & CT36  & 400 & 30 & 20 & 30  \\
                  &SE  &&     & 1997 Aug 7  & CT36  & 400 & 30 & 20 & 30  \\
                  &CTR &&     & 1997 Nov 8  & CT36  & 2\t1200 & 2\t90 & 2\t60 & 2\t90 \\
NGC 362           &CTR &14.83 &0.05 & 1997 Aug 8  & CT36  & 2\t400 & 2\t30 & 2\t20 & 2\t30  \\
                  &NE  &&     & 1997 Nov 8  & CT36  & 800 & 60 & 40 & 60  \\
                  &NW  &&     & 1997 Nov 8  & CT36  & 800 & 60 & 40 & 60  \\
                  &SW  &&     & 1997 Nov 8  & CT36  & 800 & 60 & 40 & 60  \\
                  &SE  &&     & 1997 Nov 8  & CT36  & 800 & 60 & 40 & 60  \\
                  &CTR &&     & 1997 Nov 11 & CT36  & 2\t900 & 150 & 90 & 150 \\
                  &NW  &&     & 1999 Jun 11 & CT36  & 360 & 30 & 20 & 30  \\
                  &CTR &&     & 1999 Aug 27 & CT36  & 600 & 30 & 20 & 30  \\
NGC 1261          &CTR &16.09 &0.01 & 1997 Aug 8  & CT36  & 2\t600 & 2\t90 & 2\t60 & 2\t90 \\
                  &NW  &&     & 1998 Aug 26 & CT36  & 900 & 90 & 60 & 90  \\
                  &SW  &&     & 1999 Aug 26 & CT36  & 900 & 90 & 60 & 90  \\
Pal 1             &CTR &15.70 &0.15 & 1996 Sep 21 & KP36  & 300 & 60 & 50 & 90  \\
AM 1              &CTR &20.45 &0.00 & 1995 Oct 15 & CT60  & 500 &150 & 60 &120  \\
                  &CTR &&     & 1995 Oct 16 & CT60  & 2\t500 & 2\t150 & 2\t60 & 2\t120 \\
                  &CTR &&     & 1995 Oct 16 & CT60  & \d & \d & 2\t300 & 2\t300 \\
Eridanus          &CTR &19.83 &0.02 & 1997 Nov 9  & CT36  & 2\t1000 & 2\t90 & 2\t60 & 2\t90 \\
Pal 2             &CTR &21.01 &1.24 & 1997 Oct 5  & KP4   & 500 & 30 & 20 & 30 \\
NGC 1851          &CTR &15.47 &0.02 & 1997 Aug 10 & CT36  & 400 & 60 & 45 & 60 \\
                  &NE  &&     & 1997 Nov 8  & CT36  & 400 & 60 & 40 & 60 \\ 
                  &NW  &&     & 1997 Nov 8  & CT36  & 400 & 60 & 40 & 60 \\ 
                  &SW  &&     & 1997 Nov 8  & CT36  & 400 & 60 & 40 & 60 \\ 
                  &SE  &&     & 1997 Nov 9  & CT36  & 400 & 60 & 40 & 60 \\
NGC 1904  (M79)   &CTR &15.59 &0.01 & 1997 Nov 7  & CT36  & 400 & 60 & 45 & 60 \\
NGC 2298          &CTR &15.60 &0.14 & 1997 Nov 7  & CT36  & 500 & 75 & 45 & 60 \\
NGC 2419          &CTR &19.83 &0.08 & 1994 Dec 2  & KP4   & 100 & 15 & 10 & 12 \\
                  &CTR &&     & 1994 Dec 2  & KP4   & 3\t500 & 3\t150 & 3\t100 & 3\t120 \\
Pyxis             &CTR &18.63 &0.21 & 1995 Jan 29 & CT60  & 3\t600 & 3\t180 & 3\t120 & 3\t180 \\
NGC 2808          &CTR &15.59 &0.22 & 1997 Jun 2  & CT36  & 2\t400 & 2\t40 & 2\t30 & 2\t40 \\
                  &NW  &&     & 1997 Nov 9  & CT36  & 600 & 75 & 40 & 60 \\
                  &SW  &&     & 1997 Nov 10 & CT36  & 600 & 75 & 40 & 60 \\
                  &NE  &&     & 1997 Nov 11 & CT36  & 600 & 75 & 40 & 60 \\
                  &SE  &&     & 1998 Apr 17 & CT36  & 600 & 75 & 40 & 60 \\
                  &SE  &&     & 1998 Apr 19 & CT36  & 600 & 75 & 40 & 60 \\
                  &CTR &&     & 1999 Jun 14 & CT36  & 3\t600 & 150 & 75 & 120 \\
                  &CTR &&     & 1999 Jun 15 & CT36  & 900 &150 & 75 &120 \\
                  &CTR &&     & 2001 Mar 25 & CT36  & 600 & 75 & 40 & 60 \\
E 3               &CTR &15.47 &0.30 & 1997 Nov 7  & CT36  & 300 & 40 & 20 & 30 \\
Pal 3             &CTR &19.95 &0.04 & 1995 Feb 1  & CT60  & 420 & 75 & 60 & 90 \\
NGC 3201          &CTR &14.20 &0.24 & 1995 Feb 1  & CT60  & 120 & 15 &  5 &  5 \\
                  &SW  &&     & 1998 Apr 17 & CT36  & 400 & 20 & 10 & 15 \\
                  &NW  &&     & 1998 Apr 17 & CT36  & 400 & 20 & 10 & 15 \\
                  &NE  &&     & 1998 Apr 21 & CT36  & 800 & 40 & 20 & 30 \\
                  &SE  &&     & 1998 Apr 21 & CT36  & 800 & 40 & 20 & 30 \\
                  &CTR &&     & 1998 Apr 22 & CT36  & 400 & 20 & 10 & 15 \\
Pal 4             &CTR &20.21 &0.01 & 1997 May 8  & KP36  & 2\t500 & 2\t90 & 2\t60 & 2\t100 \\
NGC 4147          &CTR &16.49 &0.02 & 1997 May 8  & KP36  & 120 & 10 & 10 & 10 \\
                  &CTR &&     & 1997 May 8  & KP36  & 600 & 60 & 60 & 60 \\
                  &CTR &&     & 1997 May 9  & KP36  & 600 & 45 & 45 & 60 \\
NGC 4372          &CTR &15.03 &0.39 & 1997 Jun 1  & CT36  & 2\t200 & 2\t20 & 2\t15 & 2\t20 \\
                  &SW  &&     & 1998 Apr 21 & CT36  & 2\t800 & 60 & 30 & 40 \\
                  &NW  &&     & 1998 Apr 21 & CT36  & 2\t800 & 60 & 30 & 40 \\
                  &NE  &&     & 1998 Apr 21 & CT36  & 800 & 60 & 30 & 40 \\
                  &SE  &&     & 1998 Apr 21 & CT36  & 800 & 60 & 30 & 40 \\
                  &CTR &&     & 1998 Apr 22 & CT36  & 800 & 30 & 15 & 20 \\
Rup 106           &CTR &17.25 &0.20 & 1995 Jan 30 & CT60  & 180 & 30 & 30 & 60 \\
NGC 4590  (M68)   &CTR &15.21 &0.05 & 1995 Feb 1  & CT60  & 180 & 25 & 10 & 10 \\
                  &NW  &&     & 1998 Apr 22 & CT36  & 600 & 30 & 20 & 30 \\
                  &SW  &&     & 1998 Apr 22 & CT36  & 600 & 30 & 20 & 30 \\
                  &SE  &&     & 1998 Apr 22 & CT36  & 600 & 30 & 20 & 30 \\
                  &NE  &&     & 1998 Apr 22 & CT36  & 600 & 30 & 20 & 30 \\
NGC 4833          &CTR &15.08 &0.32 & 1997 Jun 1  & CT36  & 2\t250 & 2\t30 & 2\t20 & 2\t30 \\
                  &CTR &&     & 1998 Apr 22 & CT36  & 800 & 40 & 20 & 20 \\
NGC 5024  (M53)   &CTR &16.32 &0.02 & 1996 Mar 12 & KP4   & 10  &  2 & 2  &  2 \\
                  &CTR &&     & 1996 Mar 12 & KP4   & 120 & 30 & 20 & 30 \\
                  &NE  &&     & 1998 Mar 23 & KP36  & 600 & 75 & 60 & 75 \\
                  &SE  &&     & 1998 Mar 23 & KP36  & 600 & 75 & 60 & 75 \\
                  &SW  &&     & 1998 Mar 23 & KP36  & 600 & 75 & 60 & 75 \\
                  &NW  &&     & 1998 Mar 23 & KP36  & 600 & 75 & 60 & 75 \\
NGC 5053          &CTR &16.23 &0.01 & 1997 May 8  & KP36  & 100 & 10 & 10 & 10 \\
                  &CTR &&     & 1997 May 8  & KP36  & 600 & 60 & 60 & 60 \\
                  &CTR &&     & 1997 May 9  & KP36  & 600 & 45 & 45 & 60 \\
NGC 5139  ($\omega$~Cen) &CTR &13.94 &0.12 & 1997 Jun 1  & CT36  & 2\t150 & 2\t15 & 2\t10 & 2\t15 \\
                  &ROA 24 &&  & 1997 Jun 1  & CT36  & 150 & 15 & 10 & 15 \\
                  &NW  &&     & 1997 Jun 2  & CT36  & 2\t150 & 2\t15 & 2\t10 & 2\t15 \\
                  &NE  &&     & 1997 Jun 2  & CT36  & 2\t150 & 2\t15 & 2\t10 & 2\t15 \\
                  &SW  &&     & 1997 Jun 2  & CT36  & 2\t150 & 2\t15 & 2\t10 & 2\t15 \\
                  &ROA 24 &&  & 1997 Jun 2  & CT36  & 150 & 15 & 10 & 15 \\
                  &SW  &&     & 1998 Apr 17 & CT36  & 240 & 20 & 12 & 20 \\
                  &W   &&     & 1998 Apr 17 & CT36  & 240 & 20 & 12 & 20 \\
                  &NW  &&     & 1998 Apr 17 & CT36  & 240 & 20 & 12 & 20 \\
                  &N   &&     & 1998 Apr 17 & CT36  & 240 & 20 & 12 & 20 \\
                  &SW  &&     & 1998 Apr 18 & CT36  & \d  & \d &300 &300 \\
                  &S   &&     & 1998 Apr 18 & CT36  & 240 & 20 & 12 & 20 \\
                  &CTR &&     & 1998 Apr 18 & CT36  & 240 & 20 & 12 & 20 \\
                  &SE  &&     & 1998 Apr 19 & CT36  & 500 & 40 & 25 & 40 \\
                  &E   &&     & 1998 Apr 19 & CT36  & 240 & 20 & 12 & 20 \\
                  &NE  &&     & 1998 Apr 21 & CT36  & 240 & 20 & 12 & 20 \\
                  &SW  &&     & 1998 Apr 22 & CT36  & \d  & \d &300 &300 \\
                  &ROA 24 &&  & 2001 Mar 25 & CT36  & 150 & 20 & 10 & 15 \\         
NGC 5272  (M3)    &CTR &15.07 &0.01 & 1997 May 8  & KP36  & 120 & 10 &  8 & 10 \\
                  &CTR &&     & 1997 May 9  & KP36  & 300 & 30 & 30 & 45 \\
                  &CTR &&     & 1997 May 10 & KP36  & 500 &100 &100 &100 \\
                  &SW  &&     & 1998 Mar 21 & KP36  & 300 & 30 & 30 & 45 \\
                  &W   &&     & 1998 Mar 21 & KP36  & 300 & 30 & 30 & 45 \\
                  &NW  &&     & 1998 Mar 21 & KP36  & 300 & 30 & 30 & 45 \\
                  &S   &&     & 1998 Mar 21 & KP36  & 300 & 30 & 30 & 45 \\
                  &SE  &&     & 1998 Mar 22 & KP36  & 300 & 30 & 30 & 45 \\
                  &N   &&     & 1998 Mar 22 & KP36  & 300 & 30 & 30 & 45 \\
                  &CTR &&     & 1998 Mar 22 & KP36  & 300 & 30 & 30 & 45 \\
                  &E   &&     & 1998 Mar 22 & KP36  & 300 & 30 & 30 & 45 \\
                  &NE  &&     & 1998 Mar 22 & KP36  & 300 & 30 & 30 & 45 \\
NGC 5286          &CTR &16.08 &0.24 & 1995 Jan 29 & CT60  & 120 & 30 & 15 & 15 \\
                  &NW  &&     & 1999 Jun 13 & CT36  & 2\t750 & 75 & 45 & 75 \\
                  &SW  &&     & 1999 Jun 13 & CT36  & 2\t750 & 75 & 45 & 75 \\
                  &CTR &&     & 2001 Mar 25 & CT36  & 2\t750 & 75 & 45 & 75 \\
AM 4              &CTR &17.69 &0.05 & 1995 Jan 30 & CT60  & 360 & 60 & 60 & 90 \\
NGC 5466          &CTR &16.02 &0.00 & 1997 May 9  & KP36  & 600 & 45 & 45 & 60 \\
NGC 5634          &CTR &17.16 &0.05 & 1997 May 8  & KP36  & 200 & 20 & 20 & 20 \\
                  &CTR &&     & 1998 Aug 26 & CT36  & 2\t600 & 120 & 90 & 120 \\
                  &CTR &&     & 2001 Mar 26 & CT36  & 2\t600 & 120 & 90 & 120 \\
NGC 5694          &CTR &18.00 &0.09 & 2001 Mar 25 & CT36  & 2\t750 & 75 & 45 & 90 \\          
IC 4499           &CTR &17.08 &0.23 & 1995 Feb 1  & CT60  & 240 & 30 & 20 & 40 \\
NGC 5824          &CTR &17.94 &0.13 & 1998 Apr 17 & CT36  & 2\t700 & 100 & 75 & 100 \\
Pal 5             &CTR &16.92 &0.03 & 1997 May 10 & KP36  & 2\t400 & 2\t60 & 2\t60 & 2\t60 \\
NGC 5897          &NW  &15.76 &0.09 & 1998 Aug 23 & CT36  & 600 & 60 & 40 & 60 \\
                  &NE  &&     & 1998 Aug 23 & CT36  & 600 & 60 & 40 & 60 \\
                  &SW  &&     & 1998 Aug 24 & CT36  & 900 & 90 & 60 & 90 \\
                  &SE  &&     & 1998 Aug 24 & CT36  & 900 & 90 & 60 & 90 \\
                  &CTR &&     & 1999 Jun 16 & CT36  & 2\t600 & 120 & 90 & 120 \\
                  &CTR &&     & 2001 Mar 28 & CT36  & 600 &120 & 90 &120 \\
NGC 5904  (M5)    &CTR &14.46 &0.03 & 1997 May 8  & KP36  & 100 & 10 &  8 & 10 \\
                  &CTR &&     & 1997 May 8  & KP36  & 400 & 90 & 75 & 90 \\
                  &SW  &&     & 1998 Mar 22 & KP36  & 300 & 30 & 20 & 30 \\
                  &NW  &&     & 1998 Mar 22 & KP36  & 300 & 30 & 20 & 30 \\
                  &NE  &&     & 1998 Mar 23 & KP36  & 300 & 30 & 20 & 30 \\
                  &SE  &&     & 1998 Mar 23 & KP36  & 300 & 30 & 20 & 30 \\
                  &CTR &&     & 1998 Apr 18 & CT36  & 400 & 20 & 15 & 20 \\
NGC 5927          &CTR &15.82 &0.45 & 1998 Aug 21 & CT36  & 2\t600 & 45 & 25 & 30 \\
                  &CTR &&     & 1998 Aug 22 & CT36  & 2\t600 & 45 & 25 & 30 \\
NGC 5986          &CTR &15.96 &0.28 & 1995 Jan 28 & CT60  & 120 & 30 & 15 & 15 \\
                  &CTR &&     & 1995 Jan 30 & CT60  & 120 & 30 & 15 & 15 \\
                  &CTR &&     & 1997 May 28 & CT36  & 2\t400 & 2\t30 & 2\t20 & 2\t40 \\
                  &CTR &&     & 1997 Jun 2  & CT36  & 400 & 30 & 20 & 40 \\
                  &CTR &&     & 1998 Aug 27 & CT36  & 900 & 75 & 45 & 75 \\
                  &CTR &&     & 2001 Mar 28 & CT36  & 300 & 75 & 45 & 75 \\
Pal 14   (AvdB)   &CTR &19.54 &0.04 & 1996 Mar 12 & KP4   & \d  &\d  &300 &300 \\
                  &CTR &&     & 1996 Mar 12 & KP4   &  45 & 20 &  8 &  8 \\
NGC 6093  (M80)   &CTR &15.56 &0.18 & 1997 Jun 1  & CT36  & 2\t300 & 2\t30 & 2\t20 & 2\t30 \\
                  &CTR &&     & 1997 Jun 2  & CT36  & 200 & 20 & 10 & 20 \\
                  &NW  &&     & 1998 Aug 20 & CT36  & 700 & 40 & 30 & 40 \\
                  &SW  &&     & 1998 Aug 20 & CT36  & 700 & 40 & 30 & 40 \\
                  &NE  &&     & 1998 Aug 22 & CT36  & 700 & 40 & 30 & 40 \\
                  &SE  &&     & 1998 Aug 24 & CT36  & 700 & 40 & 30 & 40 \\
                  &SE  &&     & 1998 Aug 27 & CT36  & 700 & 40 & 30 & 40 \\
                  &CTR &&     & 1999 Aug 26 & CT36  & 2\t400 & 45 & 30 & 45 \\
                  &CTR &&     & 2001 Mar 28 & CT36  & 700 & 40 & 30 & 40 \\
NGC 6121  (M4)    &NW  &12.82 &0.35 & 1997 Jun 2  & CT36  & 2\t120 & 2\t10 & 2\t10 & 2\t10 \\
                  &SW  &&     & 1997 Jun 2  & CT36  & 2\t120 & 10 & 10 & 10 \\
                  &NE  &&     & 1997 Jun 2  & CT36  & 120 & 10 & 10 & 10 \\
                  &SE  &&     & 1997 Jun 2  & CT36  & 120 & 10 & 10 & 10 \\
                  &CTR &&     & 1999 Aug 26 & CT36  & 180 & 12 & 10 & 10 \\
                  &CTR &&     & 2001 Mar 28 & CT36  & 180 & 20 & 15 & 15 \\
                  &CTR &&     & 2001 Mar 28 & CT36  & 900 &120 & 90 &120 \\
NGC 6101          &CTR &16.10 &0.05 & 2001 Mar 28 & CT36  & 2\t600 & 90 & 60 & 90 \\
NGC 6144          &CTR &15.86 &0.36 & 1999 Jun 12 & CT36  & 2\t750 & 75 & 45 & 60 \\
NGC 6139          &CTR &17.35 &0.75 & 1998 Aug 26 & CT36  & 2\t900 & 200 & 100 & 100 \\
NGC 6171  (M107)  &CTR &15.05 &0.33 & 1997 May 9  & KP36  & 500 & 45 & 30 & 45 \\
                  &CTR &&     & 1998 Aug 27 & CT36  & 700 & 30 & 16 & 25 \\
NGC 6205  (M13)   &CTR &14.33 &0.02 & 1996 Sep 21 & KP36  & 20  &  3 &  2 &  3 \\
                  &CTR &&     & 1996 Sep 21 & KP36  & 200 & 30 & 20 & 30 \\
                  &CTR &&     & 1996 Sep 21 & KP36  & 900 &100 & 90 &150 \\
                  &NE  &&     & 1997 Sep 21 & KP36  & 300 & 20 & 12 & 20 \\
                  &SE  &&     & 1997 Sep 21 & KP36  & 300 & 20 & 12 & 18 \\
                  &SW  &&     & 1997 Sep 21 & KP36  & 300 & 20 & 12 & 18 \\
                  &NW  &&     & 1998 Mar 22 & KP36  & 300 & 20 & 12 & 20 \\
                  &NW  &&     & 1999 Mar 13 & KP36  & 300 & 20 & 12 & 20 \\
NGC 6229          &CTR &17.45 &0.01 & 1996 Sep 23 & KP36  & 600 &180 & 75 &150 \\
                  &CTR &&     & 1997 May 8  & KP36  & 400 & 60 & 30 & 60 \\
NGC 6218  (M12)   &CTR &14.01 &0.19 & 1996 Sep 23 & KP36  & 90  & 30 & 20 & 30 \\
                  &CTR &&     & 1996 Sep 23 & KP36  & 180 & 10 &  6 & 10 \\
                  &CTR &&     & 1997 May 10 & KP36  & 2\t400 & 2\t90 & 2\t60 & 2\t90 \\
                  &NW  &&     & 1998 Mar 23 & KP36  & 250 & 25 & 12 & 12 \\
                  &NE  &&     & 1998 Mar 23 & KP36  & 250 & 25 & 12 & 12 \\
                  &SE  &&     & 1998 Mar 23 & KP36  & 250 & 25 & 12 & 12 \\
                  &SW  &&     & 1998 Mar 23 & KP36  & 250 & 25 & 12 & 12 \\
                  &SW  &&     & 1998 Aug 23 & CT36  & 300 & 20 & 15 & 15 \\
NGC 6254  (M10)   &CTR &14.08 &0.28 & 1997 May 9  & KP36  & 2\t300 & 20 & 15 & 2\t25 \\
                  &CTR &&     & 1998 Aug 21 & CT36  & 400 & 20 & 15 & 15 \\
                  &NW  &&     & 1998 Aug 27 & CT36  & 400 & 20 & 15 & 15 \\
                  &SW  &&     & 1998 Aug 27 & CT36  & 400 & 20 & 15 & 15 \\
                  &SE  &&     & 1998 Aug 27 & CT36  & 400 & 20 & 15 & 15 \\
                  &NE  &&     & 1998 Aug 27 & CT36  & 400 & 20 & 15 & 15 \\
Pal 15            &CTR &19.51 &0.40 & 1997 May 10 & KP36  & 2\t500 & 2\t100 & 2\t75 & 2\t100 \\
NGC 6266  (M62)   &CTR &15.63 &0.47 & 1998 Apr 22 & CT36  & 800 & 30 & 15 & 20 \\
NGC 6273  (M19)   &CTR &15.90 &0.38 & 1995 Feb 1  & CT60  & 120 & 30 & 15 & 15 \\
                  &NW  &&     & 1998 Apr 21 & CT36  & 2\t800 & 75 & 40 & 45 \\
                  &SW  &&     & 1998 Apr 21 & CT36  & 2\t800 & 75 & 40 & 45 \\
                  &SE  &&     & 1998 Apr 21 & CT36  & 2\t800 & 75 & 40 & 45 \\
                  &NE  &&     & 1998 Apr 21 & CT36  & 2\t800 & 75 & 40 & 45 \\
                  &CTR &&     & 1998 Apr 22 & CT36  & 400 & 75 & 40 & 45 \\
NGC 6287          &CTR &16.72 &0.60 & 1997 Jun 1  & CT36  & 2\t400 & 2\t90 & 2\t60 & 2\t90 \\
                  &CTR &&     & 1997 Jun 1  & CT36  & \d & \d & 2\t300 & 2\t300 \\
NGC 6293          &CTR &16.00 &0.36 & 1999 Aug 26 & CT36  & 2\t900 & 90 & 45 & 75 \\
NGC 6316          &CTR &16.77 &0.54 & 1998 Aug 26 & CT36  & 2\t700 & 150 & 90 & 90 \\
                  &CTR &&     & 1999 Jun 12 & CT36  & 2\t750 & 200 &100 &100 \\
NGC 6341  (M92)   &CTR &14.65 &0.02 & 1996 Sep 25 & KP36  & 300 & 30 & 20 & 45 \\
                  &CTR &&     & 1997 May 8  & KP36  & 300 & 30 & 20 & 30 \\
                  &CTR &&     & 1997 May 10 & KP36  & 2\t500 & 2\t90 & 2\t60 & 2\t90 \\
                  &NE  &&     & 1997 Sep 23 & KP36  & 300 & 30 & 20 & 30 \\
                  &SE  &&     & 1997 Sep 23 & KP36  & 300 & 30 & 20 & 30 \\
                  &SW  &&     & 1998 Mar 22 & KP36  & 300 & 30 & 20 & 30 \\
                  &NW  &&     & 1998 Mar 22 & KP36  & 300 & 30 & 20 & 30 \\
                  &SW  &&     & 1998 Mar 23 & KP36  & 300 & 30 & 20 & 30 \\
                  &NW  &&     & 1998 Mar 23 & KP36  & 300 & 30 & 20 & 30 \\
NGC 6333  (M9)    &CTR &15.67 &0.38 & 1998 Aug 23 & CT36  & 2\t500 & 60 & 40 & 45 \\
NGC 6356          &CTR &16.76 &0.28 & 1998 Aug 19 & CT36  & 800 &150 & 90 &120 \\
                  &CTR &&     & 1999 Jun 11 & CT36  & 2\t900 & 90 & 45 & 60 \\
NGC 6352          &CTR &14.43 &0.22 & 1997 Jun 2  & CT36  & 2\t300 & 2\t20 & 2\t15 & 2\t20 \\
NGC 6366          &CTR &14.94 &0.71 & 1997 May 9  & KP36  & 300 & 30 & 30 & 2\t45 \\
NGC 6362          &NE  &14.68 &0.09 & 1999 Jun 11 & CT36  & 500 & 30 & 20 & 30 \\
                  &CTR &&     & 2001 Mar 28 & CT36  & 500 & 30 & 20 & 30 \\         
NGC 6388          &CTR &16.13 &0.37 & 1998 Apr 18 & CT36  &1000 & 60 & 30 & 40 \\
NGC 6402  (M14)   &CTR &16.69 &0.60 & 1996 Sep 23 & KP36  & 600 &120 & 60 & 2\t120 \\
                  &SE  &&     & 1998 Aug 21 & CT36  & 2\t1000 & 90 & 40 & 40 \\
                  &NE  &&     & 1998 Aug 22 & CT36  & 2\t900 & 90 & 40 & 40 \\ 
                  &NW  &&     & 1999 Jun 11 & CT36  & 2\t900 & 90 & 40 & 40 \\
                  &SW  &&     & 1999 Jun 11 & CT36  & 2\t900 & 90 & 40 & 40 \\
                  &CTR &&     & 1999 Aug 25 & CT36  & 2\t900 & 180 & 75 & 75 \\
                  &CTR &&     & 2001 Mar 26 & CT36  & 2\t900 & 90 & 40 & 40 \\         
NGC 6397          &SW  &12.37 &0.18 & 1998 Apr 16 & CT36 & 100 & 10 & 6 & 8 \\
                  &NW  &&     & 1998 Apr 16 & CT36 & 100 & 10 & 6 & 8 \\
                  &NE  &&     & 1998 Apr 16 & CT36 & 100 & 10 & 6 & 8 \\
                  &SE  &&     & 1998 Apr 16 & CT36 & 100 & 10 & 6 & 8 \\
                  &CTR &&     & 1999 Jun 13 & CT36 & 150 & 10 & 10 & 10 \\
NGC 6426          &CTR &17.68 &0.36 & 1996 Sep 25 & KP36 & 600 & 120 & 90 & 120 \\
                  &CTR &&     & 1997 May 9  & KP36 & 600 & 120 & 90 & 120 \\
NGC 6441          &CTR &16.78 &0.47 & 1998 Aug 20 & CT36 & 2\t700 & 60 & 30 & 40 \\
                  &CTR &&     & 1998 Aug 23 & CT36 & 2\t700 & 2\t90 & 60 & 75 \\
NGC 6496          &CTR &15.74 &0.15 & 1998 Aug 27 & CT36 & 1000 & 60 & 35 & 60 \\
NGC 6517          &CTR &18.48 &1.08 & 1997 Sep 23 & KP36 & 2\t900 & 2\t90 & 2\t60 & 2\t90 \\
NGC 6522          &CTR &15.92 &0.48 & 1998 Apr 22 & CT36 & 250,1200 & 90 & 40 & 50 \\
NGC 6535          &CTR &15.22 &0.34 & 1996 Sep 25 & KP36 & 300 & 90 & 60 & 90 \\
                  &CTR &&     & 1997 May 9  & KP36 & 300 & 30 & 30 & 30 \\
NGC 6528          &CTR &16.17 &0.54 & 1998 Aug 22 & CT36 & 2\t600 & 120 & 60 & 75 \\
NGC 6539          &CTR &17.62 &1.02 & 1997 Sep 20 & KP36 & 2\t1200 & 2\t150 & 2\t90 & 2\t150 \\
NGC 6544          &CTR &14.71 &0.76 & 1999 Aug 26 & CT36 & 2\t500 & 45 & 20 & 25 \\
NGC 6541          &CTR &14.82 &0.14 & 1998 Apr 18 & CT36 & 360 & 25 & 15 & 25 \\
NGC 6584          &CTR &15.96 &0.10 & 2001 Mar 28 & CT36 & 2\t500 & 60 & 45 & 60 \\
NGC 6626  (M28)   &CTR &14.95 &0.40 & 1998 Aug 20 & CT36 & 2\t500 & 40 & 20 & 30 \\
NGC 6637  (M69)   &CTR &15.28 &0.18 & 1998 Aug 23 & CT36 & 600 & 45 & 30 & 45 \\
NGC 6652          &CTR &15.28 &0.09 & 1999 Jun 12 & CT36 & 750 & 45 & 30 & 45 \\
                  &CTR &&     & 1999 Jun 13 & CT36 & 750 & 45 & 30 & 45 \\
NGC 6656  (M22)   &CTR &13.60 &0.34 & 1998 Apr 22 & CT36 & 360 & 15 & 8 & 10 \\
                  &NW  &&     & 1998 Aug 24 & CT36 & 600 & 30 & 20 & 30 \\
                  &SW  &&     & 1998 Aug 24 & CT36 & 600 & 30 & 20 & 30 \\
                  &SE  &&     & 1998 Aug 25 & CT36 & 600 & 30 & 20 & 30 \\
                  &NE  &&     & 1998 Aug 25 & CT36 & 600 & 30 & 20 & 30 \\
                  &SW  &&     & 1998 Aug 26 & CT36 & 300 & 30 & 20 & 30 \\
                  &NW  &&     & 1998 Aug 26 & CT36 & 300 & 30 & 20 & 30 \\
                  &NE  &&     & 1998 Aug 26 & CT36 & 300 & 30 & 20 & 30 \\
                  &SE  &&     & 1998 Aug 26 & CT36 & 300 & 30 & 20 & 30 \\
NGC 6681  (M70)   &CTR &14.99 &0.07 & 1998 Aug 22 & CT36 & 600 & 30 & 25 & 30 \\
NGC 6712          &CTR &15.60 &0.45 & 1996 Sep 25 & KP36 & 300 & 90 & 60 & 90 \\
NGC 6715  (M54)   &CTR &17.58 &0.15 & 1998 Apr 22 & CT36 & 900 & 90 & 60 & 90 \\
NGC 6723          &CTR &14.84 &0.05 & 1997 May 31 & CT36 & 2\t200 & 2\t30 & 2\t20 & 2\t30 \\
                  &CTR &&     & 1998 Aug 20 & CT36 & 500 & 30 & 20 & 30 \\
NGC 6752          &NE  &13.13 &0.04 & 1997 Aug 10 & CT36 & 200 & 20 & 10 & 20 \\
                  &NW  &&     & 1997 Aug 10 & CT36 & 200 & 20 & 10 & 20 \\
                  &SE  &&     & 1997 Nov 8  & CT36 & 200 & 20 & 10 & 20 \\
                  &SW  &&     & 1997 Nov 8  & CT36 & 200 & 20 & 10 & 20 \\
                  &CTR &&     & 1997 Nov 9  & CT36 & 200 & 20 & 10 & 20 \\
                  &CTR &&     & 1997 Nov 10 & CT36 & 200 & 20 & 10 & 20 \\
                  &CTR &&     & 1998 Apr 18 & CT36 & 120 &  8 &  6 &  8 \\
                  &S   &&     & 1998 Apr 18 & CT36 & 120 &  8 &  6 &  8 \\
                  &SE  &&     & 1998 Apr 19 & CT36 & 240 & 16 & 12 & 16 \\
                  &E   &&     & 1998 Apr 19 & CT36 & 240 & 16 & 12 & 16 \\
                  &NE  &&     & 1998 Apr 19 & CT36 & 240 & 16 & 12 & 16 \\
                  &N   &&     & 1998 Apr 21 & CT36 & 120 &  8 &  6 &  8 \\
                  &NW  &&     & 1998 Apr 21 & CT36 & 120 &  8 &  6 &  8 \\
                  &W   &&     & 1998 Apr 21 & CT36 & 120 &  8 &  6 &  8 \\
                  &SW  &&     & 1998 Apr 22 & CT36 & 120 &  8 &  6 &  8 \\
NGC 6760          &CTR &16.72 &0.77 & 1996 Sep 25 & KP36 & 600 &120 & 90 &120 \\
                  &CTR &&     & 1997 Sep 20 & KP36 & 2\t900 & 2\t90 & 2\t60 & 2\t90 \\
NGC 6779  (M56)   &CTR &15.68 &0.26 & 1996 Sep 21 & KP36 & 200 & 10 & 10 & 10 \\
                  &CTR &&     & 1996 Sep 21 & KP36 & 400 & 60 & 60 & 60 \\
NGC 6809  (M55)   &NW  &13.89 &0.08 & 1997 Jun 1  & CT36 & 2\t150 & 2\t15 & 2\t10 & 2\t15 \\
                  &NE  &&     & 1997 Jun 2  & CT36 & 150 & 15 & 10 & 15 \\
                  &SE  &&     & 1997 Jun 2  & CT36 & 2\t150 & 2\t15 & 2\t10 & 2\t15 \\
                  &SW  &&     & 1997 Jun 2  & CT36 & 2\t150 & 15 & 10 & 15 \\
                  &CTR &&     & 1997 Nov 10 & CT36 & 200 & 20 & 10 & 20 \\
                  &SW  &&     & 1998 Apr 17 & CT36 & 300 & 12 & 10 & 15 \\
                  &W   &&     & 1998 Apr 17 & CT36 & 300 & 12 & 10 & 15 \\
                  &NW  &&     & 1998 Apr 17 & CT36 & 300 & 12 & 10 & 15 \\
                  &NW  &&     & 1998 Aug 21 & CT36 & 300 & 20 & 15 & 20 \\
                  &N   &&     & 1998 Aug 21 & CT36 & 300 & 20 & 15 & 20 \\
                  &NE  &&     & 1998 Aug 21 & CT36 & 300 & 20 & 15 & 20 \\
                  &E   &&     & 1998 Aug 21 & CT36 & 300 & 20 & 15 & 20 \\
                  &SE  &&     & 1998 Aug 21 & CT36 & 300 & 20 & 15 & 20 \\
                  &S   &&     & 1998 Aug 22 & CT36 & 300 & 20 & 15 & 20 \\
                  &S   &&     & 1998 Aug 26 & CT36 & 300 & 20 & 15 & 20 \\
Pal 11            &    &16.72 &0.35 & 1997 Sep 22 & KP36 & 2\t900 & 2\t100 & 2\t75 & 2\t100 \\
NGC 6838  (M71)   &CTR &13.80 &0.25 & 1996 Sep 21 & KP36 & 60 &  8 &  5  &  6 \\
                  &CTR &&     & 1996 Sep 21 & KP36 & 240 & 60 & 45 & 60 \\
                  &NE  &&     & 1997 Sep 23 & KP36 & 300 & 25 & 15 & 25 \\
                  &SE  &&     & 1997 Sep 23 & KP36 & 300 & 25 & 15 & 25 \\
                  &SW  &&     & 1997 Sep 23 & KP36 & 300 & 25 & 15 & 25 \\
                  &NW  &&     & 1997 Sep 23 & KP36 & 300 & 25 & 15 & 25 \\
NGC 6864  (M75)   &CTR &17.09 &0.16 & 1997 Aug 9  & CT36 & 2\t900 & 2\t100 & 2\t60 & 2\t100 \\
                  &SE  &&     & 1998 Aug 20 & CT36 & 2\t700 & 60 & 40 & 40 \\
                  &SW  &&     & 1998 Aug 23 & CT36 & 2\t700 & 90 & 60 & 60 \\
                  &NW  &&     & 1998 Aug 23 & CT36 & 2\t700 & 90 & 60 & 60 \\
                  &NE  &&     & 1998 Aug 26 & CT36 & 2\t700 & 90 & 60 & 60 \\
                  &NE  &&     & 1998 Aug 27 & CT36 & 2\t700 & 90 & 60 & 60 \\
NGC 6934          &CTR &16.28 &0.10 & 1996 Sep 25 & KP36 & 600 & 120 & 90 &120 \\
                  &CTR &&     & 1997 Sep 19 & KP36 & 2\t600 & 2\t120 & 2\t90 & 2\t120 \\
NGC 6981  (M72)   &CTR &16.31 &0.05 & 1996 Sep 25 & KP36 & 600 & 120 & 90 & 120 \\
                  &CTR &&     & 1997 Aug 11 & CT36 & 2\t600 & 2\t60 & 2\t45 & 2\t60 \\
NGC 7006          &CTR &18.23 &0.05 & 1996 Sep 19 & KP36 & 2\t300 & 2\t20 & 2\t10 & 2\t20 \\
                  &CTR &&     & 1996 Sep 21 & KP36 & 300 &100 &100 &200 \\
NGC 7078  (M15)   &CTR &15.39 &0.10 & 1996 Sep 21 & KP36 & 100 & 10 &  8 & 15 \\
                  &CTR &&     & 1996 Sep 21 & KP36 & 300 & 60 & 50 & 90 \\
                  &NE  &&     & 1997 Sep 20 & KP36 & 300 & 30 & 20 & 30 \\
                  &SE  &&     & 1997 Sep 20 & KP36 & 300 & 30 & 20 & 30 \\
                  &NW  &&     & 1997 Sep 23 & KP36 & 300 & 30 & 20 & 30 \\
                  &SW  &&     & 1997 Sep 23 & KP36 & 300 & 30 & 20 & 30 \\
NGC 7089  (M2)    &CTR &15.50 &0.06 & 1996 Sep 19 & KP36 &  60 &  5 &  5 & 10 \\
                  &CTR &&     & 1996 Sep 23 & KP36 & 240 & 60 & 45 & 75 \\
                  &CTR &&     & 1996 Sep 23 & KP36 & 600 & 2\t360 & 2\t240 & 200 \\
                  &CTR &&     & 1996 Sep 24 & KP36 & \d  & 2\t360 & 2\t240 & \d \\
                  &NE  &&     & 1997 Sep 22 & KP36 & 360 & 45 & 30 & 45 \\
                  &SE  &&     & 1997 Sep 22 & KP36 & 360 & 45 & 30 & 45 \\
                  &SW  &&     & 1997 Sep 22 & KP36 & 360 & 45 & 30 & 45 \\
                  &NW  &&     & 1997 Sep 22 & KP36 & 360 & 45 & 30 & 45 \\
                  &CTR &&     & 1997 Nov 11 & CT36 & 2\t1000 &300 &200 &300 \\
                  &CTR &&     & 1999 Jun 13 & CT36 & 900 & 45 & 30 & 45 \\ 
NGC 7099  (M30)   &CTR &14.64 &0.03 & 1997 May 31 & CT36 & 2\t200 & 2\t30 & 2\t20 & 2\t30 \\
                  &CTR &&     & 1997 Aug 4  & CT36 & 200 & 30 & 20 & 30 \\
                  &CTR &&     & 1997 Aug 10 & CT36 & 200 & 30 & 20 & 30 \\
                  &NE  &&     & 1997 Nov 7  & CT36 & 200 & 30 & 20 & 30 \\
                  &NE  &&     & 1997 Nov 9  & CT36 & 200 & 30 & 20 & 30 \\
                  &NW  &&     & 1997 Nov 9  & CT36 & 200 & 30 & 20 & 30 \\
                  &SE  &&     & 1997 Nov 9  & CT36 & 200 & 30 & 20 & 30 \\
                  &SW  &&     & 1997 Nov 10 & CT36 & 200 & 30 & 20 & 30 \\
Pal 12            &CTR &16.46 &0.02 & 1995 Oct 21 & CT60 & 180 & 60 & 60 & 60 \\
                  &CTR &&     & 1995 Oct 21 & CT60 & \d  & \d &300 &300 \\
Pal 13            &CTR &17.23 &0.05 & 1996 Sep 21 & KP36 & 900 &180 &150 &300 \\
NGC 7492          &CTR &17.10 &0.00 & 1997 Aug 4  & CT36 & 2\t600 & 2\t100 & 2\t100 & 2\t120 \\
\noalign{\vskip0.1in}
\multicolumn{9}{c}{Magellanic Clouds Old Globular Clusters} \\
\noalign{\vskip0.1in}
NGC 121           &CTR &18.65 &0.05 & 1997 Aug 9  & CT36 & 2\t600 & 2\t90 & 2\t60 & 2\t90 \\
                  &CTR &&     & 1998 Aug 19 & CT36 & 2\t900 &240 &180 &300 \\
NGC 1466          &CTR &18.78 &0.09 & 1998 Aug 27 & CT36 & 900    &120 & 90   &150 \\
                  &CTR &&     & 1999 Aug 26 & CT36 & 2\t900 &240 &180 &300 \\
NGC 1841          &CTR &19.12 &0.20 & 1997 Nov 11 & CT36 & 2\t600 &180 &120 &180 \\
NGC 1835          &CTR &18.75 &0.08 & 1997 Nov 11 & CT36 & 2\t600 &180 &120 &180 \\
NGC 1916          &CTR &18.93 &0.14 & 1998 Apr 22 & CT36 & 2\t600 &180 &120 &180 \\
NGC 2019          &CTR &18.87 &0.12 & 1997 Nov 10 & CT36 & 2\t600 &180 &120 &180 \\
NGC 2210          &CTR &18.78 &0.09 & 1998 Apr 22 & CT36 & 600    &180 &120 &180 \\
Hodge 11          &CTR &18.72 &0.07 & 1995 Jan 30 & CT60 & 180    & 30 & 30 & 60 \\
NGC 2257          &CTR &18.50 &0.00 & 2001 Mar 28 & CT36 & 2\t900 &240 &180 &300 \\
\enddata
\tablenotetext{a}{Field codes: CTR = pointing centered on cluster; if listed fields are NW, NE, SW, and SE, this indicates a $2\times2$ mosaic; if N, S, W, and E are also listed, this indicates a $3\times3$ mosaic. ROA 24 = pointing centered on this yellow PAGB star in \oCen.}
\tablenotetext{b}{$V$-band distance modulus and interstellar reddening, taken from H10 for Galactic clusters; see text for sources for Magallanic Cloud clusters.}
\tablenotetext{c}{Telescope codes: KP4 = KPNO Mayall 4-m; KP36 =
KPNO 0.9-m; CT60 = CTIO 1.5-m; CT36 = CTIO 0.9-m}
\end{deluxetable*}
}

\clearpage

\onecolumngrid

\section{Figure sets}

Figures 20 and 21 show examples of the CMDs and CDDs for the clusters observed in this program. The complete set of figures will be available with the published journal paper.

\begin{figure*}
\figurenum{20}
\plotone{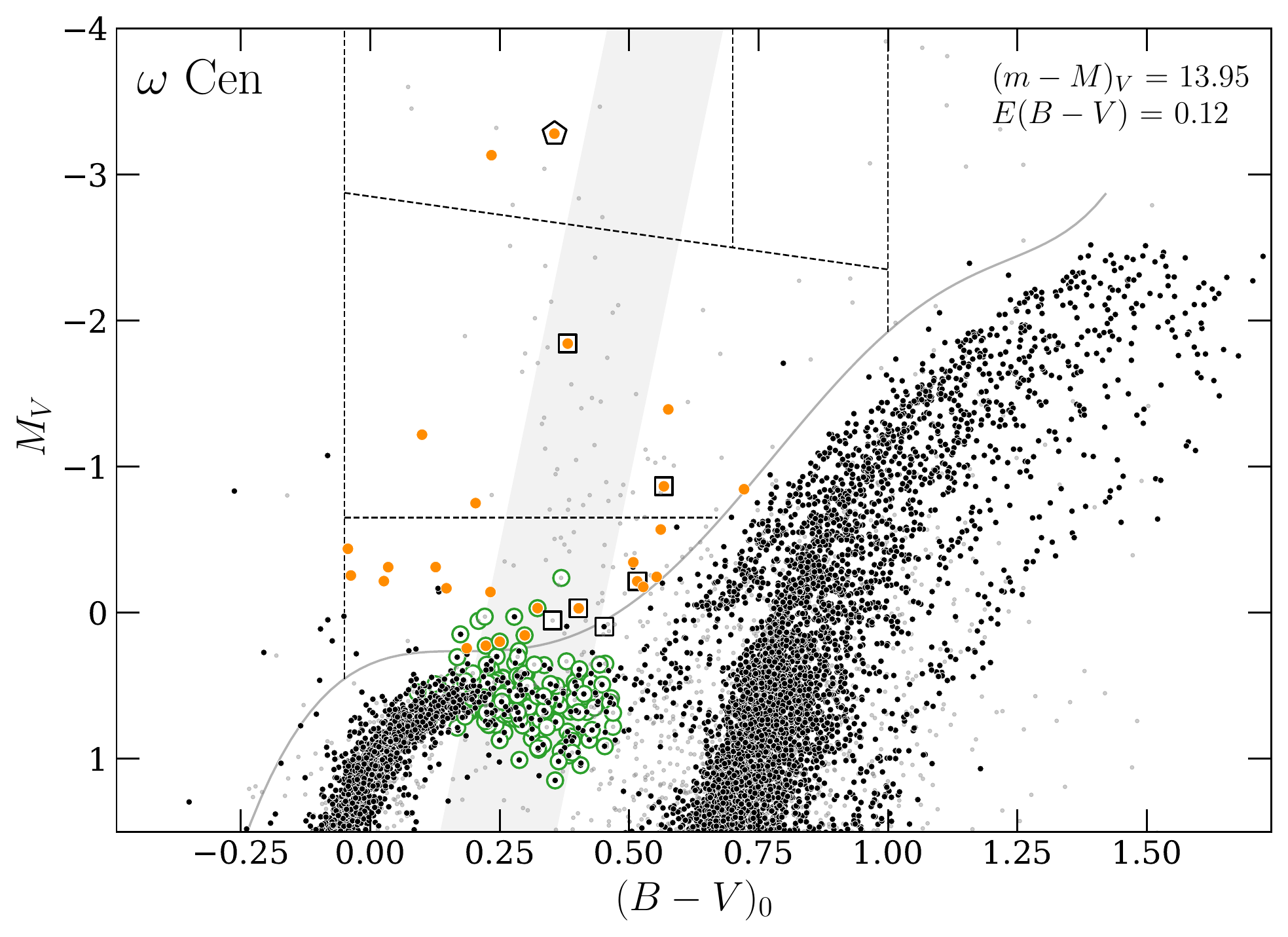}
\caption{Color-magnitude diagram for $\omega$ Cen. Black points show stars with membership probabilities $> 0.8$, while the gray points denote field stars. Orange points show those member stars that passed all of our AHB search criteria. RR~Lyrae variables are encircled in green, with black squares and black pentagons enclosing the Type~II Cepheids and RV Tauri stars, respectively. The AHB classification boxes from Figure~\ref{fig:classification} are marked with dashed lines and the continuous curve at the bottom. The adopted $V$-band distance modulus and reddening from H10 are indicated in the legend.}
\label{fig:CMDs}
\end{figure*}

\begin{figure*}
\figurenum{21}
\plotone{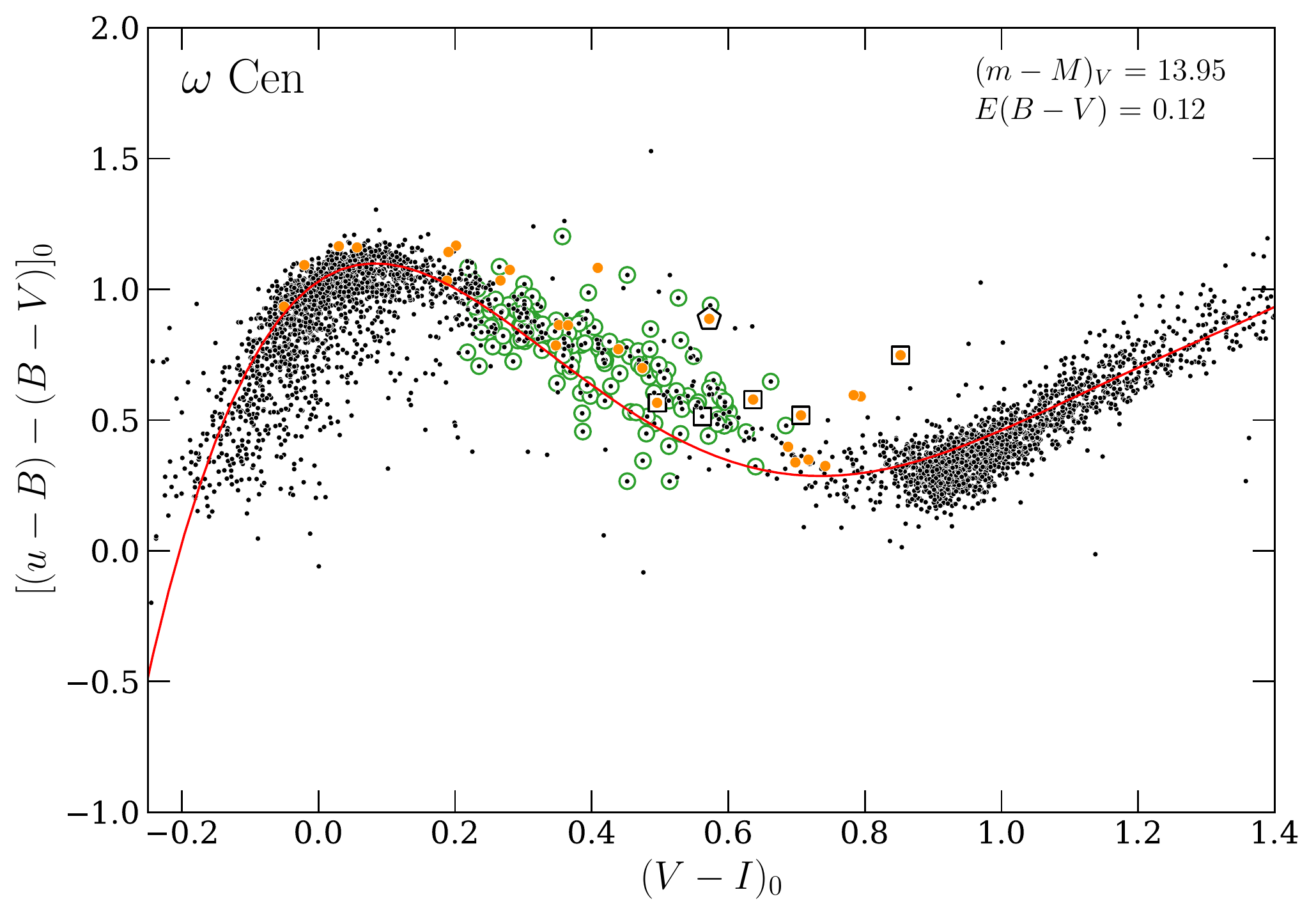}
\caption{Color-difference diagram for $\omega$~Cen. Black points show stars on and brighter than the horizontal branch, with membership probabilities $> 0.8$ (see text \S\ref{subsec:cluster_diagrams}).  Orange points show those members that passed all of our AHB search criteria. RR~Lyrae variables are encircled in green, with black squares and black pentagons enclosing the Type~II Cepheids and RV~Tauri variables, respectively. The red curve marks our template fit from Equation~\ref{eq:cdd}. The adopted $V$-band distance modulus and reddening from H10 are indicated in the legend.}
\label{fig:CDDs}
\end{figure*}

\clearpage

\twocolumngrid

\bibliography{ahb.bib}

\end{document}